\documentclass[aps, prd, reprint, showpacs, preprintnumbers, twocolumn, nofootinbib, superscriptaddress]{revtex4-1}
\pdfoutput=1 
\usepackage{amssymb,amsmath,hyperref}
\usepackage{color}
\usepackage{graphicx}
\usepackage{bm}

\newcommand{\pol}{\varepsilon}  
\newcommand{\Dic}{\text{Dic}}  
\newcommand{\C}{\text{C}}

\begin{document}

\title{Helicity operators for mesons in flight on the lattice}
\date{4 August 2011}
\author{Christopher E. Thomas}
\email{thomasc@jlab.org}
\affiliation{Jefferson Laboratory, 12000 Jefferson Avenue, Newport News, VA 23606, USA}
\author{Robert G. Edwards}
\affiliation{Jefferson Laboratory, 12000 Jefferson Avenue,  Newport News, VA 23606, USA}
\author{Jozef J. Dudek}
\affiliation{Jefferson Laboratory, 12000 Jefferson Avenue,  Newport News, VA 23606, USA}
\affiliation{Department of Physics, Old Dominion University, Norfolk, VA 23529, USA}
\collaboration{for the Hadron Spectrum Collaboration}

\preprint{JLAB-THY-11-1368}

\begin{abstract}
Motivated by the desire to construct meson-meson operators of definite relative momentum in order to study resonances in lattice QCD, we present a set of single-meson interpolating fields at non-zero momentum that respect the reduced symmetry of a cubic lattice in a finite cubic volume.  These operators follow from the subduction of operators of definite helicity into irreducible representations of the appropriate little groups.  We show their effectiveness in explicit computations where we find that the spectrum of states interpolated by these operators is close to diagonal in helicity, admitting a description in terms of single-meson states of identified $J^{PC}$.  The variationally determined optimal superpositions of the operators for each state give rapid relaxation in Euclidean time to that state, ideal for the construction of meson-meson operators and for the evaluation of matrix elements at finite momentum.
\end{abstract}

\maketitle

\section{Introduction}
\label{sec:intro}

Hadron spectroscopy is concerned with the study of resonances which appear as poles of hadron scattering amplitudes. These scattering amplitudes are ultimately determined by the underlying theory of quarks and gluons known as QCD. Our best current general-purpose technique for computation of quantities in QCD is lattice QCD, which considers the field theory numerically on a grid of Euclidean space-time points of finite extent. Scattering amplitudes are not directly accessible in the Euclidean theory, but computations of the discrete spectrum of states in a finite volume offer a way to access scattering amplitudes through a formalism presented by L\"uscher\cite{Luscher:1990ux}. Within this framework, narrow meson resonances appear through admixtures of localised single-hadron states, which are interpolated well from the vacuum by local $q\bar{q}$-like operators, and meson-meson states having definite relative momentum determined by the finite-volume boundary conditions.  

In order to accurately extract a complete finite-volume meson spectrum, it is necessary to include in a basis of interpolating fields not just local operators resembling single-hadron states, but also some which resemble pairs of mesons projected into definite relative momentum.  Our desire then is to form operators that can interpolate the meson-meson pair $M_A\, M_B$ which will be of the form ${\cal O}_{M_A}(\vec{p}) {\cal O}_{M_B}(\vec{p}\,')$, where ${\cal O}_{M_A}(\vec{p})$ efficiently interpolates meson $M_A$ with momentum $\vec{p}$ (and similarly ${\cal O}_{M_B}(\vec{p}\,')$ efficiently interpolates meson $M_B$). An optimal operator for $M_A$ with momentum $\vec{p}$ follows from variational diagonalisation of a matrix of single-hadron correlators evaluated with a basis of suitable operators. The subsequent reduction of the contribution of excited $M_A^\star$ mesons to the $M_A M_B$ correlators will mean that the relevant energy levels, and hence signals for scattering, can be extracted at earlier Euclidean times where statistical noise is smaller. 

This suggests that we need to develop a basis of single-hadron interpolating fields having non-zero momentum. In lattice QCD this cannot be achieved by simply Lorentz boosting operators at rest, as one is required to take into account the reduced symmetry of a discrete grid of finite extent. The lattice discretisation breaks rotational symmetry at small distances, while the finite volume breaks rotational symmetry at large distances.  The latter is of fundamental importance for states at non-zero momentum because the allowed momenta are determined by the boundary conditions which implement the symmetry of the ``box".  The relationship between the discrete finite-volume energy spectrum and the infinite volume scattering amplitudes is also determined by the symmetry of the boundary.  In most lattice QCD calculations, including those presented in this paper, both the lattice discretisation and the box have the same cubic symmetry, but this need not necessarily be so.  For mesons in flight, the symmetry is further reduced to the \emph{little group} of allowed cubic rotations that leave the meson momentum invariant and it is operators transforming irreducibly under this reduced symmetry group that we will derive.

As well as their use in constructing meson-meson operators, in-flight meson interpolators play a significant role in computations of meson matrix elements of certain currents, such as those used to measure electromagnetic and weak form-factors.  Including operators which interpolate mesons at non-zero momentum increases the number of kinematic points available, allowing the form factor to be calculated at more $Q^2$ values.  These quantities can reveal information about the quark-gluon structure of hadrons~\cite{Dudek:2009kk,Collins:2011mk}.

We will build on recent success\cite{Dudek:2010wm} in constructing a basis of meson interpolators at rest that transform irreducibly under the cubic symmetry of the lattice.  It was found that on the dynamical lattices of Refs.~\cite{Edwards:2008ja,Lin:2008pr}, the effect of the finite lattice spacing was relatively small; operators \emph{subduced} from definite continuum spin into irreducible representations of the octahedral group overlapped with states in a manner compatible with an effectively restored rotational symmetry.  The observed spectrum of states was explained in terms of single-meson states of determined $J^{PC}$ with $J_z$ values subduced across cubic irreps, and, as expected for a calculation featuring only fermion-bilinear operators, there was no clear observation of a spectrum of meson-meson states whose distribution across irreps is determined by the symmetry of the finite-volume boundary conditions. 

In this article we follow a similar philosophy, constructing fermion-bilinear operators for mesons in flight which, having a clear interpretation in an infinite volume continuum, allow the continuum quantum numbers of single-hadron states to be identified.  We then demonstrate the effectiveness of these constructions by using them to extract an extensive spectrum of isovector mesons in flight at various momenta, with the continuum spins and parities reliably identified.  With the extracted spectrum viewed in the finest detail we will observe effects that may be compatible with the admixture of meson-meson states.  We propose that by including meson-meson operators constructed from products of the in-flight operators derived herein, in future calculations we will be able to resolve this admixture precisely and extract meson-meson scattering information.  While Ref.~\cite{Luscher:1990ux} considered only the case of the total system at rest, it has been shown that more information can be gained from calculations that work in a moving frame\cite{Rummukainen:1995vs, Feng:2010es}.

\vspace{.2cm}

We begin in Section \ref{sec:ops} by considering an infinite volume continuum, constructing helicity operators and discussing operator-state overlaps.  In Section \ref{sec:finitevol} we move to a finite volume, showing how these operators subduce into lattice little group irreducible representations to give subduced helicity operators.  After giving a brief description of the lattice data sets used and the computational method in Section \ref{sec:lattices}, in Section \ref{sec:spin_assign} we describe the method for identifying the $J^{PC}$ of extracted states and show some examples of its application.  The extracted spin-identified spectra of mesons in flight are given in Section \ref{sec:results} and we conclude in Section \ref{sec:conclusions}.

\section{Mesons in flight in an infinite volume}
\label{sec:ops}

\subsection{Helicity operators for mesons in flight}
\label{sec:helops}

We begin by describing the construction of \emph{helicity operators} which enable the energies of mesons in flight to be extracted in an infinite volume continuum, before considering the overlap of these operators with states.  In Section \ref{sec:finitevol} we will show how these operators can be subduced into the irreducible representations relevant for a cubic lattice in a finite cubic volume, and how, because they have a definite infinite volume interpretation, they enable the identification of the spin and parity of states.  

As a starting point, consider the fermion-bilinear operators constructed in Ref.~\cite{Dudek:2010wm}, referring to that reference for more details,
\begin{multline}
\mathcal{O}^{J,M}(\vec{p}) \sim \sum_{m_1,m_2,m_3,\ldots} \text{CGs}(m_1,m_2,m_3,\ldots)  \sum_{\vec{x}} e^{i \vec{p} \cdot \vec{x}} \\ \times \bar{\psi}(\vec{x},t) \Gamma_{m_1} \overleftrightarrow{D}_{m_2}\overleftrightarrow{D}_{m_3} \ldots \psi(\vec{x},t)  ~.
\label{equ:restops}
\end{multline}
Here $\overleftrightarrow{D} \equiv \overleftarrow{D} - \overrightarrow{D}$ is a gauge-covariant derivative, $\Gamma$ is any product of Dirac gamma matrices, $\psi$ is a quark field (smeared using distillation in our implementation) and spin, flavour and colour indices have been suppressed for clarity.  As described in \cite{Dudek:2010wm}, the vector-like gamma matrices and derivatives are expressed in a circular basis and then coupled together using standard $SU(2)$ Clebsch Gordan coefficients (represented by ``CGs'' in Eq.~\ref{equ:restops}) to form the operator $\mathcal{O}^{J,M}(\vec{p})$.  However, for consistency between the operator constructions below and conventional definitions for helicity states, here we use a slightly different circular basis compared to that reference,
\begin{eqnarray}
\overleftrightarrow{D}_{\pm1} &=& i \sum_{i} \pol^*_{i}(\vec{0},\pm1) \overleftrightarrow{D}_i = \mp \frac{i}{\sqrt{2}} \left(\overleftrightarrow{D}_x \mp i \overleftrightarrow{D}_y \right)   ~, \nonumber \\
\overleftrightarrow{D}_{0} &=& i \sum_{i} \pol^*_{i}(\vec{0},0) \overleftrightarrow{D}_i = i \overleftrightarrow{D}_z  ~,
\end{eqnarray}  
and similarly for the vector-like gamma matrices.  Here $\pol_{i}(\vec{0},m)$ is the polarisation vector, given in Appendix \ref{app:polvectors}, for a spin-one meson at rest with spin $z$-component $m$.

With full continuum rotational symmetry and $\vec{p}=\vec{0}$, these operators have definite spin ($J$), spin $z$-component ($M$) and parity, $P$ (determined by the choice of $\Gamma$ and the number of derivatives), i.e. they only overlap with states having these quantum numbers.  In addition, if $\psi$ and $\bar{\psi}$ correspond to quarks of the same flavour, these operators have definite charge conjugation parity $C$ (determined by the choice of $\Gamma$, the number of derivatives and how these derivatives are coupled together), generalising to $G$-parity as appropriate.  In the continuum the overlap of these operators with states is given by
\begin{equation}
\bigl<0 \bigl| \mathcal{O}^{J,M}(\vec{p}=\vec{0}) \bigr| \vec{p}=\vec{0}; J', M' \bigr> = Z^{[J]} \delta_{J,J'} \delta_{M,M'} ~,
\label{equ:restoverlap}
\end{equation}
where the conserved parity and charge conjugation parity quantum numbers have been suppressed.

At non-zero momentum, the spin $z$-component $M$ is not a good quantum number unless the momentum is directed along the $z$-axis; in general, Eq.~\ref{equ:restoverlap} is not applicable to $\vec{p} \neq \vec{0}$.  It is more convenient to consider operators with definite helicity, the projection of the spin component along the direction of $\vec{p}$.  In Appendix \ref{app:helicity} we show that, in analogy to the transformation from basis states with definite spin $z$-component to helicity states, \emph{helicity operators} can be constructed by
\begin{equation}
\mathbb{O}^{J,\lambda}(\vec{p}) = \sum_{M} \mathcal{D}^{(J)*}_{M \lambda}(R) ~ \mathcal{O}^{J,M}(\vec{p})  ~,
\label{equ:helicityops}
\end{equation}
where $\mathbb{O}^{J,\lambda}$ is a helicity operator with helicity $\lambda$, $\mathcal{D}^{(J)}_{M \lambda}(R)$ is a Wigner-$\mathcal{D}$ matrix, and $R$ is the (active) transformation that rotates $(0,0,|\vec{p}|)$ to $\vec{p}$.  Note that in general there are an infinite number of $R$ which rotate $(0,0,|\vec{p}|)$ to $\vec{p}$.\footnote{If $(\phi,\theta,\psi)$ represents a rotation around the $z$-axis by $\psi$ followed by a rotation around the $y$-axis by $\theta$ and finally a rotation around the $z$-axis by $\phi$, one convention is $R=(\phi, \theta, 0)$~\cite{Chung:spin} and another, the Jacob-Wick convention~\cite{Jacob:1959at}, is $R=(\phi, \theta, -\phi)$.  Including an additional arbitrary initial rotation $\psi$ will still give a rotation from $(0,0,|\vec{p}|)$ to $\vec{p}$.  Different conventions will lead to different phases in the definition of states and operators.}  There is some subtlety in the appropriate choice of $R$ when the symmetry is reduced (e.g. on a finite volume lattice with a finite lattice spacing) and we discuss this below in Section \ref{sec:finitevol}.

\hfill

An equivalent way to construct these operators would be to choose an initial circular basis for the vector-like gamma matrices and derivatives defined in terms of the component of spin along $\vec{p}$ instead of the spin $z$-component.  These vectors could then be coupled together using the standard $SU(2)$ Clebsch Gordan coefficients as described above.  This emphasises that by constructing helicity operators we've just effected a basis change.  As will be illustrated below, the resulting basis is much more convenient for studying mesons with non-zero momenta because it respects the symmetries of a system with a meson in flight and helicity is a good quantum number (in an infinite volume continuum).

\subsection{Helicity operator overlaps}
\label{sec:overlaps}

Here we discuss the overlap of helicity states onto the helicity operators constructed above, $\bigl<0 \bigl| \mathbb{O}^{J,P,\lambda}(\vec{p}) \bigr| \vec{p}; J'^{P'},\lambda' \bigr>$, still considering an infinite volume continuum.

The overlaps for operators with $\vec{p} = \vec{0}$ have a simple form, Eq.~\ref{equ:restoverlap}, $J'=J$, $P'=P$ and $\lambda'=\lambda$.  These constraints arise from the rotational symmetry in 3 spatial dimensions of a system containing a particle at rest, and no further constraints are gained by imposing Lorentz symmetry\footnote{Although Lorentz symmetry constraints would relate operators containing the temporal and spatial pieces of Lorentz vectors, e.g. $\gamma^0$ and $\gamma^i$.}.  This can be seen explicitly by comparing the Lorentz covariant parameterisations given in Appendix A of Ref.~\cite{Dudek:2007wv} and the 3-rotation covariant parameterisations given in Appendix \ref{app:decompositions} herein.  However, at finite momentum, 3-rotation symmetry, a subgroup of full Lorentz symmetry, provides less stringent constraints, something that can again be seen explicitly by comparing those two different parameterisations.  Since in the operator constructions we treat space and time asymmetrically, we will consider in the detail the restricted 3-rotational symmetry constraints.

Using only the constraints arising from 3-rotation symmetry, the overlap of a state of definite $J^P$ ($P,P'$ are the parities at rest) and helicity onto a helicity operator is given by
\begin{equation}
\bigl<0 \bigl| \mathbb{O}^{J,P,\lambda}(\vec{p}) \bigr| \vec{p}; J'^{P'},\lambda' \bigr> = Z^{[J,J',P,P',\lambda]} \delta_{\lambda,\lambda'}  ~.
\label{equ:helopoverlap}
\end{equation}
The helicity of the state, $\lambda'$, is constrained to be the same of that of the operator, $\lambda$, but an operator constructed to have integer spin $J$ at rest can in general overlap onto states of any integer spin $J'$ when boosted to non-zero momentum.  This is because at non-zero momentum the 3-dimensional rotational group is broken to a subgroup, called the little group, made up of the rotations and reflections which leave the momentum, $\vec{p}$, invariant.  In an infinite volume continuum the little group is U(1)~\cite{Moore:2005dw,Moore:2006ng}, the group of rotations and reflections in two dimensions, regardless of the momentum direction.  The irreducible representations of this little group are labelled by the magnitude of helicity (and also, for $\lambda=0$, a `parity', $\tilde{\eta}$, described below): $|\lambda|^{\tilde{\eta}} = 0^+, 0^-, \frac{1}{2}, 1, \frac{3}{2}, 2, \frac{5}{2}, 3, \dots$.  Apart from the two one-dimensional irreps with $\lambda=0$, these irreps are all two-dimensional.  Another consequence of this reduced symmetry is that an operator with $|\lambda| \neq 0$ can overlap with states of both parities, and the overlap factors, $Z$, can depend on $\lambda$.  In Appendix \ref{app:decompositions} we give the most general parameterisations of these operator-state overlaps and these features can be seen explicitly.  Note that any flavour quantum number, such as charge conjugation parity or its generalisation, is still a good quantum number if it is so at rest.

Helicity states at non-zero momentum are \emph{not} eigenstates of parity because a parity transformation, $\hat{P}$, reverses the direction of the momentum, $\vec{p} \rightarrow -\vec{p}$.  A reflection in a plane containing the momentum direction, $\hat{\Pi}$ (a parity transformation followed by a rotation to bring the momentum direction back to the original direction), preserves $\vec{p}$.  However, helicity states are also in general \emph{not} eigenstates of $\hat{\Pi}$ because under such a transformation $\lambda \rightarrow -\lambda$.  If we consider a state with $\vec{p} = \vec{p_z}$ along the $z$-axis and a reflection in the $yz$ plane ($x \rightarrow -x$), $\hat{\Pi}_{yz}$, then we have (Appendix \ref{app:parity})
\begin{equation}
\hat{\Pi}_{yz} \left|\vec{p_z}; J^P, \lambda \right> = \tilde{\eta} \left|\vec{p_z}; J^P, -\lambda \right>  \nonumber ~,
\end{equation}
where $J^P$ are the quantum numbers of the state at rest and $\tilde{\eta} \equiv P(-1)^J$.  It can be seen that for $\lambda = 0$ the helicity states \emph{are} eigenstates of $\hat{\Pi}$ with eigenvalue $\tilde{\eta}$ and therefore\footnote{and because $\lambda=0$ there is no additional phase depending on the choice of reflection plane (Appendix \ref{app:parity})}
\begin{eqnarray}
\bigl<0 \bigl| \mathbb{O}^{J,P,\lambda=0}(\vec{p}) \bigr| \vec{p}; J'^{P'},\lambda' \bigr> =  \nonumber \\ 
 Z^{[J,J',P,P',\lambda=0]} \delta_{\tilde{\eta},\tilde{\eta}'} \delta_{\lambda',0} ~.
\label{equ:parityoverlap}
\end{eqnarray}
For $\lambda \neq 0$ the helicity states are not eigenstates of $\hat{\Pi}$ \footnote{although we can form eigenstates by taking linear combinations, $\sim |\vec{p}; J, \lambda\rangle \pm \tilde{\eta}|\vec{p}; J, -\lambda\rangle$, they are of limited use here.}.

\hfill

If we had started with Lorentz 4-vector gamma matrices and derivatives instead of 3-vectors, we could have constructed operators which only overlap onto states with one $J^P$ at non-zero momentum.  Performing a Lorentz boost on the state and operator on the left-hand side of Eq.~\ref{equ:restoverlap} would leave the right-hand side unchanged, or putting it slightly differently, we can always boost back to the rest frame where Eq.~\ref{equ:restoverlap} is valid.  However, our use of only spatial derivatives precludes us from doing this.  We do not include temporal derivatives in our operator constructions because in distillation the quark fields in the operators are only smeared in the spatial directions.  This means that we can not form a lattice-discretised temporal derivative which is simply a rotation of a lattice-discretised spatial derivative.  In addition, we have used anisotropic lattices and smeared differently in the temporal direction in the action, and these lead to similar problems.  If we naively included temporal derivatives, these would be on a different footing to the spatial derivatives and so the resulting operators would not be Lorentz covariant and would overlap onto more than one $J^P$ at non-zero momentum.

Although our operators (and hence the operator-state overlaps) are not Lorentz covariant because they have been built from 3-vectors, in a theory with Lorentz symmetry there are still more stringent constraints on the overlaps than those given in Eq.~\ref{equ:helopoverlap}.  In Appendix \ref{app:decompositions} we show the restrictions on state-operator overlaps which arise from the constraints of a Lorentz symmetric theory applied to our helicity operators.  The $J^P$ of the states with which the operator has non-zero overlap depends on which gamma matrices and derivatives the operator is constructed from and how these have been coupled together.  An operator which at rest overlaps with only one $J^P$ can in general overlap with states of many $J^P$ at non-zero momentum, but the set of allowed $J^P$ is reduced by Lorentz symmetry constraints compared to the infinite set allowed by 3-rotation symmetry, Eq.~\ref{equ:helopoverlap}.

\hfill

To illustrate the above points, consider a simple example where the fermion bilinear operator consists of a spatial gamma matrix in the Cartesian basis and there are no derivatives.  Projected onto zero momentum, this operator, $\mathcal{O}_i = \bar{\psi} \gamma_i \psi$, has non-zero overlap with only the vector state, $J^{PC} = 1^{--}$,
\begin{equation}
\bigl<0 \bigl| \mathcal{O}_i(\vec{p}=\vec{0}) \bigr| \vec{p}=\vec{0}; 1^{--}(M) \bigr> = Z \pol_i(\vec{p}=\vec{0},M) ~,
\end{equation}
where $\pol_i$ is the polarisation vector of the state and $M$ is its spin $z$-component.  Now consider a non-zero momentum along the positive $z$-axis\footnote{chosen to be along the $z$-axis so that the helicity state/operator is the same $J_z$ state/operator}, $\vec{p_z} = (0,0,p)$.  In a theory with only 3-rotation symmetry and \emph{not} full Lorentz symmetry, the state-operator overlaps are given in the last column of Table \ref{table:paramoverlaps:1} in Appendix \ref{app:decompositions}, i.e. the operator has non-zero overlap with states of all integer $J$.  For states with $J=0,1$ the overlaps are\footnote{note that the overlap onto the $M=0$ piece of the $1^{+-}$ vanishes, a consequence of Eq.~\ref{equ:parityoverlap}}
\begin{eqnarray}
\bigl<0 \bigl| \mathcal{O}_i(\vec{p_z}) \bigr| \vec{p}=\vec{p_z}; 0^{+-} \bigr> &=& Z_0 p_i ~,  \nonumber \\
\bigl<0 \bigl| \mathcal{O}_i(\vec{p_z}) \bigr| \vec{p}=\vec{p_z}; 1^{--}(M) \bigr> &=& Z_1 \pol_i + Z'_1 (\pol^j p_j) p_i ~,  \nonumber \\
\bigl<0 \bigl| \mathcal{O}_i(\vec{p_z}) \bigr| \vec{p}=\vec{p_z}; 1^{+-}(M) \bigr> &=& Z_2 \epsilon_{ijk} p^j \pol^k ~.  \nonumber 
\end{eqnarray}

However, in a theory with Lorentz symmetry most of these overlaps are constrained to be zero.  For example, there are no Lorentz covariant structures that reduce to the $Z'_1$ [$(\pol^j p_j) p_i$] or $Z_2$ [$\epsilon_{ijk} p^j \pol^k$] terms above.  In this case the operator has non-zero overlap with only two states (shown in the column labelled ``L1'' in the aforementioned table),
\begin{eqnarray}
\bigl<0 \bigl| \mathcal{O}_i(\vec{p_z}) \bigr| \vec{p}=\vec{p_z}; 0^{+-} \bigr> &\propto& p_i ~,  \nonumber \\
\bigl<0 \bigl| \mathcal{O}_i(\vec{p_z}) \bigr| \vec{p}=\vec{p_z}; 1^{--}(M) \bigr> &\propto& \pol_i(\vec{p},M) ~.  \nonumber 
\end{eqnarray}

If instead we consider the Lorentz covariant operator $\mathcal{O}_{\mu} = \bar{\psi} \gamma_{\mu} \psi$, the non-zero overlaps are
\begin{eqnarray}
\bigl<0 \bigl| \mathcal{O}_{\mu}(\vec{p_z}) \bigr| \vec{p}=\vec{p_z}; 0^{+-} \bigr> &\propto& p_{\mu} ~,  \nonumber \\
\bigl<0 \bigl| \mathcal{O}_{\mu}(\vec{p_z}) \bigr| \vec{p}=\vec{p_z}; 1^{--}(M) \bigr> &\propto& \pol_{\mu}(\vec{p},M) ~.  \nonumber 
\end{eqnarray}
In this case we can take linear combinations of the four components to project onto just one of the states: taking the inner product with $p^{\mu}$ projects onto the $0^{+-}$ state, whereas taking the inner product with $\pol^{*\mu}(\vec{p},M')$ projects onto the $M'=M$ component of the $1^{--}$ state only.

In summary, although simplified somewhat by using helicity operators, the pattern of operator-state overlaps at non-zero momentum is more complicated than at zero momentum.  In particular, because we construct operators out of 3-vectors rather than Lorentz 4-vectors, we can not in general form operators that have non-zero overlap with states of only one $J^P$.  However, as we shall show in Section \ref{sec:spin_assign}, the remaining constraints prove to be enough for us to be able to identify the states' spins and parities.  In the following section we consider complications arising from a finite cubic volume.

\section{Mesons in flight and helicity operators on the lattice}
\label{sec:finitevol}

In general, the symmetry of the lattice discretisation (which breaks rotational symmetry at small distances) need not be the same as the symmetry of the finite volume, the boundary conditions (which break rotational symmetry at large distances).  However, in this work we only consider a cubic lattice in a finite cubic box with periodic boundary conditions, and so both the lattice and the boundary have the same symmetry.  In this case, the full rotational symmetry of the continuum is broken to the (double cover) of the octahedral group, $\text{O}_h^D$, equivalent to the (double cover) of the symmetry group of the cube.  Spin is no longer a good quantum number and states are not classified by $J$ but instead by the irreducible representations, \emph{irreps} ($\Lambda$), of $O_h^D$.  In Refs.~\cite{Dudek:2009qf,Dudek:2010wm}, it was found that on the dynamical lattices of Refs.~\cite{Edwards:2008ja,Lin:2008pr}, the effect of the finite lattice spacing was relatively small.  Operators with definite continuum spin (Eq.~\ref{equ:restops}) were subduced into octahedral group irreps; the observed spectrum of states and operator-state overlaps were compatible with an effective restoration of rotational symmetry, and the spectrum was explained in terms of single-meson states of determined $J^{PC}$ with $J_z$ values subduced across these irreps.

In going from zero momentum to non-zero momentum the symmetry is broken further.  Whereas, in an infinite volume continuum the little group is the same regardless of the momentum direction, $\vec{p}$, in a finite volume the particular lattice little group depends on the \emph{star} of $\vec{p}$~\cite{Moore:2005dw}.  The allowed $\vec{p}$ are determined by the boundary conditions and the star of $\vec{p}$ is the set of all $\vec{p}$ related by allowed lattice rotations; as a shorthand, we refer to these different stars as different \emph{momentum types}.  Momenta are quantised by the periodic boundary conditions on the cubic box and we give all momenta in units of $\frac{2\pi}{L_s a_s}$; $a_s$ and $L_s$ are respectively the spatial lattice spacing and spatial extent of the lattice in lattice units.  The relevant little groups are given in Refs.~\cite{Moore:2005dw,Moore:2006ng}.  We summarise the allowed lattice momenta and the corresponding little groups and irreps in Table \ref{table:littlegroups}.

\begin{table}
\begin{tabular}{c|c|l}
\textbf{Lattice} & \textbf{Little Group} & \textbf{Irreps} ($\Lambda$ or $\Lambda^P$) \\
\textbf{Momentum} & (double cover) & (for single cover) \\
\hline
$(0,0,0)$ & $\text{O}_h^D$ & $A_1^{\pm}$, $A_2^{\pm}$, $E^{\pm}$, $T_1^{\pm}$, $T_2^{\pm}$ \\
$(n,0,0)$ & $\Dic_4$ & $A_1$, $A_2$, $B_1$, $B_2$, $E_2$ \\
$(n,n,0)$ & $\Dic_2$ & $A_1$, $A_2$, $B_1$, $B_2$ \\
$(n,n,n)$ & $\Dic_3$ & $A_1$, $A_2$, $E_2$ \\
$(n,m,0)$ & $\C_4$ & $A_1$, $A_2$ \\
$(n,n,m)$ & $\C_4$ & $A_1$, $A_2$ \\
$(n,m,p)$ & $\C_2$ & $A$ \\
\hline
\end{tabular}
\caption{Allowed lattice momenta on a cubic lattice in a finite cubic box, along with the corresponding little groups (the double covers relevant for integer and half-integer spin) from Ref.~\cite{Moore:2005dw,Moore:2006ng}.  We list only the single cover irreps relevant for integer spin.  Lattice momenta are given in units of $2\pi / (L_s a_s)$ where $n,m,p \in \mathbb{Z}^*$ are non-zero integers with $n \neq m \neq p$.  The $A$ and $B$ irreps have dimension one, $E$ two and $T$ three.  $\Dic_n$ is the dicyclic group of order $4n$.}
\label{table:littlegroups}
\end{table}

In analogy to Ref.~\cite{Dudek:2010wm}, we consider subduction coefficients, $\mathcal{S}_{\Lambda,\mu}^{\tilde{\eta},\lambda}$, which specify how the helicity, $\lambda$, subduces into a little group irrep $\Lambda$ (row $\mu=1 \ldots \text{dim}[\Lambda]$).  Using these we can construct a little group operator, a \emph{subduced helicity operator}, from a helicity operator:
\begin{equation}
\mathbb{O}^{[J,P,|\lambda|]}_{\Lambda,\mu}(\vec{p}) = \sum_{\hat{\lambda}=\pm|\lambda|} \mathcal{S}_{\Lambda,\mu}^{\tilde{\eta},\hat{\lambda}} \mathbb{O}^{J,P,\hat{\lambda}}(\vec{p}) ~,
\end{equation}
where $\tilde{\eta} \equiv P(-1)^J$ with $J$ and $P$ the spin and parity of the operator $\mathbb{O}^{J,P,\lambda}(\vec{p}=\vec{0})$.  The subduced helicity operators are different orthogonal combinations of the two signs of helicity, $+|\lambda|$ and $-|\lambda|$.  These subduction coefficients can be calculated using the group theoretic projection formula (for example, see Appendix A of Ref.~\cite{Dudek:2010wm}).

In Table \ref{table:subductions} we give these subduction coefficients\footnote{Note that each subduction coefficient could be multiplied by an arbitrary phase and it would still give a subduction from the helicity to the little group irrep.} for momenta of the form $(n,0,0)$, $(n,n,0)$ and $(n,n,n)$ ($n \in \mathbb{Z}^*$) for $|\lambda| \leq 4$.  More details of our conventions are given in Appendix \ref{app:latticerotations}.  For all these momenta, $\lambda^{\tilde{\eta}}$ = $0^+$ and $0^-$ subduce onto the $A_1$ and $A_2$ irreps respectively.  However, the other $\lambda$ contained in those irreps (i.e. which $\lambda$ can mix with $\lambda=0$) depends on the momentum type: $(n,0,0)$ ($\Dic_4$) also has $\lambda = \pm4,\dots$; $(n,n,0)$ ($\Dic_2$) has $\lambda = \pm2,\pm4,\dots$; $(n,n,n)$ ($\Dic_3$) has $\lambda = \pm3,\dots$.  Note that, although the pattern of subductions does not depend on conventions, the relative phases in subduction coefficients can be convention dependent, determined by the helicity operator construction, the rotations used (discussed below) and the particular representation matrices chosen for the two-dimensional irreps.

\begin{table}
\begin{tabular}{c|c|c|c}
\textbf{Group} & $|\lambda|^{\tilde{\eta}}$ & \textbf{$\Lambda(\mu)$} & \textbf{$\mathcal{S}_{\Lambda,\mu}^{\tilde{\eta},\lambda}$} \\
\hline
$\Dic_{4}$ 
 & $0^+$ & $A_1(1)$ & $1$ \\
$(n,0,0)$
 & $0^-$ & $A_2(1)$ & $1$ \\
 & $1$   & $E_2\left(\begin{smallmatrix}1 \\ 2\end{smallmatrix}\right)$ & $(\delta_{s,+} \pm \tilde{\eta} \delta_{s,-})/\sqrt{2}$ \\
 & $2$   & $B_1(1)$ & $(\delta_{s,+} + \tilde{\eta} \delta_{s,-})/\sqrt{2}$ \\
 & $2$   & $B_2(1)$ & $(\delta_{s,+} - \tilde{\eta} \delta_{s,-})/\sqrt{2}$ \\
 & $3$   & $E_2\left(\begin{smallmatrix}1 \\ 2\end{smallmatrix}\right)$ & $(\pm\delta_{s,+} + \tilde{\eta} \delta_{s,-})/\sqrt{2}$ \\
 & $4$   & $A_1(1)$ & $(\delta_{s,+} + \tilde{\eta} \delta_{s,-})/\sqrt{2}$ \\
 & $4$   & $A_2(1)$ & $(\delta_{s,+} - \tilde{\eta} \delta_{s,-})/\sqrt{2}$ \\
\hline
$\Dic_{2}$
 & $0^+$ & $A_1(1)$ & $1$ \\
$(n,n,0)$
 & $0^-$ & $A_2(1)$ & $1$ \\
 & $1$   & $B_1(1)$ & $(\delta_{s,+} + \tilde{\eta} \delta_{s,-})/\sqrt{2}$ \\
 & $1$   & $B_2(1)$ & $(\delta_{s,+} - \tilde{\eta} \delta_{s,-})/\sqrt{2}$ \\
 & $2$   & $A_1(1)$ & $(\delta_{s,+} + \tilde{\eta} \delta_{s,-})/\sqrt{2}$ \\
 & $2$   & $A_2(1)$ & $(\delta_{s,+} - \tilde{\eta} \delta_{s,-})/\sqrt{2}$ \\
 & $3$   & $B_1(1)$ & $(\delta_{s,+} + \tilde{\eta} \delta_{s,-})/\sqrt{2}$ \\
 & $3$   & $B_2(1)$ & $(\delta_{s,+} - \tilde{\eta} \delta_{s,-})/\sqrt{2}$ \\
 & $4$   & $A_1(1)$ & $(\delta_{s,+} + \tilde{\eta} \delta_{s,-})/\sqrt{2}$ \\
 & $4$   & $A_2(1)$ & $(\delta_{s,+} - \tilde{\eta} \delta_{s,-})/\sqrt{2}$ \\
\hline
$\Dic_{3}$
 & $0^+$ & $A_1(1)$ & $1$ \\
$(n,n,n)$
 & $0^-$ & $A_2(1)$ & $1$ \\
 & $1$   & $E_2\left(\begin{smallmatrix}1 \\ 2\end{smallmatrix}\right)$ & $(\delta_{s,+} \pm \tilde{\eta} \delta_{s,-})/\sqrt{2}$ \\
 & $2$   & $E_2\left(\begin{smallmatrix}1 \\ 2\end{smallmatrix}\right)$ & $(\pm\delta_{s,+} - \tilde{\eta} \delta_{s,-})/\sqrt{2}$ \\
 & $3$   & $A_1(1)$ & $(\delta_{s,+} - \tilde{\eta} \delta_{s,-})/\sqrt{2}$ \\
 & $3$   & $A_2(1)$ & $(\delta_{s,+} + \tilde{\eta} \delta_{s,-})/\sqrt{2}$ \\
 & $4$   & $E_2\left(\begin{smallmatrix}1 \\ 2\end{smallmatrix}\right)$ & $(\delta_{s,+} \mp \tilde{\eta} \delta_{s,-})/\sqrt{2}$ \\
\hline
\end{tabular}
\caption{Subduction coefficients, $\mathcal{S}_{\Lambda,\mu}^{\tilde{\eta},\lambda}$, for $|\lambda| \leq 4$ with $s \equiv \text{sign}(\lambda)$; other notation is defined in the text.}
\label{table:subductions}
\end{table}

Instead of constructing little group operators via helicity operators which are then subduced into little group irreps, we could have projected directly from our continuum operators (in the $J_z$ basis) into these little group irreps, but the subduction coefficients would then in general depend on the momentum direction and not just the momentum type.  Alternatively, we could have worked only in terms of lattice irreps, subducing from octahedral group irreps to lattice little group irreps.  However, the method we have described makes it clear how the operators can be physically interpreted in the limit of an infinite volume continuum and, as we show below, enables the continuum quantum numbers of states to be identified.

\hfill

As noted above, in a finite volume there is some subtlety in the choice of the $R$ in Eq.~\ref{equ:helicityops} (out of an infinite number of possibilities) which rotates $(0,0,|\vec{p}|)$ (not necessarily an allowed lattice momentum) to $\vec{p}$ (an allowed lattice momentum).  In an infinite volume continuum the choice is not important as long as one convention is chosen because the two directions perpendicular to $\vec{p}$ are interchangeable; they are related by a rotation.  However, on a finite lattice the two transverse directions are \emph{not} in general interchangeable; they are not related by an allowed lattice rotation.  Consider, for example, the two (shortest) lattice vectors perpendicular to $(0,1,1)$, namely $(1,0,0)$ and $(0,1,-1)$ -- these have different lengths.  Therefore, if the $R$'s are not chosen appropriately there is the possibility of an inconsistency between correlators from different momentum directions within the same momentum type.  For example, if $R$ used for $(0,1,1)$ and $R'$ used for $(1,1,0)$ are not related by a lattice rotation, this could lead to effectively different definitions of the little group irreps and different bases for little group rows in $(0,1,1)$ compared to $(1,1,0)$; e.g. $B_1$ ($B_2$) from $(0,1,1)$ could correspond to $B_2$ ($B_1$) from $(1,1,0)$.  This would prove problematic if averaging over different momentum directions to increase statistics and when constructing multi-meson operators which contain a sum over momentum directions with various weights.  If $R$ and $R'$ are related by an allowed lattice rotation, these inconsistencies do not arise because there is no allowed lattice rotation that rotates from direction $(1,0,0)$ to direction $(0,1,-1)$, and so the different irreps or irrep rows cannot be mixed up.

We ensure consistency between different momentum directions by breaking down $R$ into two stages: $R = R_{\text{lat}} R_{\text{ref}}$.  First, $R_{\text{ref}}$, rotate from $(0,0,|\vec{p}|)$ to $\vec{p}_{\text{ref}}$, where $\vec{p}_{\text{ref}}$ is a reference direction for momenta of type $\vec{p}$ (i.e. for the star of $\vec{p}$).  Note that $R_{\text{ref}}$ is \emph{not} in general an allowed lattice rotation, but this is permissible because all we are effecting by this rotation is a basis transformation.  For $R_{\text{lat}}$ we choose a lattice rotation which rotates from $\vec{p}_{\text{ref}}$ to $\vec{p}$.  There are in general a finite number of possible choices of lattice rotations for $R_{\text{lat}}$; the particular choice is not important as long as we make the choice consistently for each momentum direction (for example when later constructing multi-meson operators).  More details of our implementation are given in Appendix \ref{app:latticerotations}.

To illustrate this, consider momentum directions with $p^2 = 2$ and choose $\vec{p}_{\text{ref}} = (0,1,1)$.  If we consider $\vec{p} = (0,1,1)$ then $R = R_{\text{ref}}$ is just a rotation from $(0,0,p)$ to $\vec{p}$.  On the other hand, if we consider $\vec{p} = (1,1,0)$ then $R = R_{\text{lat}} R_{\text{ref}}$ where $R_{\text{ref}}$ is that just described and $R_{\text{lat}}$ is a lattice rotation that rotates $(0,1,1)$ to $(1,1,0)$.

Now that we have described the construction of subduced helicity operators, in the next section we briefly describe some of the lattice and computational details, before explaining our method for determining the continuum spin of extracted states and giving some examples in the following section.

\section{Computational details}
\label{sec:lattices}

We use anisotropic dynamical Clover lattices with lattice parameters described in detail in Refs.~\cite{Edwards:2008ja,Lin:2008pr}, having a spatial lattice spacing, $a_s \sim 0.12~\text{fm}$, and a temporal lattice spacing approximately $\xi = a_s/a_t = 3.5$ times smaller, corresponding to $a_t^{-1} \sim 5.6~\text{GeV}$.  In Table \ref{table:datasets} we give details of the data sets used; in all cases we increase statistics by averaging over equivalent momentum directions within a given momentum type.  In the main body of results shown here, we use the $N_f=3$ data set with lattice size $L_s^3 \times T = 16^3 \times 128$ in lattice units.  This ensemble has three degenerate dynamical quark flavours, i.e. has SU(3) flavour symmetry, and corresponds to $m_{\pi} = m_K = m_{\eta} \approx 700$ MeV.  Some comparisons are performed using the lattices of different volumes and quark masses given in Table \ref{table:datasets}.  We consider isovector mesons and so only connected Wick contractions contribute.  

Correlators are constructed using the \emph{distillation} method~\cite{Peardon:2009gh} which, in combination with these anisotropic lattices, has proved fruitful in a number of applications \cite{Dudek:2009qf,Dudek:2010wm,Bulava:2010yg,Dudek:2010ew,Dudek:2011tt,Edwards:2011jj}.  For studying mesons in flight, an important aspect of distillation is that it allows a large basis of operators at the source and the sink, each projected onto a definite momentum.

\begin{table*}
\begin{tabular}{c|c|c|c|c|c|c|c|c}
\textbf{$m_{\ell}$} & \textbf{$m_{s}$} & \textbf{$m_{\pi} / \text{MeV}$} & \textbf{Volume} & $N_{\text{vecs}}$ & \textbf{Mom. Type} & \textbf{$N_{\text{mom}}(N_{\text{possible mom}})$} & \textbf{$N_{\text{cfgs}}$} & \textbf{$N_{t_{\text{srcs}}}$} \\
\hline
 -0.0743 & -0.0743 & 702 & $16^3 \times 128$ & 64 & (1,0,0) & 6(6)  & 200 & 3 \\
 -0.0743 & -0.0743 & 702 & $16^3 \times 128$ & 64 & (1,1,0) & 6(12) & 200 & 3 \\
 -0.0743 & -0.0743 & 702 & $16^3 \times 128$ & 64 & (1,1,1) & 8(8)  & 200 & 3 \\
 -0.0743 & -0.0743 & 702 & $16^3 \times 128$ & 64 & (2,0,0) & 6(6)  & 200 & 3 \\
\hline
 -0.0743 & -0.0743 & 702 & $20^3 \times 128$ & 128 & (1,0,0) & 6(6)  & 198 & 3 \\
\hline
 -0.0840 & -0.0743 & 396 & $20^3 \times 128$ & 128 & (1,0,0) & 6(6)  & 600 & 6 \\
 -0.0840 & -0.0743 & 396 & $20^3 \times 128$ & 128 & (1,1,0) & 6(12)  & 600 & 6 \\
\hline
 -0.0840 & -0.0743 & 396 & $24^3 \times 128$ & 162 & (1,0,0) & 6(6)  & 553 & 4 \\
 -0.0840 & -0.0743 & 396 & $24^3 \times 128$ & 162 & (1,1,0) & 6(12)  & 553 & 4 \\
\hline
\end{tabular}
\caption{The data sets used.  $m_{\ell}$ is the bare mass of the degenerate up and down quarks, and $m_s$ is the bare strange quark mass.  $N_{\text{vecs}}$ is the number of distillation vectors used, $N_{\text{mom}}$ is the number of momentum directions averaged over out of a possible $N_{\text{possible mom}}$, $N_{\text{cfgs}}$ is the number of configurations and $N_{t_{\text{srcs}}}$ the number of time-sources.}
\label{table:datasets}
\end{table*}

Our operator basis consists of all possible combinations of gamma matrices and zero, one or two lattice-discretised gauge-covariant derivatives, coupled together and subduced into lattice little group irreps, to form subduced helicity operators as described above.  The notation follows that of Ref.~\cite{Dudek:2010wm}: an operator $(\Gamma \times D^{[N]}_{J_D})^{J}_{\lambda}$ contains a gamma matrix, $\Gamma$, with the naming scheme given in Table \ref{table:gamma}, and $N$ derivatives coupled to spin $J_D$, altogether coupled to spin $J$; $\lambda$ refers to the helicity component.

\begin{table}
    \begin{tabular}{r|cccccccc}
         &      $a_0$ & $\pi$ &         $\pi_2$ &       $b_0$ &     $\rho$ &    $\rho_2$    &           $a_1$           & $b_1$\\ 
        \hline
        $\Gamma$ &  $1$ &   $\gamma_5$ &    $\gamma_0\gamma_5$ &    $\gamma_0$ &$\gamma_i$& $\gamma_i \gamma_0$ &   $\gamma_5 \gamma_i$ & $\epsilon_{ijk} \gamma_j \gamma_k$
    \end{tabular}
\caption{Gamma matrix naming scheme. \label{table:gamma}}
\end{table}

The number of operators in each irrep is given in Table \ref{table:numops}.  For the two-dimensional irreps we increase statistics by averaging the correlators over the two irrep rows.  We consider isovector mesons which are eigenstates of charge conjugation (or $G$-parity) and so the irreps are labelled by charge conjugation parity, $C$, as well as little group irrep, $\Lambda$.  Correlators were analysed using our implementation of the variational method\cite{Michael:1985ne,Luscher:1990ck} described in \cite{Dudek:2010wm}.  We use the $\Omega$ baryon mass to set the scale and quote all energies as ratios, $E/m_{\Omega}$, where $a_t m_{\Omega} = 0.353(3)$ on the lattices with $m_{\pi} \approx 700$ MeV~\cite{Lin:2008pr}.

\begin{table}
\begin{tabular}{c|c|c}
\textbf{Group} & \textbf{$\Lambda^C(\text{dim})$} & \textbf{Number of ops} \\
\hline
$\Dic_4$  & $A_1^+(1)$ & 14 \\
$(n,0,0)$ & $A_2^+(1)$ & 20 \\
          & $E_2^+(2)$ & 23 \\
          & $B_1^+(1)$ & 11 \\
          & $B_2^+(1)$ & 11 \\
          & $A_1^-(1)$ & 18 \\
          & $A_2^-(1)$ & 12 \\
          & $E_2^-(2)$ & 29 \\
          & $B_1^-(1)$ & 9  \\
          & $B_2^-(1)$ & 9  \\
\hline
$\Dic_2$  & $A_1^+(1)$ & 25 \\
$(n,n,0)$ & $A_2^+(1)$ & 31 \\
          & $B_1^+(1)$ & 23 \\
          & $B_2^+(1)$ & 23 \\
          & $A_1^-(1)$ & 27 \\
          & $A_2^-(1)$ & 21 \\
          & $B_1^-(1)$ & 29 \\
          & $B_2^-(1)$ & 29 \\
\hline
$\Dic_3$  & $A_1^+(1)$ & 15 \\
$(n,n,n)$ & $A_2^+(1)$ & 21 \\
          & $E_2^+(2)$ & 33 \\
          & $A_1^-(1)$ & 21 \\
          & $A_2^-(1)$ & 15 \\
          & $E_2^-(2)$ & 35 \\
\hline
\end{tabular}
\caption{The number of operators in our basis for each little group irrep $\Lambda^C$ ($C$ is the charge conjugation parity); also shown is the dimension of each irrep.}
\label{table:numops}
\end{table}

As a first test of the subduced helicity operator constructions, for each possible momentum with $|\vec{p}|^2 = 1$, $2$, $3$ and $4$, all possible correlators were computed, including those off-diagonal in irrep and/or irrep row.  This allowed the orthogonality of correlators between different irreps, and between different rows in the same irrep, to be verified, along with the positivity of diagonal correlators and, within each irrep, the hermiticity of the correlator matrix and the consistency of different rows in the same irrep.  In addition, the consistency of correlators from different momenta directions with the same momentum type was verified, something which would not necessarily be true if the rotation matrices, $R$, had not been chosen appropriately (see Section \ref{sec:finitevol}).

\section{Spin determination}
\label{sec:spin_assign}

For mesons with non-zero momentum, the identification of the continuum spin of a state extracted in a lattice calculation is more complicated than for mesons at rest; even in an infinite volume continuum, an operator in general can have non-zero overlap with states of many $J^P$, as discussed in Section \ref{sec:overlaps}.  The reduced symmetry compared to the zero momentum case means that more states appear in each irrep leading to many degeneracies in the spectrum; some are dynamical degeneracies of physical states in the true spectrum and others are caused by the reduced symmetry.   Trying to identify the continuum quantum numbers by matching degenerate levels in different little group irreps (after taking the continuum limit) will be even more problematic than for mesons at rest.  It would appear difficult or impossible to disentangle the states shown in Figs.~\ref{fig:Dic4_001} to \ref{fig:Dic4_002} without using some more information beyond solely their energies.

For mesons stable against hadronic decay, a general idea as to where to expect different energy levels can be obtained from using the mass of the mesons at rest (extracted on the same lattice) and the dispersion relation,
\begin{equation}
\left(a_t E(|\vec{p}|)\right)^2 = (a_t m)^2 + \frac{1}{\xi^2} \left(\frac{2\pi}{L_s} \right)^2 |\vec{p}|^2 ~ .
\label{equ:dispersion}
\end{equation} 
To determine a precise value for $\xi = a_s/a_t$ on this lattice, we use our measured $\pi$ energies, $a_t E_{\pi}(|\vec{p}|)$, at $|\vec{p}|^2=1,2,3,4$ and the $\pi$ mass from Ref.~\cite{Dudek:2009qf}, and fit the dispersion relation\footnote{Fitting to a more general form of the dispersion relation with $p^2$ and $p^4$ terms gives a consistent result, $\xi = 3.441(13)$, and does not improve the goodness of fit.} to obtain a good fit with $\xi = 3.441(8)$.  However, because of the degeneracies (within statistical uncertainties) in the spectrum, some further information is needed to determine the spin of extracted states.  Therefore, following a similar procedure to that developed for mesons at rest in Refs.~\cite{Dudek:2009qf,Dudek:2010wm}, we consider the state-operator overlaps.  

After subducing to lattice little group irreps, and neglecting the effects of a finite volume and finite lattice spacing, Eq.~\ref{equ:helopoverlap} becomes
\begin{equation}
\hspace{-0.12cm} \bigl<0 \bigl| \mathbb{O}^{[J,P,|\lambda|]}_{\Lambda,\mu}(\vec{p}) \bigr| \vec{p}; J'^{P'},\lambda' \bigr> 
 = Z^{[J,J',P,P',\lambda]} \mathcal{S}_{\Lambda,\mu}^{\tilde{\eta},\lambda} \delta_{\lambda,\lambda'}  ~,
\label{equ:helopoverlap_subduced}
\end{equation}
where $\tilde{\eta} \equiv P(-1)^J$ and $J$ and $P$ are respectively the spin and parity of the operator at rest.  As discussed above, even in an infinite volume continuum, at non-zero momentum an operator can in general overlap onto states of many different spins and both parities.  We use the constraints on operator-state overlaps which arise from Lorentz symmetry applied to our helicity operators to determine the continuum spin and parity of states.  These constraints can be read off from columns ``L1'', ``L2'' and ``L3'' of Tables \ref{table:overlaps:0}, \ref{table:overlaps:1}, \ref{table:overlaps:2} and \ref{table:overlaps:2b} in Appendix \ref{app:decompositions}.

\begin{figure}[tb]
\includegraphics[width=0.4\textwidth]{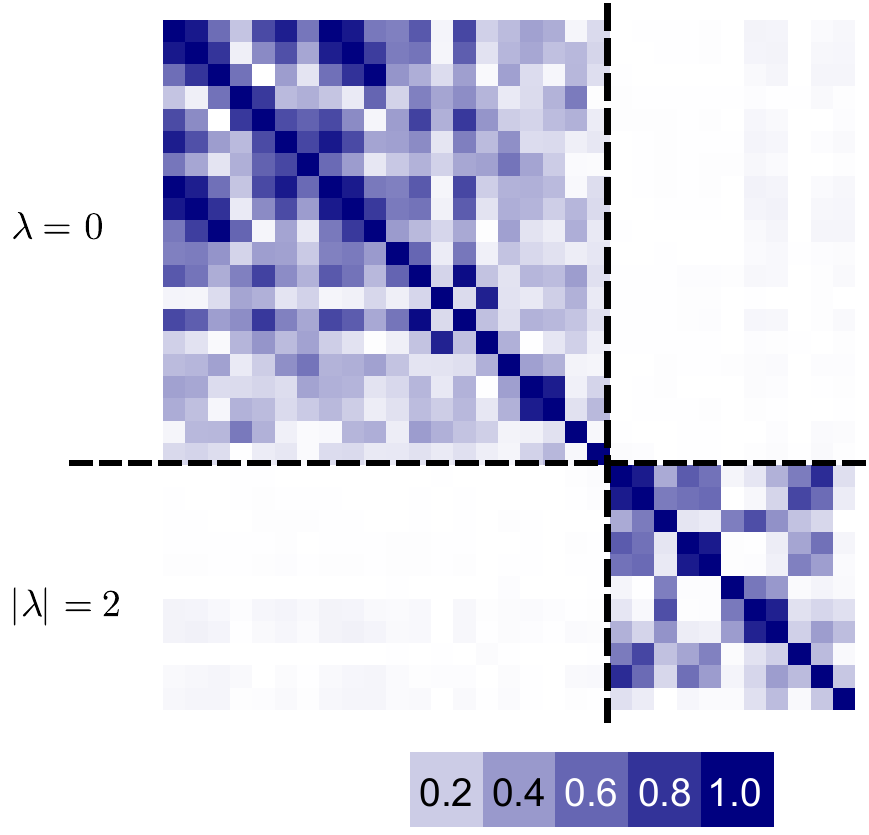}
\caption{Normalised correlation matrix ($C_{ij}/\sqrt{C_{ii}C_{jj}}$) on time-slice 5 in the $A_2^+$ irrep with $|\vec{p}|^2 = 2$ ($\Dic_2$).  The first 20 operators are subduced from $\lambda = 0$ (top and left), the remaining 11 operators are subduced from $|\lambda| = 2$ (bottom and right).}
\label{fig:hel_corr_matrix_plot}
\end{figure}

We observe that mixings due to the reduced symmetry are small; a subduced operator from a helicity operator with helicity magnitude $|\lambda|$ only overlaps significantly onto states of helicity magnitude $|\lambda|$ and not with other helicities in that irrep.  In Fig.~\ref{fig:hel_corr_matrix_plot} we show the normalised correlation matrix on time-slice 5 for the $A_2^+$ irrep with $|\vec{p}|^2 = 2$ (including zero, one and two-derivative operators) ordered such that those subduced from $\lambda=0$ come before those from $|\lambda|=2$.  The correlation matrix is observed to be approximately block diagonal, correlators mixing $\lambda = 0$ and $|\lambda| = 2$ are small.  

\begin{figure*}[tb]
\includegraphics[width=1.0\textwidth]{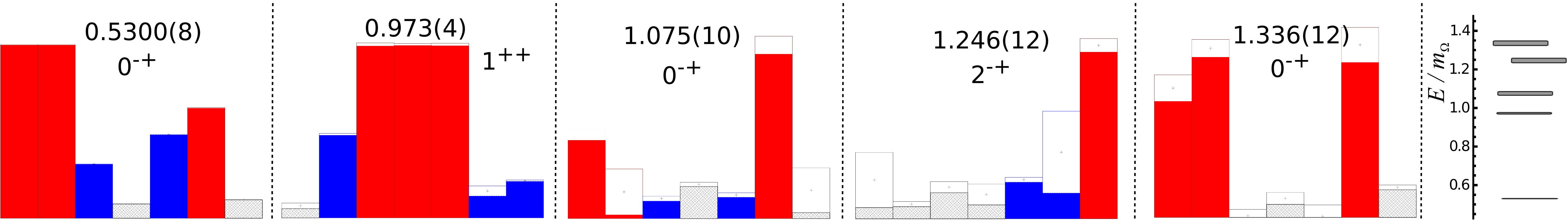}
\caption{Overlaps, $|Z|$, of operators onto states labelled by $E/m_{\Omega}$ in the $\Lambda^C = A_2^+$ irrep with $|\vec{p}|^2 = 1$ ($\Dic_4$); the extracted spectrum is shown on the right.  The $Z$'s are normalised so that the largest value for that operator across all states is equal to 1.  The unshaded area at the head of each bar represents the one sigma uncertainty on either side of the mean.  For each state the bars represent, from left to right, the operators $(\pi)^{J=0}_{\lambda=0}$, $(\pi_2)^{J=0}_{\lambda=0}$, $(a_1)^{J=1}_{\lambda=0}$,  $(\rho \times D^{[1]}_{J=1})^{J=1}_{\lambda=0}$, $(\rho_2 \times D^{[1]}_{J=1})^{J=1}_{\lambda=0}$, $(b_1 \times D^{[1]}_{J=1})^{J=0}_{\lambda=0}$ and $(b_1 \times D^{[1]}_{J=1})^{J=2}_{\lambda=0}$.  The colour coding is explained in the text.}
\label{fig:Z_histograms_Dic4_A2p}
\end{figure*}

This lack of mixing between different $|\lambda|$ is also seen in the $Z$ values.  For example, consider the $\Lambda^C = A_2^+$ irrep.  In Fig.~\ref{fig:Z_histograms_Dic4_A2p} we show the $Z$'s for the lightest five states in this irrep with $|\vec{p}|^2 = 1$, along with the $J^{PC}$ assignment for each state; for simplicity of presentation, we use a smaller operator basis, only including the seven zero and one-derivative operators.  The operator naming scheme is described in Section \ref{sec:lattices}.  Also shown on the right of the figure is the spectrum of energy levels extracted in this irrep (the spin-identified spectrum is shown in the left hand panel of Fig.~\ref{fig:Dic4_001}). 

The same operators are shown for each state but the colour coding varies according to the $J^{PC}$ hypothesis: red bars correspond to operators which overlap onto states with that $J^{PC}$ at rest, blue bars correspond to operators which overlap with such states only at non-zero momentum and hatched grey bars correspond to operators which do not overlap with such states (using the constraints arising from Lorentz symmetry).  As a concrete example, consider the first two operators shown, $(\pi)^{J=0}_{\lambda=0}$ and $(\pi_2)^{J=0}_{\lambda=0}$.  These contain, respectively, $\gamma_5$ and $\gamma_0\gamma_5$ in spin space, both with no derivatives, and at zero momentum both overlap only with $J^{PC}=0^{-+}$ states.  Both contain no vector indices (i.e. no $\gamma^i$ or $D^i$) and so Table \ref{table:overlaps:0} is relevant but, as explained in Appendix \ref{app:decompositions}, because these operators overlap with $J^P=0^-$ at rest, we must flip all the state parities in that table.  The operator $(\pi)^{J=0}_{\lambda=0}$ is of the form given in the ``L1'' column and from there we see that, using Lorentz covariance constraints, it only overlaps onto $J^P=0^-$.  In the figures it is therefore coloured red for $0^-$ states and grey for other $J^P$.  On the other hand, $(\pi_2)^{J=0}_{\lambda=0}$ is of the form given in the ``L2'' column and we see that at non-zero momentum it can overlap with both $0^-$ and $1^+$.  It is therefore coloured red for $0^-$ states, blue for $1^+$ states and grey for other $J^P$.

For all the states shown, it can be seen that the red bars are largest, although these overlaps can be small if there is a mismatch in structure between the operator and the state.  The blue bars can be significant, these should be proportional to some power of $|\vec{p}|/M$ and so are suppressed compared to the red bars at low momentum, and the grey bars are small.  From these figures it can be seen that in each case the $J^{PC}$ can be determined unambiguously.  For $|\vec{p}|^2 = 1$ ($\Dic_4$) the continuum helicities which subduce into the $A_2$ irrep are $0, 4, \ldots$ (Table \ref{table:subductions}).  The restricted operator basis used here does not contain helicities greater than two and so we can not test helicity mixing here.  In any case, the lightest spin-four state is expected to be significantly higher in energy than the heaviest states we extract in this analysis.

\begin{figure*}[tb]
\includegraphics[width=1.0\textwidth]{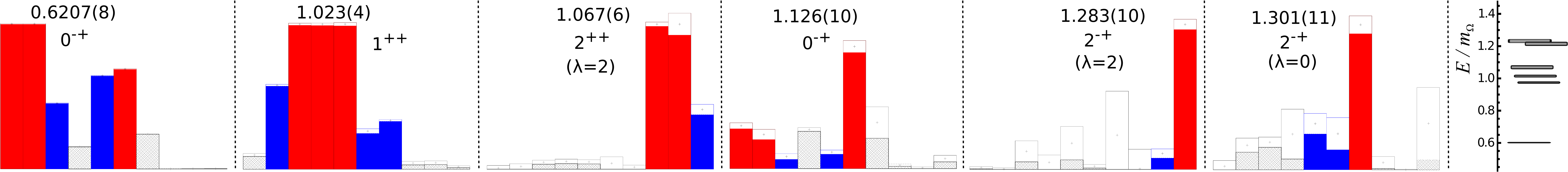}
\caption{As Fig.~\ref{fig:Z_histograms_Dic4_A2p} but for the $\Lambda^C = A_2^+$ irrep with $|\vec{p}|^2 = 2$ ($\Dic_2$).  For each state the bars represent, from left to right, the operators $(\pi)^{J=0}_{\lambda=0}$, $(\pi_2)^{J=0}_{\lambda=0}$, $(a_1)^{J=1}_{\lambda=0}$,  $(\rho\times D^{[1]}_{J=1})^{J=1}_{\lambda=0}$, $(\rho_2\times D^{[1]}_{J=1})^{J=1}_{\lambda=0}$, $(b_1\times D^{[1]}_{J=1})^{J=0}_{\lambda=0}$, $(b_1\times D^{[1]}_{J=1})^{J=2}_{\lambda=0}$, $(\rho\times D^{[1]}_{J=1})^{J=2}_{|\lambda|=2}$, $(\rho_2\times D^{[1]}_{J=1})^{J=2}_{|\lambda|=2}$ and $(b_1\times D^{[1]}_{J=1})^{J=2}_{|\lambda|=2}$.}
\label{fig:Z_histograms_Dic2_A2p}
\end{figure*}

In Fig.~\ref{fig:Z_histograms_Dic2_A2p} we show the $Z$'s for the overlap of the lightest six states in the $A_2^+$ irrep with $|\vec{p}|^2 = 2$ onto the ten zero and one-derivative operators in that irrep, along with the spin assignment hypothesis for each state.  Again, for each state, the continuum $J^{PC}$ can be unambiguously determined.  For $|\vec{p}|^2 = 2$ ($\Dic_2$) the continuum helicities which subduce into the $A_2$ irrep are $0, 2, 4, \ldots$ (Table \ref{table:subductions}) and the $A_2^+$ operator basis contains operators coming from $\lambda=0$ (the first seven operators) and $|\lambda|=2$ (the last three operators).  It can be seen that for each state only $\lambda=0$ \emph{or} $|\lambda|=2$ overlaps are statistically significantly non-zero, further supporting the observation at the level of the correlation matrix, Fig.~\ref{fig:hel_corr_matrix_plot}.  Furthermore, these overlaps are generally smaller than the overlaps onto operators with the correct continuum helicity but which are constrained to be zero by Lorentz symmetry.

\hfill

A second stage of the spin identification procedure is to compare the $Z$'s for the overlap of a given state onto a given continuum operator when this operator is subduced into two or more different little group irreps.  From Eq.~\ref{equ:helopoverlap_subduced}, neglecting any finite volume or discretisation effects, these overlap factors should be equal, something that is empirically found to be true for mesons at rest extracted on this lattice in Refs.~\cite{Dudek:2009qf,Dudek:2010wm}.  For mesons in flight the opportunities for such comparisons are more limited.  The only directly analogous situation is when the same $|\lambda|$ subduces into two one-dimensional irreps (as opposed to subducing into one two-dimensional irrep); for example, $\lambda = \pm2$ with $|\vec{p}|^2 = 1$ subduces into $B_1$ and $B_2$, and $\lambda = \pm1$ with $|\vec{p}|^2 = 2$ subduces into $B_1$ and $B_2$.  These comparisons can only be used to confirm the continuum helicity rather than the continuum spin (although determining the continuum helicity does give a minimum value of the continuum spin) and so such comparisons are less useful for mesons in flight than for mesons at rest.

\begin{figure}[tb]
\includegraphics[width=0.49\textwidth]{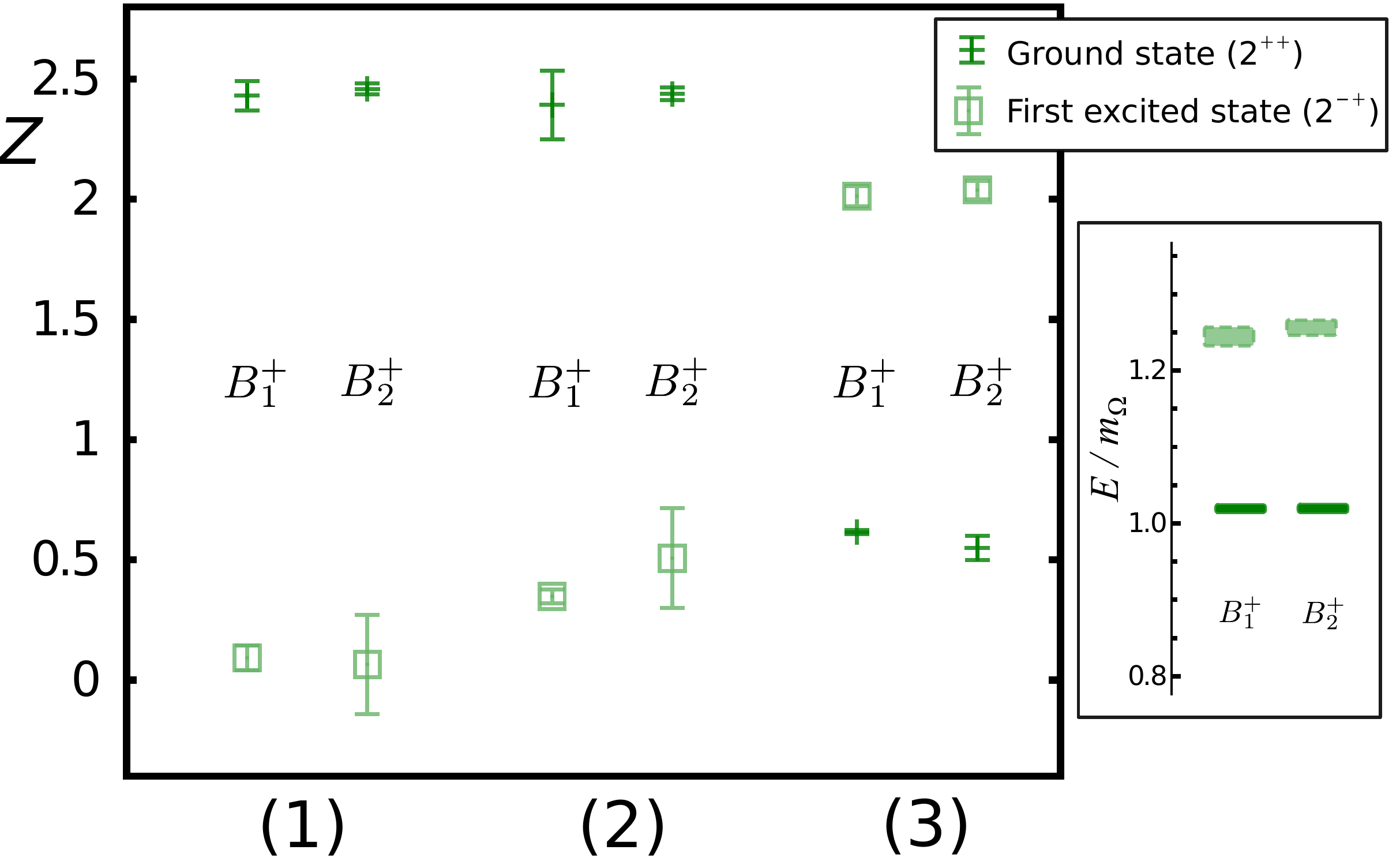}
\caption{$|Z|$ values for the $\lambda = \pm2$ components of $J^{PC}=2^{\pm+}$ states with $|\vec{p}|^2 = 1$ ($\Dic_4$) subduced into the $\Lambda^C = B_1^+$ and $B_2^+$ irreps.  The inset shows the extracted lowest-lying energy levels.  The operators are: (1) $(\rho\times D^{[1]}_{J=1})^{J=2}_{|\lambda|=2}$, (2) $(\rho_2\times D^{[1]}_{J=1})^{J=2}_{|\lambda|=2}$ and (3) $(b_1 \times D^{[1]}_{J=1})^{J=2}_{|\lambda|=2}$; the $Z$'s from the $B_1^+$ irrep (shifted to the left) and the $B_2^+$ (shifted to the right) are plotted.}
\label{fig:Z_comp_Dic4_B1pB2p}
\end{figure}

\begin{figure}[tb]
\includegraphics[width=0.49\textwidth]{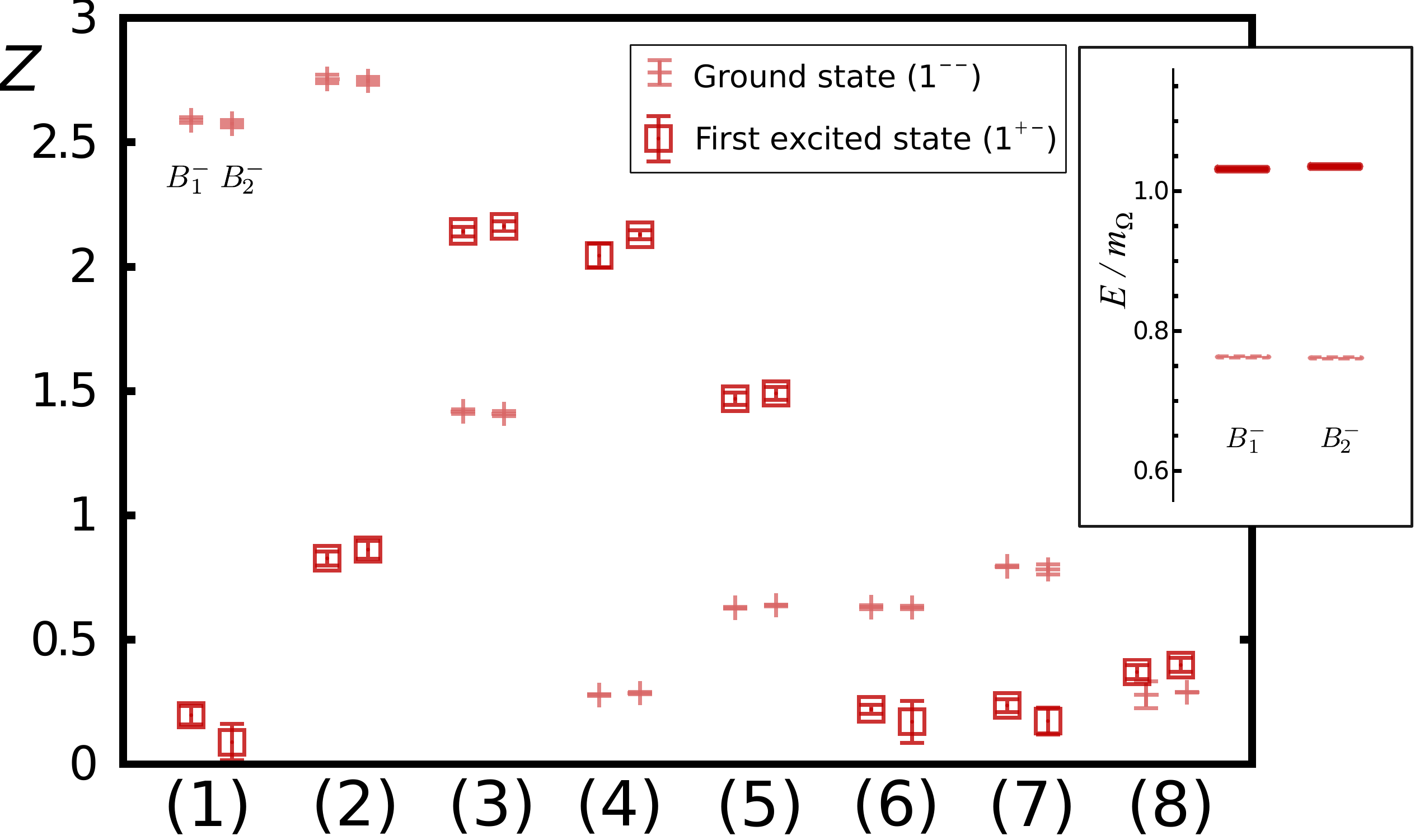}
\caption{$|Z|$ values for the $\lambda = \pm1$ components of $J^{PC}=1^{\pm-}$ states with $|\vec{p}|^2 = 2$ ($\Dic_2$) subduced into the $\Lambda^C = B_1^-$ and $B_2^-$ irreps.  The inset shows the extracted lowest-lying energy levels.  The operators are: (1) $(\rho)^{J=1}_{|\lambda|=1}$, (2)  $(\rho_2)^{J=1}_{|\lambda|=1}$, (3) $(b_1)^{J=1}_{|\lambda|=1}$, (4) $(\pi \times D^{[1]}_{J=1})^{J=1}_{|\lambda|=1}$, (5) $(\pi_2 \times D^{[1]}_{J=1})^{J=1}_{|\lambda|=1}$, (6) $(a_0 \times D^{[1]}_{J=1})^{J=1}_{|\lambda|=1}$, (7) $(a_1 \times D^{[1]}_{J=1})^{J=1}_{|\lambda|=1}$ and (8) $(a_1 \times D^{[1]}_{J=1})^{J=2}_{|\lambda|=1}$; the $Z$'s from the $B_1^-$ irrep (shifted to the left) and the $B_2^-$ (shifted to the right) are plotted.}
\label{fig:Z_comp_Dic2_B1mB2m}
\end{figure}

As an example, in Fig.~\ref{fig:Z_comp_Dic4_B1pB2p} we show the $Z$'s for the lowest two $J^{PC}=2^{\pm+}$ states with $|\vec{p}|^2=1$ overlapping onto the three one-derivative operators subduced into the $\Lambda^C = B_1^+$ and $B_2^+$ irreps.  The good agreement between $Z$ values with the same operator subduced into different irreps is apparent.  In Fig.~\ref{fig:Z_comp_Dic2_B1mB2m} we show the $Z$'s for the lowest two $J^{PC}=1^{\pm-}$ states with $|\vec{p}|^2=2$ overlapping onto the eight zero and one-derivative operators subduced into the $\Lambda^C = B_1^-$ and $B_2^-$ irreps.  Again, there is good agreement between the $Z$'s in different irreps.

The consistency of these $Z$ values for the same $|\lambda|$ subduced into different little group irreps relied on three-rotation symmetry, assuming that the effects of a finite volume and finite lattice spacing are small for these quantities.  Using only constraints arising from three-rotation symmetry, the $Z$'s for different $|\lambda|$ components overlapping onto the same continuum operator are not related, but some of these can be related using Lorentz symmetry constraints\footnote{if our operators were built out of Lorentz vectors we could relate all the helicities in every case, but our use of operators built from 3-vectors limits where we can relate different $|\lambda|$}.  These constraints can be read off from Tables \ref{table:overlaps:0}, \ref{table:overlaps:1}, \ref{table:overlaps:2} and \ref{table:overlaps:2b} in Appendix \ref{app:decompositions}.  Because the constraints depend on which gamma matrices and how many derivatives the operator was constructed out of, and how these were coupled together, comparisons are much more involved compared to those for mesons at rest.  Therefore, in general we do not consider these comparisons as part of our spin-identification procedure.

\begin{figure}[tb]
\includegraphics[width=0.49\textwidth]{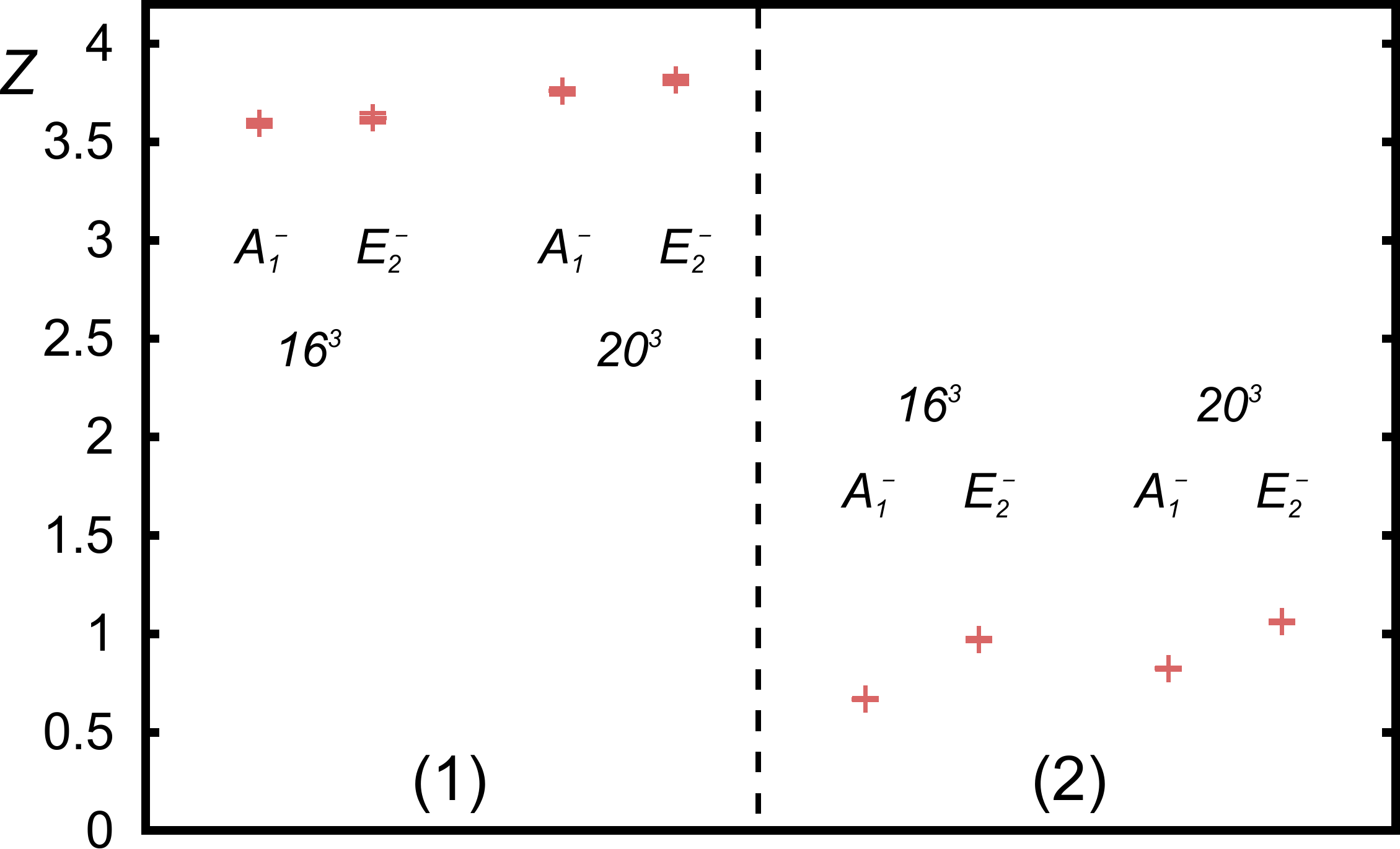}
\caption{$|Z|$ values for the $J^{PC}=1^{--}$ ground state with $|\vec{p}|^2 = 1$ ($\Dic_4$) subduced into the $\Lambda^C = A_1^-$ ($\lambda = 0$) and $E_2^-$ ($\lambda = \pm 1$) irreps.  The operators are: (1) $(\rho)^{J=1} \sim \gamma^i$, (2) $(a_0 \times D^{[1]}_{J=1})^{J=1} \sim D^i$; the $Z$'s from the $A_1^-$ irrep and $Z's$ \emph{scaled by $E/M$} from the $E_2^-$ irrep are plotted.  For comparison, also shown are the $Z$ values, \emph{all scaled by ($\sqrt{16^3/20^3}$)}, for a larger lattice with volume $20^3$ in lattice units.}
\label{fig:Z_comp_Dic4_E2mA1m}
\end{figure}

As an example, consider operators containing a $\gamma^i$ and zero derivatives, along with operators containing one derivative ($\overleftrightarrow{D^i}$) and the identity in spin-space.  Reading off from Table \ref{table:overlaps:1} (``L1'' column), we see that the overlap onto a $J^P=1^-$ state is $i Z_1$ for $\lambda = \pm1$ and $i \frac{E}{M}Z_1$ for $\lambda = 0$.  Note that if we only imposed 3-rotation covariance, the $\lambda = 0$ overlap has two independent constants, $Z_1$ and $Z_2$, and so cannot be related to the $\lambda = \pm1$ overlaps.  In Fig.~\ref{fig:Z_comp_Dic4_E2mA1m} we show the overlap factor for both these operators for the lowest $1^{--}$ state with $|\vec{p}|^2 = 1$.  There is good agreement between the different irreps for the $\gamma^i$ operator (1); for the $\overleftrightarrow{D^i}$ operator (2) the $Z$ values are similar but disagree significantly outside their statistical uncertainties.  Sources for this discrepancy could be violations of hypercubic symmetry arising from the action or the result of mixing with multi-meson states which have a different distribution across irreps compared to single-meson states.  To check whether this discrepancy could be a finite volume effect, in Fig.~\ref{fig:Z_comp_Dic4_E2mA1m} we also plot the $Z$-values \emph{scaled by $\sqrt{16^3/20^3}$} for the corresponding state extracted on a lattice with the same quark masses but a larger volume, $20^3$ in lattice units.  Although the $Z$ values for operator (2) get slightly closer together on the larger volume, there is no significant change in the pattern. 

\vspace{.2cm}

In summary, we have shown how operator overlaps can be used to identify the continuum $J^{PC}$ of mesons with non-zero momentum extracted in lattice calculations.  In Appendix \ref{app:tests} we show some tests of the robustness of our extracted spectra and in the following section we apply this method to identify continuum spin in our full extracted spectra.

\section{Spectra of mesons in flight}
\label{sec:results}

Using the data sets given in Table \ref{table:datasets} with $m_{\pi} \approx 700$ MeV and volume $16^3 \times 128$, and the operator basis described above (including zero, one and two derivatives, and enumerated in Table \ref{table:numops}), for each irrep and momentum type we computed a matrix of correlators.   These correlators were analysed using the implementation of the variational method\cite{Michael:1985ne,Luscher:1990ck} described in Ref.~\cite{Dudek:2010wm} and we identified continuum spins using the methodology described above.

\begin{figure*}[ptb]
\includegraphics[width=0.7\textwidth]{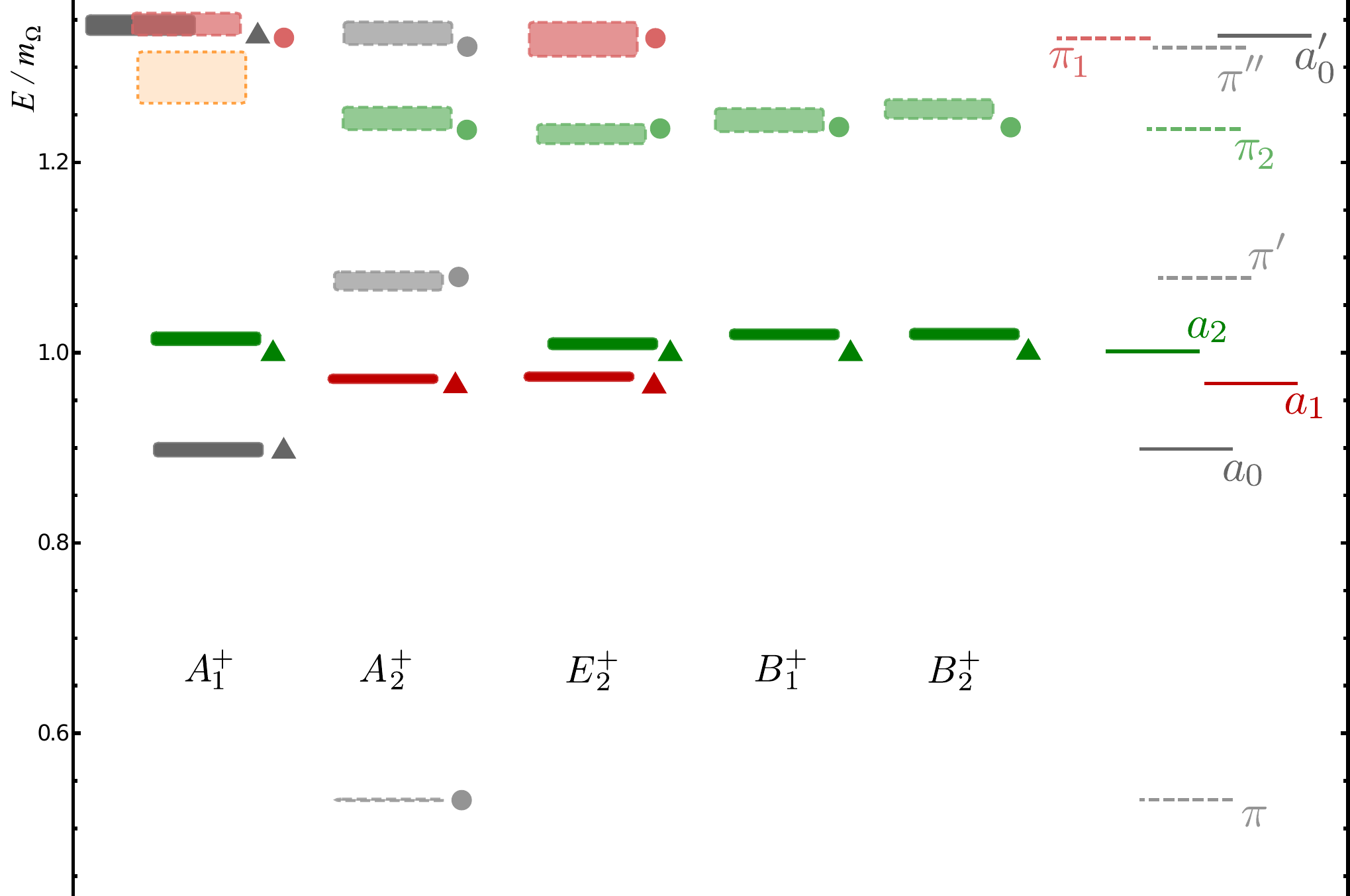}
\caption{Spectrum of lowest-lying isovector mesons with positive charge conjugation parity in each irrep, $\Lambda^C$, for $|\vec{p}|^2=1$ ($\Dic_4$), a subset of the spectrum shown in Fig.~\ref{fig:Dic4_001}.  The box height shows the one sigma statistical uncertainty above and below the central value; the colour coding, indicating the $J^P$, is described in the text.  As explained in the text, the lines in the right hand column show expected energies for the lower-lying states; the triangles (circles) show the distribution of the components of these positive (negative) parity states across the irreps.}
\label{fig:Dic4_001_large}
\end{figure*}

Before we show the full extracted spectra for each $|\vec{p}|^2$, we illustrate some features by considering one case in more detail, the lowest lying states with positive charge conjugation parity for $|\vec{p}|^2=1$ ($\Dic_4$), shown in Fig.~\ref{fig:Dic4_001_large}.  The colour coding indicates the identified $J^P$: black/grey for $J=0$, red for $J=1$, green for $J=2$ and blue for\footnote{there are no $J=3$ states in Fig.~\ref{fig:Dic4_001_large} but these appear in later figures} $J=3$; darker shades with a solid outline correspond to positive parity and lighter shades with a dashed outline correspond to negative parity.  Orange boxes with a dotted outline indicate that the state's $J^P$ could not be unambiguously determined in this analysis.  The lines in the right hand column show expected energies (without statistical uncertainty) for the lower-lying states using masses from Ref.~\cite{Dudek:2009qf} and the dispersion relation, Eq.~\ref{equ:dispersion}, with $\xi = 3.441$ (Section \ref{sec:spin_assign}).  The triangles (circles) show the distribution of the components of these expected states with definite $J^{P=+}$ ($J^{P=-}$) across the irreps, using Table \ref{table:subductions}.

Spin assignment was initially carried out independently for each irrep and, as can be seen, the different components of the same continuum state can be matched up between different irreps with good agreement between the energies.  The pattern of states across irreps (the boxes in the figure) is the same as the pattern expected for states of definite continuum $J^P$ subduced across irreps (the triangles and circles in the figure).  We note that it would be extremely challenging to identify the different states with $E/m_{\Omega} \sim 1.34$ if we did not use information from the $Z$ values.

In the $A_1^+$ irrep the state with $E/m_{\Omega} \sim 0.90$ (dark grey, solid outline) is identified as having $J^{PC} = 0^{++}$, has no partner in the other irreps and lies where the $a_0$ is expected.  The next state in this irrep $\sim 1.01$ (dark green, solid outline) is identified as the zero helicity piece of $2^{++}$\footnote{Recall that the $\lambda = 0$ piece of a state with $\tilde{\eta} \equiv P(-1)^J = +1$ or $-1$ subduces into the $A_1$ or $A_2$ irrep respectively, but the $|\lambda|>0$ pieces of a state subduce into the same irreps regardless of parity.}.  It can be matched up with states identified as $2^{++}$ in the $E_2^+$ ($\lambda = \pm1$) and the $B_1^+$ and $B_2^+$ ($\lambda = \pm2$) irreps, and lies where the $a_2$ is expected.  The state at $\sim 1.29$ (orange, dotted outline) is not well determined in this analysis.  There are two states $\sim 1.34$, one (dark grey, solid outline) is identified as $0^{++}$ and appears to be the $a_0'$.  The other (light red, dashed outline) is identified as the zero helicity piece of the exotic $1^{-+}$ and can be matched up with the $1^{-+}$ in the $E_2^+$ irrep ($\lambda = \pm1$).  This exotic state is extracted with the same precision as the non-exotic states; its considerable overlap onto operators proportional to the commutator of two covariant derivatives, requiring a non-trivial gluonic field configuration, suggests a hybrid interpretation~\cite{Dudek:2010wm,Dudek:2011bn}.

Moving to the $A_2^+$ irrep, there are three states (light grey, dashed outline) which have been identified as $J^{PC} = 0^{-+}$ at $E/m_{\Omega} \sim 0.53, 1.08, 1.34$; these do not have partners in other irreps and lie where the pion and excited pions are expected.  The state at $\sim 0.97$ (dark red, solid outline) is identified as the zero helicity piece of $1^{++}$, matches up with the $1^{++}$ in the $E_2^+$ irrep ($\lambda = \pm1$), and lies where the $a_1$ is expected.  Finally, the state $\sim 1.25$ (light green, dashed outline) is identified as the zero helicity piece of $2^{-+}$, can be matched up with the $2^{-+}$ states in the $E_2^+$ ($\lambda = \pm1$) and the $B_1^+$ and $B_2^+$ ($\lambda = \pm2$) irreps, and lies where the $\pi_2$ is expected.

\hfill

\begin{figure*}[ptb]
\includegraphics[width=0.41\textwidth]{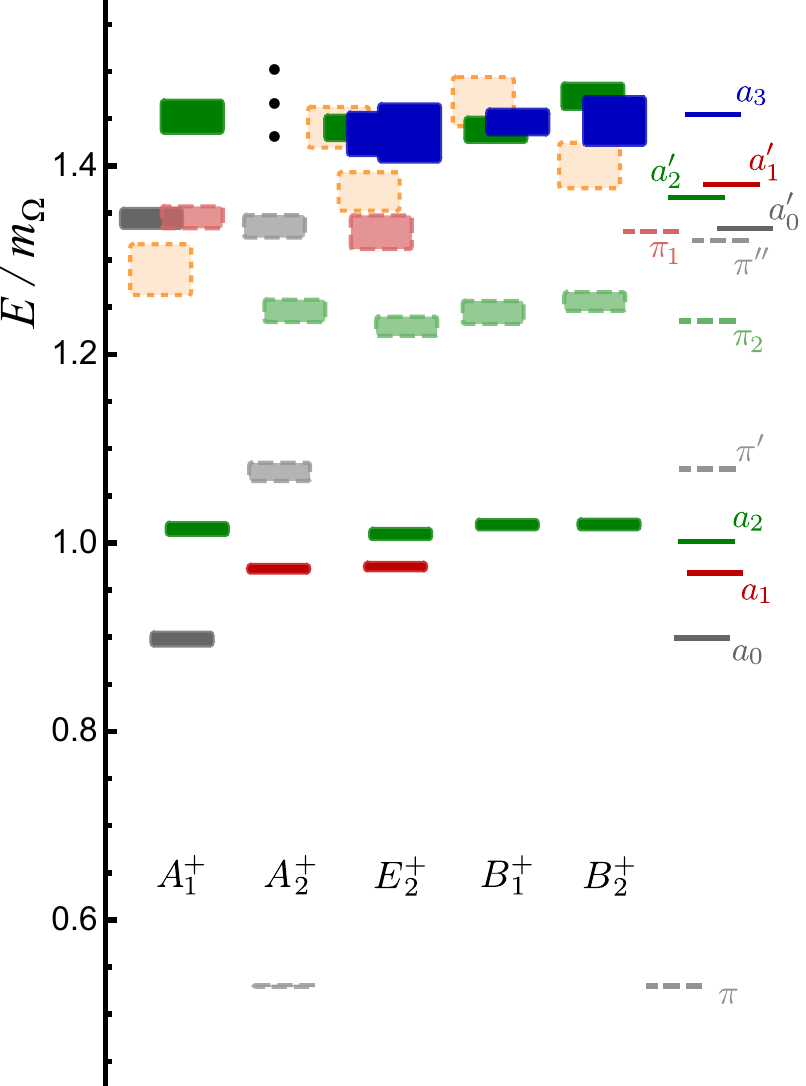}
\includegraphics[width=0.41\textwidth]{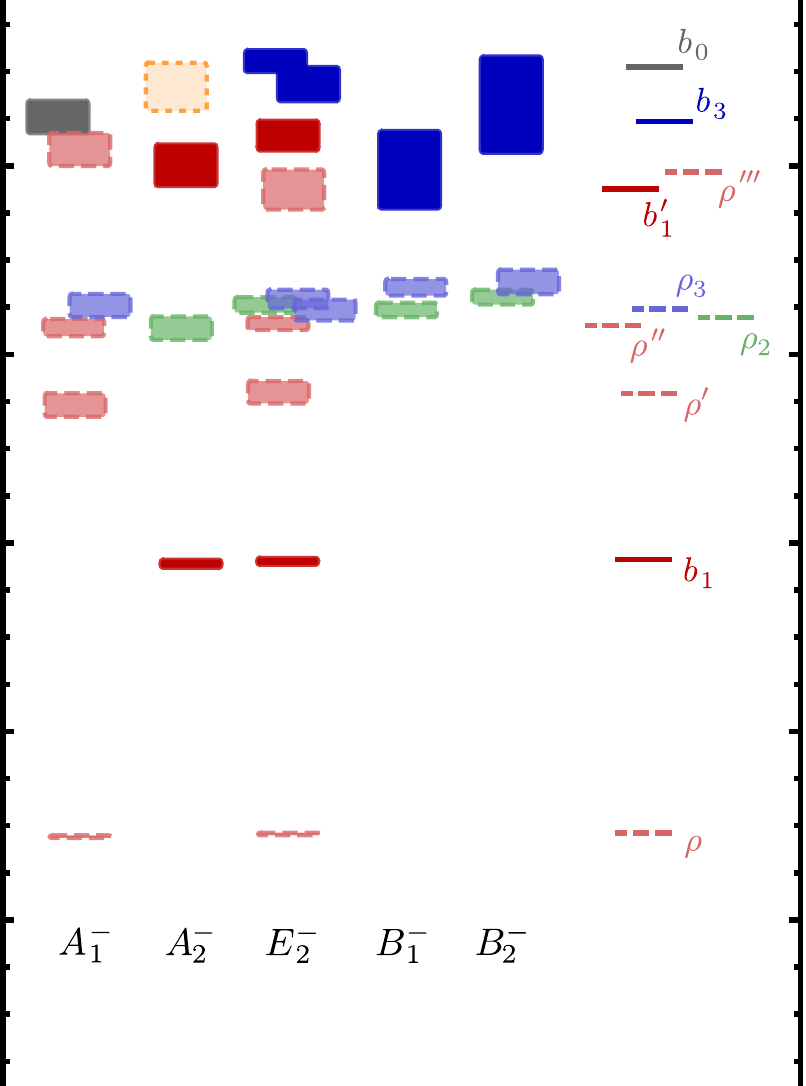}
\caption{Spectrum of low lying isovector mesons with positive (left hand panel) and negative (right hand panel) charge conjugation parity in each irrep, $\Lambda^C$, for $|\vec{p}|^2=1$ ($\Dic_4$).  The box height shows the one sigma statistical uncertainty above and below the central value; the colour coding, indicating the $J^P$, is described in the text.  Ellipses indicate that there are additional states within a given irrep in that energy range but that they are not well determined in this calculation.  The lines in the right hand column show expected energies for the lower-lying states as described in the text.}
\label{fig:Dic4_001}
\end{figure*}

\begin{figure*}[ptb]
\includegraphics[width=0.41\textwidth]{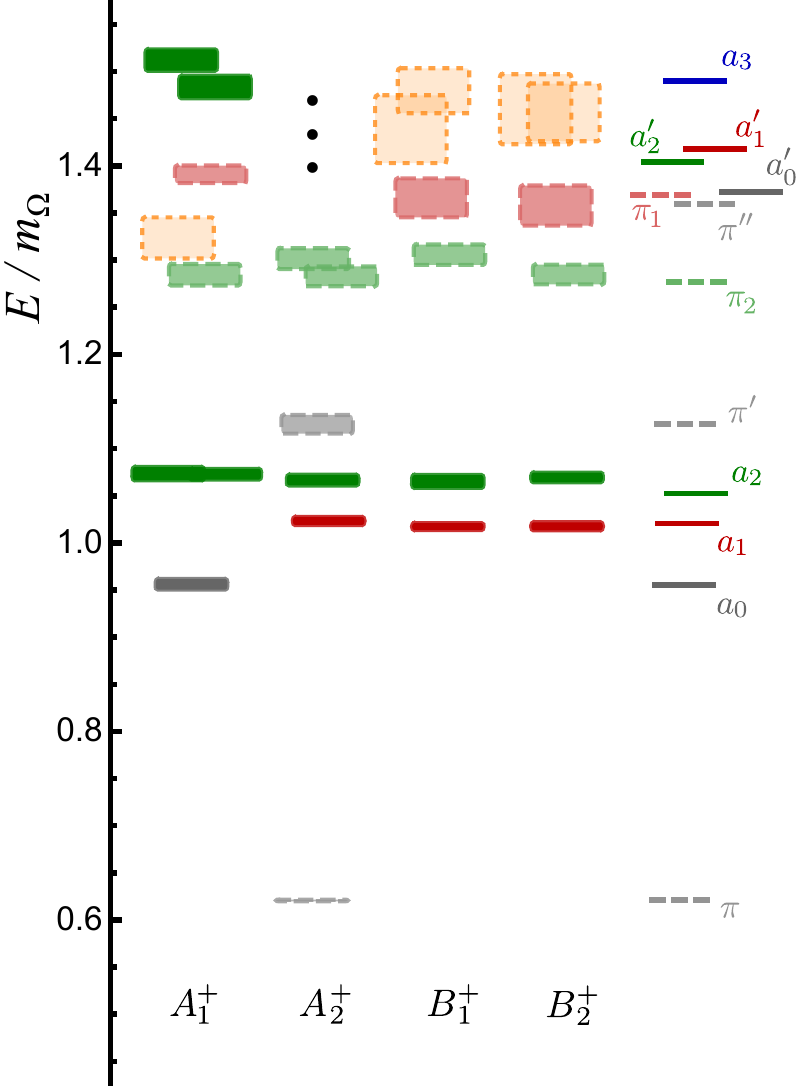}
\includegraphics[width=0.41\textwidth]{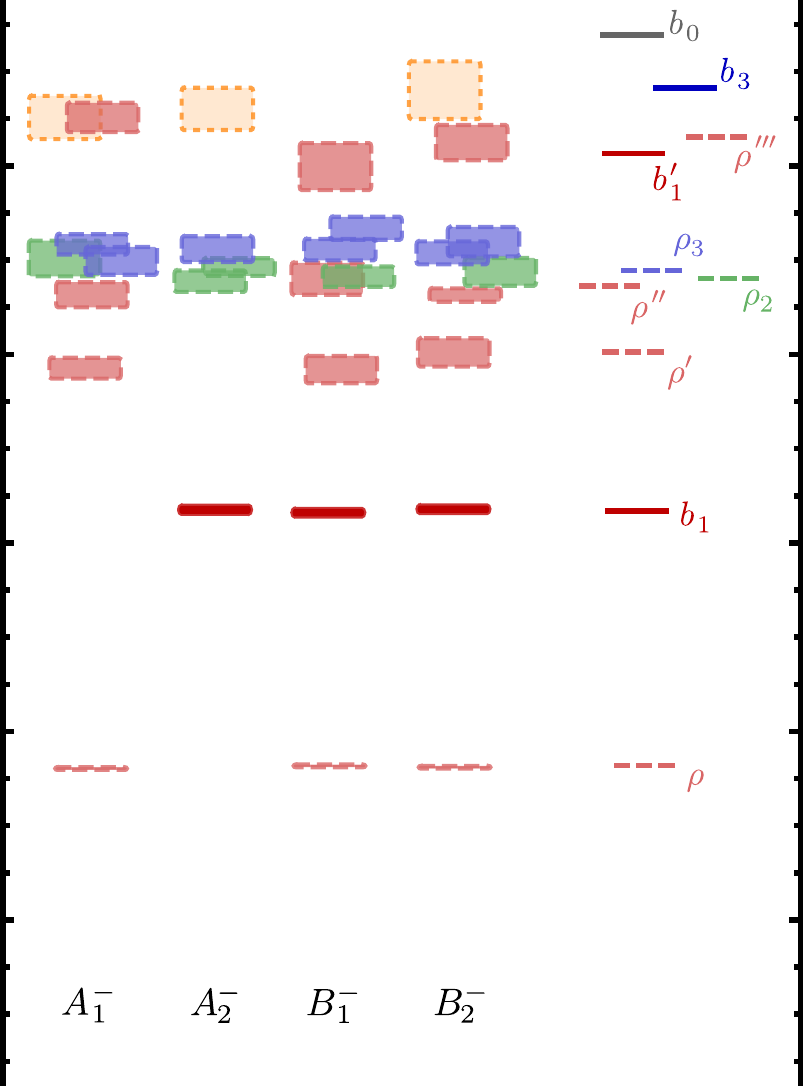}
\caption{As Fig.~\ref{fig:Dic4_001} for $|\vec{p}|^2 = 2$ ($\Dic_2$). }
\label{fig:Dic2_011}
\end{figure*}

\begin{figure*}[ptb]
\includegraphics[width=0.41\textwidth]{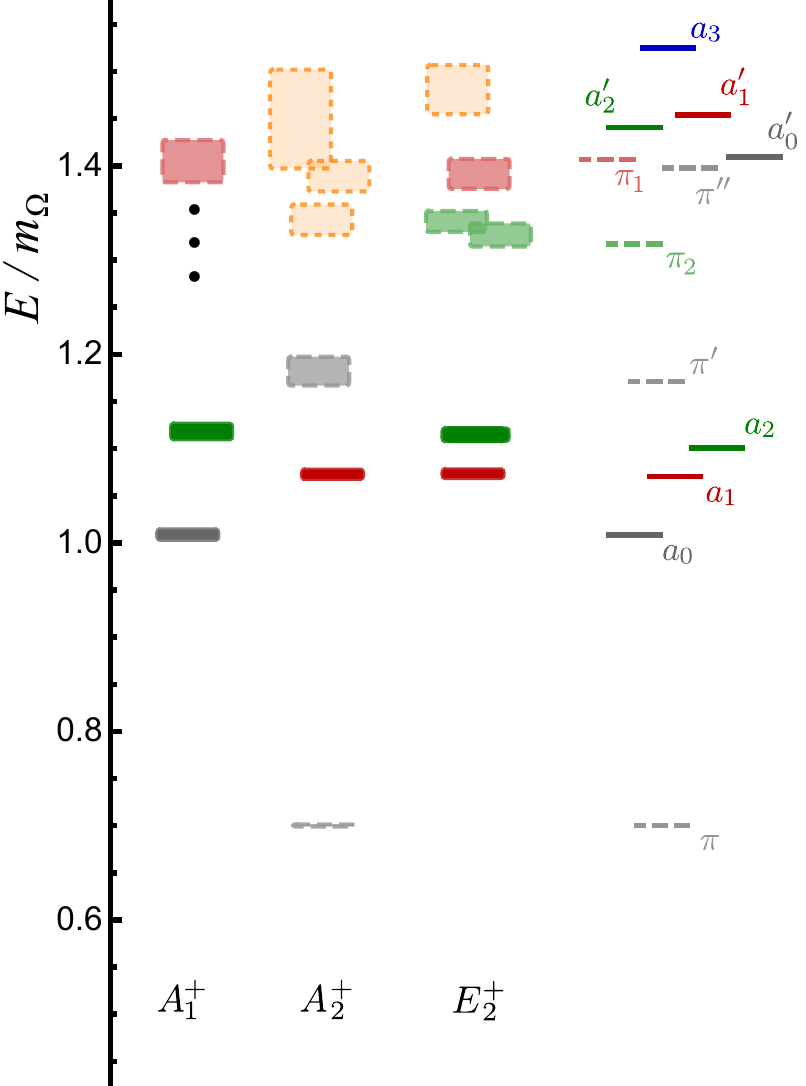}
\includegraphics[width=0.41\textwidth]{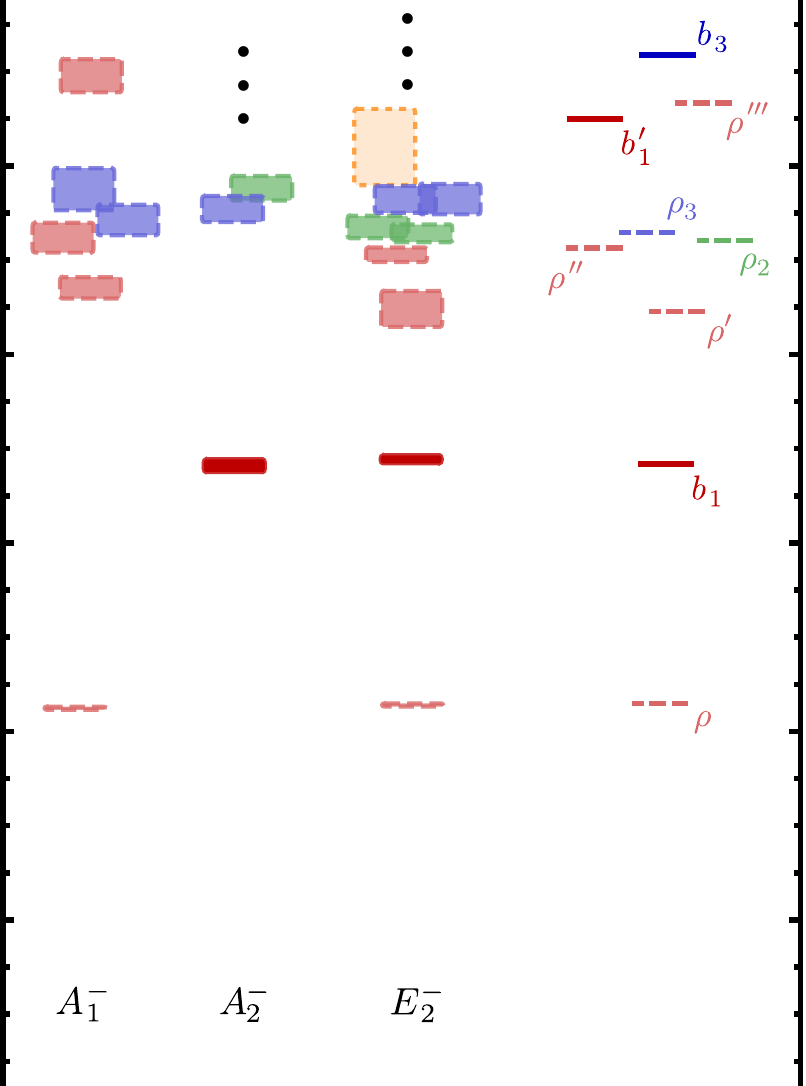}
\caption{As Fig.~\ref{fig:Dic4_001} for $|\vec{p}|^2 = 3$ ($\Dic_3$). }
\label{fig:Dic3_111}
\end{figure*}

\begin{figure*}[ptb]
\includegraphics[width=0.41\textwidth]{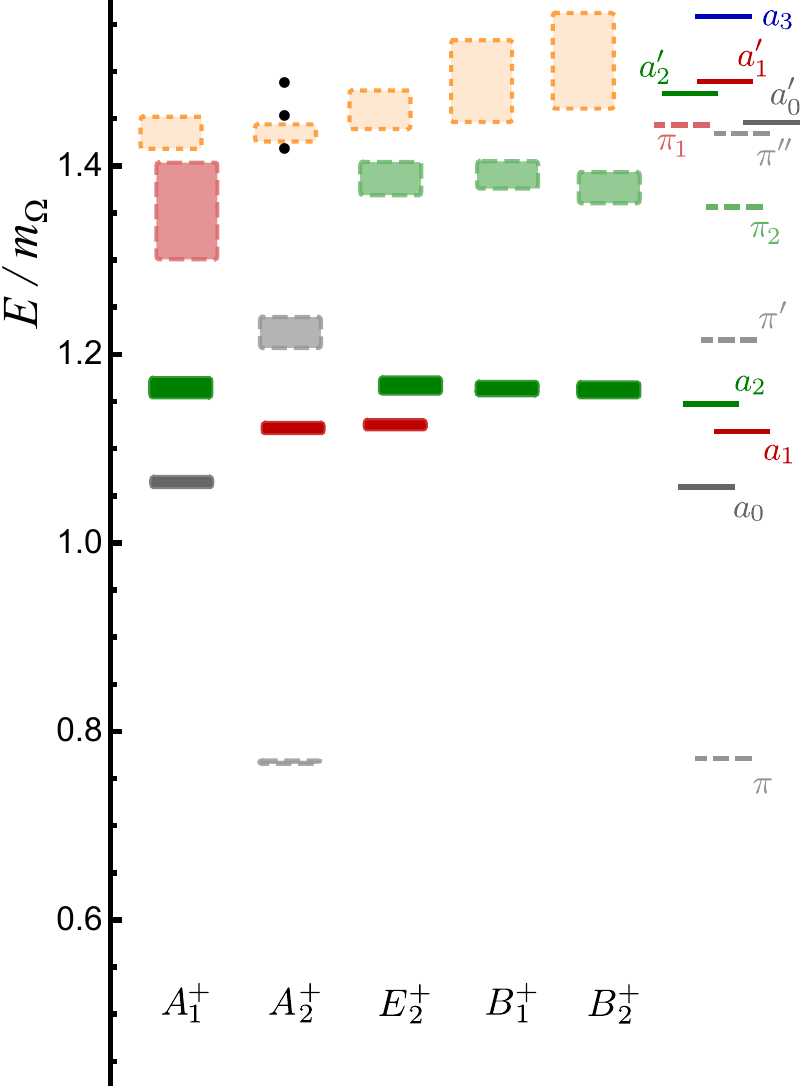}
\includegraphics[width=0.41\textwidth]{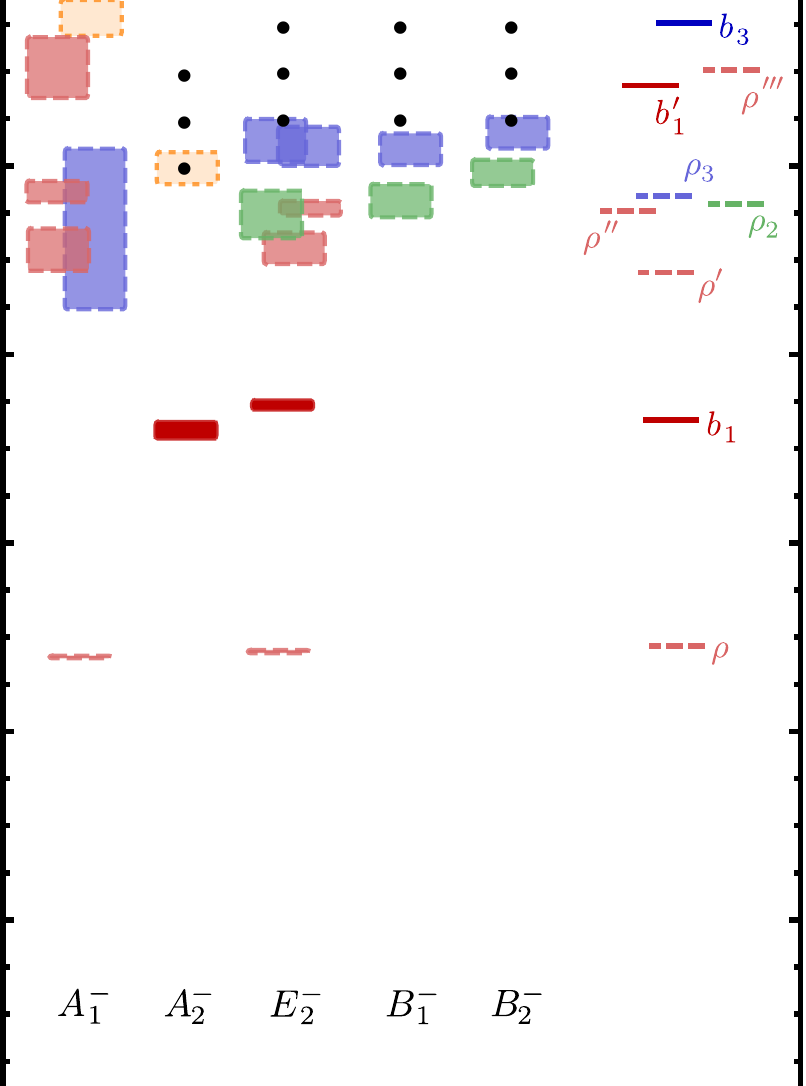}
\caption{As Fig.~\ref{fig:Dic4_001} for $|\vec{p}|^2 = 4$ ($\Dic_4$). }
\label{fig:Dic4_002}
\end{figure*}

The full spectra with $|\vec{p}|^2 = 1,2,3$ and $4$ are shown in Figs.~\ref{fig:Dic4_001}, \ref{fig:Dic2_011}, \ref{fig:Dic3_111} and \ref{fig:Dic4_002} respectively, and the patterns can be understood in a similar way.  The expected distribution of the components of states with definite continuum $J^P$ across the little group irreps can be determined from Table \ref{table:subductions} and the extracted states are observed to match this pattern.  

We are able to extract states with exotic quantum numbers, identified as hybrid mesons.  In addition, the third excited $1^{--}$ state ($\rho'''$), $E/m_{\Omega} \sim 1.4$ at $|\vec{p}|^2=1$, has considerable overlap onto operators proportional to the commutator of two covariant derivatives, requiring a non-trivial gluonic field configuration.  This suggests that we identify this state as a non-exotic hybrid meson\footnote{or a mixture of hybrid and conventional mesons with a significant hybrid component} and corresponds to the non-exotic hybrid identified at zero momentum~\cite{Dudek:2010wm,Dudek:2011bn}.

The reduced symmetry of the $\Dic_2$ and $\Dic_3$ little groups compared to $\Dic_4$ leads to even more degeneracies in the spectrum.  For example, in the left hand panel of Fig.~\ref{fig:Dic2_011} there are two degenerate states in $A_1^+$ irrep with $E/m_{\Omega} \sim 1.07$ identified as $J^{PC}=2^{++}$; one is the $\lambda = 0$ piece and one is a $\lambda = \pm2$ piece, identified from their $Z$ values.  The other $\lambda = \pm2$ piece is in the $A_2^+$ irrep and the $\lambda = \pm1$ pieces are in the $B_1^+$ and $B_2^+$ irreps.  A similar situation arises in the $A_2^+$ irrep with two states $\sim 1.29$ identified as $2^{-+}$, one corresponding to $\lambda = 0$ and the other to $\lambda = \pm2$.

We note that, around $E/m_{\Omega} \sim 1.4$ there start to be states that we can not reliably extract or identify in some irreps.  For example, in $A_2^+$ and $A_2^-$ irreps with $|\vec{p}|^2 = 1$, we are missing the $\lambda = 0$ pieces of the $3^{++}$ and $3^{+-}$ states respectively (Fig.~\ref{fig:Dic4_001}).  This could be because, although we have an extensive operator basis, there is only one operator which overlaps with $J=3$ at rest in each of those two irreps.  To reliably extract a spin-three state and increase the number of other states we can extract, we would need to extend our operator basis to include three-derivative operators.

\begin{figure}[ptb]
\includegraphics[width=0.48\textwidth]{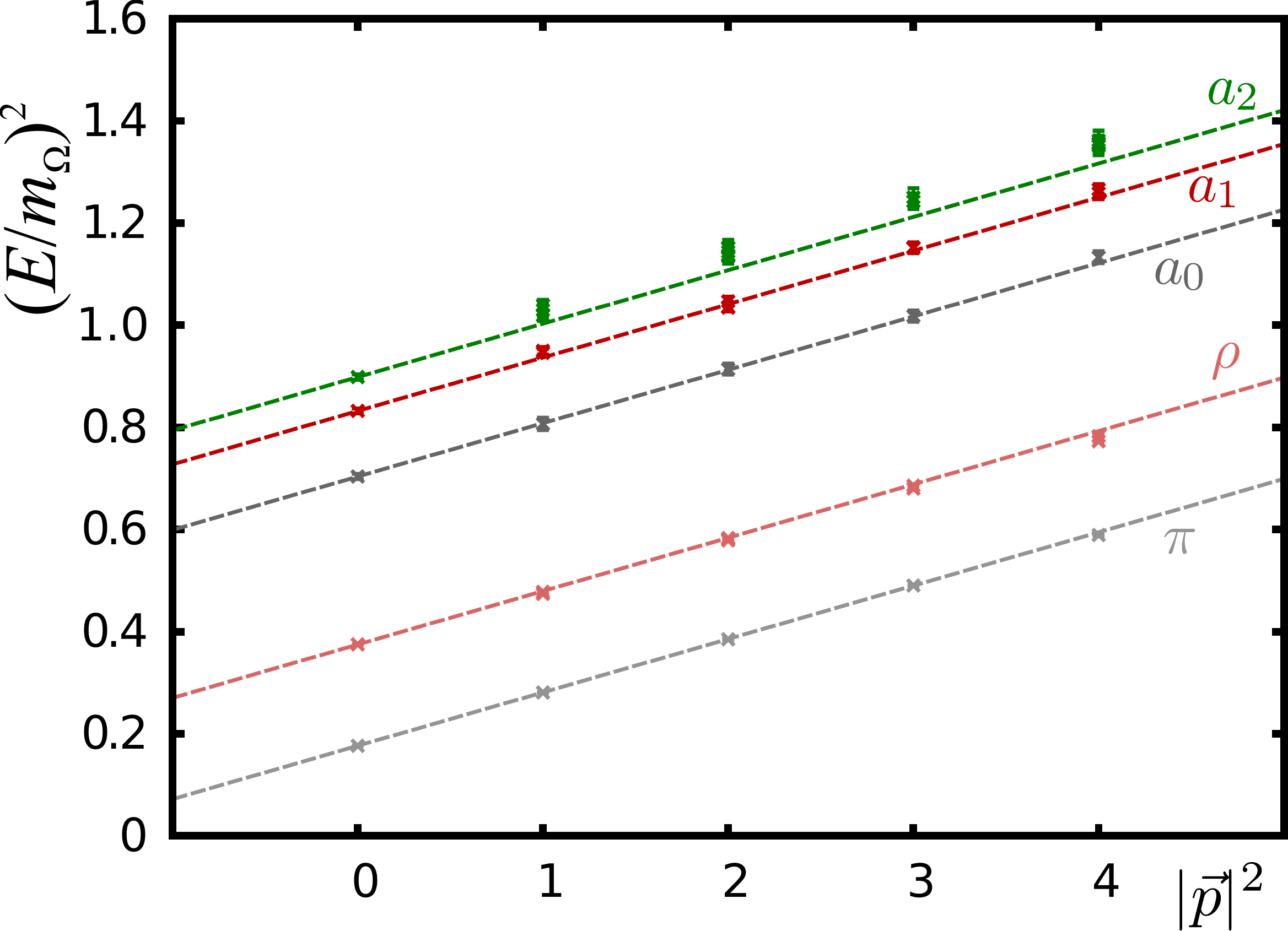}
\caption{Squared energies as a function of $|\vec{p}|^2$ for a selection of the lower-lying states.  The points (one for each irrep) are extracted energies; the lines are given by the dispersion relation, Eq.~\ref{equ:dispersion}, using $\xi = 3.441$ and the rest mass extracted in Ref.~\cite{Dudek:2010wm} (not showing any statistical uncertainty). }
\label{fig:dispersion}
\end{figure}

In Fig.~\ref{fig:dispersion} we show the extracted energies as a function of $|\vec{p}|^2$ for some of the lower-lying states.  Also shown are the expected energies given by the dispersion relation, Eq.~\ref{equ:dispersion}, using $\xi = 3.441$ and the rest mass extracted in Ref.~\cite{Dudek:2010wm}.  It can be seen that the extracted energies are in reasonably good agreement with the dispersion relation, highlighting the observation that the extracted energies are in good agreement with the expected states in Figs.~\ref{fig:Dic4_001}, \ref{fig:Dic2_011}, \ref{fig:Dic3_111} and \ref{fig:Dic4_002}, and indicating that with the dynamical tuning of parameters in the anisotropic lattice action, symmetry between space and time has been restored to good accuracy.

In summary, for all the $|\vec{p}|^2$ considered, we are able to identify the $J^{PC}$ of an extensive spectrum of extracted states across all irreps.  The observed pattern of states across irreps is the same as the pattern expected for the components of states with definite continuum $J^{PC}$ distributed across irreps (Table \ref{table:subductions}).  The energies of the different helicity components can be matched up between irreps and generally agree well.  However, the components in different irreps do not have to be degenerate in a finite volume and a possible interpretation for this is in terms of mixing with multi-particle states, something we discuss in the following section.

\section{Multi-meson states and helicities}
\label{sec:interp_helicity}

In an infinite volume continuum with rotational symmetry in three dimensions, a state's two helicity components with the same $|\lambda|$, $+|\lambda|$ and $-|\lambda|$, are constrained to have equal energy; they correspond to two rows of the same two-dimensional irrep.  However, components with different magnitudes of helicity, $|\lambda|$, of a state are \emph{not} constrained to have equal energy; they correspond to different irreps.  It is the additional constraints imposed by Lorentz symmetry, discussed in Section \ref{sec:overlaps}, which require that all the different helicity components have the same energy; it is in this sense that they correspond to one state, a state at rest which has been boosted to non-zero momentum.

In a finite volume, the reduced symmetry means that there are even fewer constraints and, depending on the little group corresponding to the particular momentum type, the two helicity components $+|\lambda|$ and $-|\lambda|$ may be split in energy.  For example, with $|\vec{p}|^2 = 1$ ($\Dic_4$), the two $\lambda = \pm 1$ components both subduce into the two-dimensional $E_2$ irrep and so are constrained to have equal energies.  However, with $|\vec{p}|^2 = 2$ ($\Dic_2$), the two $\lambda = \pm 1$ components are split between the $B_1$ and $B_2$ irreps and so are \emph{not} constrained to be degenerate.

\begin{figure*}[tb]
\includegraphics[width=0.43\textwidth]{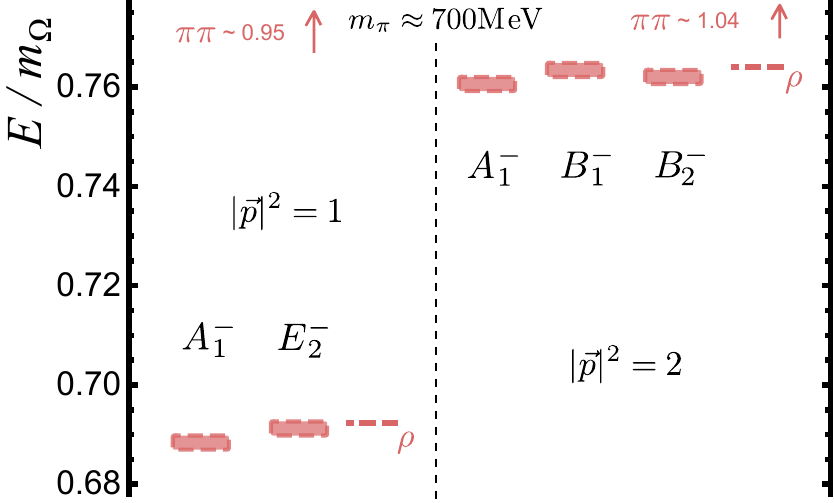}
\includegraphics[width=0.56\textwidth]{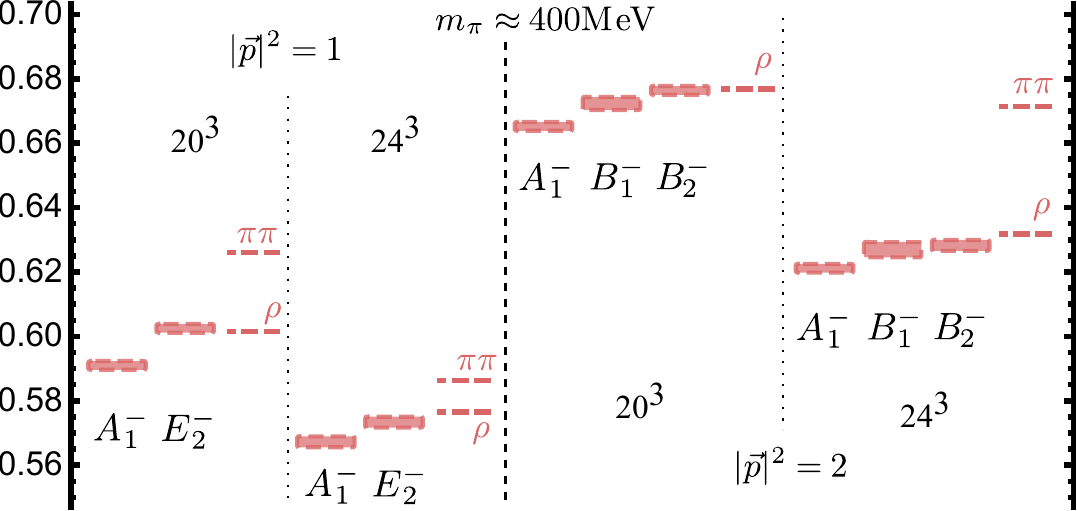}
\caption{Ground state energies with $|\vec{p}|^2=1$ and $2$ across different irreps, $\Lambda^C$.  The expected energies from the dispersion relation are shown as dashed lines.  The left hand panel shows results from the $16^3$ lattice with $m_{\pi} \approx 700$ MeV and the right hand panel shows a comparison with results from the $20^3$ and $24^3$ lattices with $m_{\pi} \approx 400$ MeV, along with the lowest $\pi\pi$ energy level in the absence of interactions (described in the text).}
\label{fig:rho_irrep_comp}
\end{figure*}

In the spectra we have extracted, in general we see very little splitting of energies between the different irreps.  As an example, in the left hand panel of Fig.~\ref{fig:rho_irrep_comp} we show the lowest energies extracted in the $A_1^-$ and $E_1^-$ irreps with $|\vec{p}|^2=1$, and in the $A_1^-$, $B_1^-$ and $B_2^-$ irreps with $|\vec{p}|^2 = 2$, along with the expected energies for the $\rho$ ($J^{PC}=1^{--}$) in flight from the dispersion relation.  There is no statistically significant splitting between different $|\lambda|$ or between the $B_1^-$ and $B_2^-$ energy levels, and these energies lie where expected from the dispersion relation.  We note that the lowest energy $\pi\pi$ states with overall momenta $\vec{p}=(0,0,1)$ and $\vec{p}=(0,1,1)$ are $\pi(0,0,0)\pi(0,0,1)$ and $\pi(0,0,0)\pi(0,1,1)$ respectively.  In the absence of interactions, these would have energies $E/m_{\Omega} \approx 0.95$ and $1.04$ respectively, significantly heavier than the corresponding ground state energies we extract in the relevant irreps.  This suggests that any mixing with $\pi\pi$ states will be small.  In fact, this is one reason for testing the helicity operator formalism on lattices with a relatively heavy pion.

For comparison, in the right hand panel of Fig.~\ref{fig:rho_irrep_comp} we show the corresponding energies extracted on $20^3$ and $24^3 \times 128$ lattices with lighter up and down quarks ($N_f=2+1$), corresponding to $m_{\pi} \approx 400$ MeV and having $a_t m_{\Omega} = 0.2951(22)$~\cite{Bulava:2010yg} and $\xi = 3.433(5)$ (from a fit to the $\pi$ dispersion relations across all volumes)\footnote{For the analyses on the lattices with $m_{\pi} \approx 400~\text{MeV}$, the operator basis consisted of only zero and one-derivative operators.}.  Here we observe more significant splittings between the energies in $A_1^-$ compared to $E_2^-$ or $B_1^-$,$B_2^-$, and a smaller splitting between $B_1$ and $B_2$; the pattern is the same on both volumes.  These splittings could arise from an imperfect restoration of Lorentz symmetry; however, comparison between the two lattices suggests an explanation in terms of mixing with multi-meson states in a finite volume. 

On the lattice with smaller $m_{\pi}$ and volume $20^3$ [$24^3$], the lowest energy $\pi\pi$ state with overall momentum $\vec{p}=(0,0,1)$ is $\pi(0,0,0)\pi(0,0,1)$ and in the absence of interactions this has energy $E/m_{\Omega} \approx 0.63$ [$0.59$]; the lowest $\pi\pi$ state with overall momentum $\vec{p}=(0,1,1)$ is $\pi(0,0,0)\pi(0,1,1)$ at $\approx 0.73$ [$0.67$] in the absence of interactions.  The relative closeness of these $\pi\pi$ energy levels to the energies we extract suggests that the $\pi\pi$ admixture in will be larger here compared to the lattices with heavier pions.

For $\vec{p}=(0,0,1)$, in the absence of interactions the $\pi(0,0,0)\pi(0,0,1)$ has lowest energy ($E/m_{\Omega} \approx 0.63$ or $0.59$ on the $20^3$ or $24^3$ respectively) and it subduces only into the $A_1$ irrep.  The next lowest energy state is $\pi(0,-1,0)\pi(0,1,1)$ ($\approx 0.89$ or $0.79$) and this subduces into both the $A_1$ and $E_2$ irreps\footnote{and also the $B_1$ irrep; this will be expanded upon in future work, but for the present discussion it suffices to observe that the pattern of subductions into $A_1$ is different from $E_2$.}.  Therefore, the pattern of $\pi\pi$ energy levels in the $A_1$ irrep is different from the pattern in the $E_2$ irrep and so the effect of multi-meson mixing in the $\rho$ will be different in the two irreps leading to a splitting of energies.

A similar interpretation applies to $\vec{p}=(0,1,1)$: the state $\pi(0,0,0)\pi(0,1,1)$ has lowest energy ($E/m_{\Omega} \approx 0.73$ or $0.67$) in the absence of interactions and it subduces only into the $A_1$ irrep.  The next lowest energy state is $\pi(0,1,0)\pi(0,0,1)$ ($\approx 0.78$ or $0.70$), subducing into $A_1$ and $B_1$, and the next is $\pi(0,1,0)\pi(1,1,1)$ ($\approx 0.98$ or $0.86$), subducing into $A_1$ and $B_2$.  Therefore, we can again qualitatively explain the splitting of the energy levels in terms of mixing with multi-meson states.

In summary, multi-particle states have a different pattern of energy levels across irreps compared to a single-particle state.  The pattern of degeneracies seen in Figs.~\ref{fig:Dic4_001}, \ref{fig:Dic2_011}, \ref{fig:Dic3_111} and \ref{fig:Dic4_002} do not show any clear evidence for multi-particle states.  However, there are suggestions that as we reduce the pion mass mixing with such states will become more important.  

In the full spectrum, we should observe additional states, the orthogonal mixture of single and multi-meson basis states.  However, we do not expect to be able to properly resolve multi-particle states with only fermion bilinear operators.  As discussed in Ref.~\cite{Dudek:2010wm}, if the operator basis is not sufficient, the variational method can not separate the states and a conservative approach is to suggest that our mass values are only accurate up to the order of the hadronic width of the state.  By supplementing our operator basis with two-meson operators (built from the helicity operators described herein), we can extract the full spectrum of single and two-meson states.  We will then be able to extract phase shifts and, if appropriate, resonance parameters using the method of L\"uscher\cite{Luscher:1990ux}, generalised to an overall non-zero momentum\cite{Rummukainen:1995vs}, utilised explicitly in \cite{Feng:2010es}; this will be the subject of future work.

\section{Conclusions}
\label{sec:conclusions}

We have shown how to construct fermion-bilinear operators for mesons in flight which are subduced from operators of definite helicity and respect the reduced symmetry of a cubic lattice in a finite cubic volume.  We have demonstrated the effectiveness of these operator constructions by using them to extract an extensive spectrum of isovectors mesons with various momenta.  We find that the spectrum can be described in terms of single-meson states of determined $J^{PC}$ and have reliably identified the quantum numbers of the extracted states, including states with exotic quantum numbers, states of high spin and excited states.

In order to study resonances and scattering in lattice QCD, it is necessary to extract a complete finite-volume meson spectrum, and to do this we must supplement our bases of operators resembling single-hadron states with operators resembling pairs of mesons projected onto definite relative momentum.  Such operators, interpolating the meson-meson pair $M_A\, M_B$, will be of the form ${\cal O}_{M_A}(\vec{p}) {\cal O}_{M_B}(\vec{p}\,')$, where ${\cal O}_{M_A}(\vec{p})$ efficiently interpolates meson $M_A$ with momentum $\vec{p}$.  Using large bases of operators of the form described herein, we can variationally diagonalise the matrix of correlators and hence obtain the optimal operator for $M_A$ with momentum $\vec{p}$ (and similarly for $M_B$).  The subsequent reduction of the contribution of excited $M_A^\star$ mesons to the $M_A M_B$ correlators will mean that the relevant energy levels can be extracted at earlier Euclidean times where statistical noise is smaller, leading to more precise determinations of signals for scattering.

With the extracted spectrum at light pion masses viewed in the finest detail, we observe effects that may be compatible with the admixture of meson-meson states.  In future calculations, meson-meson operators will be constructed from products of the in-flight operators derived herein, allowing us to resolve this admixture precisely and so to extract information on meson-meson scattering and any possible resonance~\cite{Luscher:1990ux,Rummukainen:1995vs,Feng:2010es}.

As well as their application to constructing multi-meson operators, these in-flight interpolators play a significant role in calculations of matrix elements at finite momentum, such as those used to measure form factors.  The inclusion of meson operators with non-zero momentum in such analyses gives further kinematic points, enabling the form factor to be calculated at more $Q^2$ points and so the $Q^2$ dependence to be mapped out more fully\cite{Dudek:2009kk,Collins:2011mk}.  Again, having an optimal operator to interpolate a meson will reduce the contribution of excited mesons, meaning that quantities can be extracted at earlier Euclidean times where statistical noise is smaller.

\begin{acknowledgments}
We thank our colleagues within the Hadron Spectrum Collaboration, in particular Mike Peardon and David Richards for useful discussions.  {\tt Chroma}~\cite{Edwards:2004sx} and {\tt QUDA}~\cite{Clark:2009wm,Babich:2010mu} were used to perform this work on clusters at Jefferson Laboratory under the USQCD Initiative and the LQCD ARRA project. Gauge configurations were generated using resources awarded from the U.S. Department of Energy INCITE program at Oak Ridge National Lab, the NSF Teragrid at the Texas Advanced Computer Center and the Pittsburgh Supercomputer Center, as well as at Jefferson Lab.

Authored by Jefferson Science Associates, LLC under U.S. DOE Contract No.
DE-AC05-06OR23177. The U.S. Government retains a non-exclusive, paid-up,
irrevocable, world-wide license to publish or reproduce this manuscript for
U.S. Government purposes. 
\end{acknowledgments}

\bibliography{MesonsInFlight}

\begin{thebibliography}{26}%
\makeatletter
\providecommand \@ifxundefined [1]{%
 \@ifx{#1\undefined}
}%
\providecommand \@ifnum [1]{%
 \ifnum #1\expandafter \@firstoftwo
 \else \expandafter \@secondoftwo
 \fi
}%
\providecommand \@ifx [1]{%
 \ifx #1\expandafter \@firstoftwo
 \else \expandafter \@secondoftwo
 \fi
}%
\providecommand \natexlab [1]{#1}%
\providecommand \enquote  [1]{``#1''}%
\providecommand \bibnamefont  [1]{#1}%
\providecommand \bibfnamefont [1]{#1}%
\providecommand \citenamefont [1]{#1}%
\providecommand \href@noop [0]{\@secondoftwo}%
\providecommand \href [0]{\begingroup \@sanitize@url \@href}%
\providecommand \@href[1]{\@@startlink{#1}\@@href}%
\providecommand \@@href[1]{\endgroup#1\@@endlink}%
\providecommand \@sanitize@url [0]{\catcode `\\12\catcode `\$12\catcode
  `\&12\catcode `\#12\catcode `\^12\catcode `\_12\catcode `\%12\relax}%
\providecommand \@@startlink[1]{}%
\providecommand \@@endlink[0]{}%
\providecommand \url  [0]{\begingroup\@sanitize@url \@url }%
\providecommand \@url [1]{\endgroup\@href {#1}{\urlprefix }}%
\providecommand \urlprefix  [0]{URL }%
\providecommand \Eprint [0]{\href }%
\providecommand \doibase [0]{http://dx.doi.org/}%
\providecommand \selectlanguage [0]{\@gobble}%
\providecommand \bibinfo  [0]{\@secondoftwo}%
\providecommand \bibfield  [0]{\@secondoftwo}%
\providecommand \translation [1]{[#1]}%
\providecommand \BibitemOpen [0]{}%
\providecommand \bibitemStop [0]{}%
\providecommand \bibitemNoStop [0]{.\EOS\space}%
\providecommand \EOS [0]{\spacefactor3000\relax}%
\providecommand \BibitemShut  [1]{\csname bibitem#1\endcsname}%
\let\auto@bib@innerbib\@empty
\bibitem [{\citenamefont {Luscher}(1991)}]{Luscher:1990ux}%
  \BibitemOpen
  \bibfield  {author} {\bibinfo {author} {\bibfnamefont {M.}~\bibnamefont
  {Luscher}},\ }\href {\doibase 10.1016/0550-3213(91)90366-6} {\bibfield
  {journal} {\bibinfo  {journal} {Nucl. Phys.}\ }\textbf {\bibinfo {volume}
  {B354}},\ \bibinfo {pages} {531} (\bibinfo {year} {1991})}\BibitemShut
  {NoStop}%
\bibitem [{\citenamefont {Dudek}\ \emph
  {et~al.}(2009{\natexlab{a}})\citenamefont {Dudek}, \citenamefont {Edwards},\
  and\ \citenamefont {Thomas}}]{Dudek:2009kk}%
  \BibitemOpen
  \bibfield  {author} {\bibinfo {author} {\bibfnamefont {J.~J.}\ \bibnamefont
  {Dudek}}, \bibinfo {author} {\bibfnamefont {R.}~\bibnamefont {Edwards}}, \
  and\ \bibinfo {author} {\bibfnamefont {C.~E.}\ \bibnamefont {Thomas}},\
  }\href {\doibase 10.1103/PhysRevD.79.094504} {\bibfield  {journal} {\bibinfo
  {journal} {Phys. Rev.}\ }\textbf {\bibinfo {volume} {D79}},\ \bibinfo {pages}
  {094504} (\bibinfo {year} {2009}{\natexlab{a}})},\ \Eprint
  {http://arxiv.org/abs/0902.2241} {arXiv:0902.2241 [hep-ph]} \BibitemShut
  {NoStop}%
\bibitem [{\citenamefont {Collins}\ \emph {et~al.}(2011)\citenamefont {Collins}
  \emph {et~al.}}]{Collins:2011mk}%
  \BibitemOpen
  \bibfield  {author} {\bibinfo {author} {\bibfnamefont {S.}~\bibnamefont
  {Collins}} \emph {et~al.},\ }\href@noop {} {\  (\bibinfo {year} {2011})},\
  \Eprint {http://arxiv.org/abs/1106.3580} {arXiv:1106.3580 [hep-lat]}
  \BibitemShut {NoStop}%
\bibitem [{\citenamefont {Dudek}\ \emph {et~al.}(2010)\citenamefont {Dudek},
  \citenamefont {Edwards}, \citenamefont {Peardon}, \citenamefont {Richards},\
  and\ \citenamefont {Thomas}}]{Dudek:2010wm}%
  \BibitemOpen
  \bibfield  {author} {\bibinfo {author} {\bibfnamefont {J.~J.}\ \bibnamefont
  {Dudek}}, \bibinfo {author} {\bibfnamefont {R.~G.}\ \bibnamefont {Edwards}},
  \bibinfo {author} {\bibfnamefont {M.~J.}\ \bibnamefont {Peardon}}, \bibinfo
  {author} {\bibfnamefont {D.~G.}\ \bibnamefont {Richards}}, \ and\ \bibinfo
  {author} {\bibfnamefont {C.~E.}\ \bibnamefont {Thomas}},\ }\href {\doibase
  10.1103/PhysRevD.82.034508} {\bibfield  {journal} {\bibinfo  {journal} {Phys.
  Rev.}\ }\textbf {\bibinfo {volume} {D82}},\ \bibinfo {pages} {034508}
  (\bibinfo {year} {2010})},\ \Eprint {http://arxiv.org/abs/1004.4930}
  {arXiv:1004.4930 [hep-ph]} \BibitemShut {NoStop}%
\bibitem [{\citenamefont {Edwards}\ \emph {et~al.}(2008)\citenamefont
  {Edwards}, \citenamefont {Joo},\ and\ \citenamefont {Lin}}]{Edwards:2008ja}%
  \BibitemOpen
  \bibfield  {author} {\bibinfo {author} {\bibfnamefont {R.~G.}\ \bibnamefont
  {Edwards}}, \bibinfo {author} {\bibfnamefont {B.}~\bibnamefont {Joo}}, \ and\
  \bibinfo {author} {\bibfnamefont {H.-W.}\ \bibnamefont {Lin}},\ }\href
  {\doibase 10.1103/PhysRevD.78.054501} {\bibfield  {journal} {\bibinfo
  {journal} {Phys. Rev.}\ }\textbf {\bibinfo {volume} {D78}},\ \bibinfo {pages}
  {054501} (\bibinfo {year} {2008})},\ \Eprint {http://arxiv.org/abs/0803.3960}
  {arXiv:0803.3960 [hep-lat]} \BibitemShut {NoStop}%
\bibitem [{\citenamefont {Lin}\ \emph {et~al.}(2009)\citenamefont {Lin} \emph
  {et~al.}}]{Lin:2008pr}%
  \BibitemOpen
  \bibfield  {author} {\bibinfo {author} {\bibfnamefont {H.-W.}\ \bibnamefont
  {Lin}} \emph {et~al.} (\bibinfo {collaboration} {Hadron Spectrum}),\ }\href
  {\doibase 10.1103/PhysRevD.79.034502} {\bibfield  {journal} {\bibinfo
  {journal} {Phys. Rev.}\ }\textbf {\bibinfo {volume} {D79}},\ \bibinfo {pages}
  {034502} (\bibinfo {year} {2009})},\ \Eprint {http://arxiv.org/abs/0810.3588}
  {arXiv:0810.3588 [hep-lat]} \BibitemShut {NoStop}%
\bibitem [{\citenamefont {Rummukainen}\ and\ \citenamefont
  {Gottlieb}(1995)}]{Rummukainen:1995vs}%
  \BibitemOpen
  \bibfield  {author} {\bibinfo {author} {\bibfnamefont {K.}~\bibnamefont
  {Rummukainen}}\ and\ \bibinfo {author} {\bibfnamefont {S.~A.}\ \bibnamefont
  {Gottlieb}},\ }\href {\doibase 10.1016/0550-3213(95)00313-H} {\bibfield
  {journal} {\bibinfo  {journal} {Nucl. Phys.}\ }\textbf {\bibinfo {volume}
  {B450}},\ \bibinfo {pages} {397} (\bibinfo {year} {1995})},\ \Eprint
  {http://arxiv.org/abs/hep-lat/9503028} {arXiv:hep-lat/9503028} \BibitemShut
  {NoStop}%
\bibitem [{\citenamefont {Feng}\ \emph {et~al.}(2011)\citenamefont {Feng},
  \citenamefont {Jansen},\ and\ \citenamefont {Renner}}]{Feng:2010es}%
  \BibitemOpen
  \bibfield  {author} {\bibinfo {author} {\bibfnamefont {X.}~\bibnamefont
  {Feng}}, \bibinfo {author} {\bibfnamefont {K.}~\bibnamefont {Jansen}}, \ and\
  \bibinfo {author} {\bibfnamefont {D.~B.}\ \bibnamefont {Renner}},\ }\href
  {\doibase 10.1103/PhysRevD.83.094505} {\bibfield  {journal} {\bibinfo
  {journal} {Phys. Rev.}\ }\textbf {\bibinfo {volume} {D83}},\ \bibinfo {pages}
  {094505} (\bibinfo {year} {2011})},\ \Eprint {http://arxiv.org/abs/1011.5288}
  {arXiv:1011.5288 [hep-lat]} \BibitemShut {NoStop}%
\bibitem [{\citenamefont {Chung}()}]{Chung:spin}%
  \BibitemOpen
  \bibfield  {author} {\bibinfo {author} {\bibfnamefont {S.~U.}\ \bibnamefont
  {Chung}},\ }\href@noop {} {\ }\bibinfo {note} {CERN-71-08, BNL-QGS-02-0900,
  updated version March 2, 2004}\BibitemShut {NoStop}%
\bibitem [{\citenamefont {Jacob}\ and\ \citenamefont
  {Wick}(1959)}]{Jacob:1959at}%
  \BibitemOpen
  \bibfield  {author} {\bibinfo {author} {\bibfnamefont {M.}~\bibnamefont
  {Jacob}}\ and\ \bibinfo {author} {\bibfnamefont {G.~C.}\ \bibnamefont
  {Wick}},\ }\href {\doibase 10.1016/0003-4916(59)90051-X} {\bibfield
  {journal} {\bibinfo  {journal} {Ann. Phys.}\ }\textbf {\bibinfo {volume}
  {7}},\ \bibinfo {pages} {404} (\bibinfo {year} {1959})}\BibitemShut {NoStop}%
\bibitem [{\citenamefont {Dudek}\ \emph {et~al.}(2008)\citenamefont {Dudek},
  \citenamefont {Edwards}, \citenamefont {Mathur},\ and\ \citenamefont
  {Richards}}]{Dudek:2007wv}%
  \BibitemOpen
  \bibfield  {author} {\bibinfo {author} {\bibfnamefont {J.~J.}\ \bibnamefont
  {Dudek}}, \bibinfo {author} {\bibfnamefont {R.~G.}\ \bibnamefont {Edwards}},
  \bibinfo {author} {\bibfnamefont {N.}~\bibnamefont {Mathur}}, \ and\ \bibinfo
  {author} {\bibfnamefont {D.~G.}\ \bibnamefont {Richards}},\ }\href {\doibase
  10.1103/PhysRevD.77.034501} {\bibfield  {journal} {\bibinfo  {journal} {Phys.
  Rev.}\ }\textbf {\bibinfo {volume} {D77}},\ \bibinfo {pages} {034501}
  (\bibinfo {year} {2008})},\ \Eprint {http://arxiv.org/abs/0707.4162}
  {arXiv:0707.4162 [hep-lat]} \BibitemShut {NoStop}%
\bibitem [{\citenamefont {Moore}\ and\ \citenamefont
  {Fleming}(2006{\natexlab{a}})}]{Moore:2005dw}%
  \BibitemOpen
  \bibfield  {author} {\bibinfo {author} {\bibfnamefont {D.~C.}\ \bibnamefont
  {Moore}}\ and\ \bibinfo {author} {\bibfnamefont {G.~T.}\ \bibnamefont
  {Fleming}},\ }\href {\doibase 10.1103/PhysRevD.73.014504} {\bibfield
  {journal} {\bibinfo  {journal} {Phys. Rev.}\ }\textbf {\bibinfo {volume}
  {D73}},\ \bibinfo {pages} {014504} (\bibinfo {year} {2006}{\natexlab{a}})},\
  \Eprint {http://arxiv.org/abs/hep-lat/0507018} {arXiv:hep-lat/0507018}
  \BibitemShut {NoStop}%
\bibitem [{\citenamefont {Moore}\ and\ \citenamefont
  {Fleming}(2006{\natexlab{b}})}]{Moore:2006ng}%
  \BibitemOpen
  \bibfield  {author} {\bibinfo {author} {\bibfnamefont {D.~C.}\ \bibnamefont
  {Moore}}\ and\ \bibinfo {author} {\bibfnamefont {G.~T.}\ \bibnamefont
  {Fleming}},\ }\href {\doibase 10.1103/PhysRevD.74.054504} {\bibfield
  {journal} {\bibinfo  {journal} {Phys. Rev.}\ }\textbf {\bibinfo {volume}
  {D74}},\ \bibinfo {pages} {054504} (\bibinfo {year} {2006}{\natexlab{b}})},\
  \Eprint {http://arxiv.org/abs/hep-lat/0607004} {arXiv:hep-lat/0607004}
  \BibitemShut {NoStop}%
\bibitem [{\citenamefont {Dudek}\ \emph
  {et~al.}(2009{\natexlab{b}})\citenamefont {Dudek}, \citenamefont {Edwards},
  \citenamefont {Peardon}, \citenamefont {Richards},\ and\ \citenamefont
  {Thomas}}]{Dudek:2009qf}%
  \BibitemOpen
  \bibfield  {author} {\bibinfo {author} {\bibfnamefont {J.~J.}\ \bibnamefont
  {Dudek}}, \bibinfo {author} {\bibfnamefont {R.~G.}\ \bibnamefont {Edwards}},
  \bibinfo {author} {\bibfnamefont {M.~J.}\ \bibnamefont {Peardon}}, \bibinfo
  {author} {\bibfnamefont {D.~G.}\ \bibnamefont {Richards}}, \ and\ \bibinfo
  {author} {\bibfnamefont {C.~E.}\ \bibnamefont {Thomas}},\ }\href {\doibase
  10.1103/PhysRevLett.103.262001} {\bibfield  {journal} {\bibinfo  {journal}
  {Phys. Rev. Lett.}\ }\textbf {\bibinfo {volume} {103}},\ \bibinfo {pages}
  {262001} (\bibinfo {year} {2009}{\natexlab{b}})},\ \Eprint
  {http://arxiv.org/abs/0909.0200} {arXiv:0909.0200 [hep-ph]} \BibitemShut
  {NoStop}%
\bibitem [{\citenamefont {Peardon}\ \emph {et~al.}(2009)\citenamefont {Peardon}
  \emph {et~al.}}]{Peardon:2009gh}%
  \BibitemOpen
  \bibfield  {author} {\bibinfo {author} {\bibfnamefont {M.}~\bibnamefont
  {Peardon}} \emph {et~al.} (\bibinfo {collaboration} {Hadron Spectrum}),\
  }\href {\doibase 10.1103/PhysRevD.80.054506} {\bibfield  {journal} {\bibinfo
  {journal} {Phys. Rev.}\ }\textbf {\bibinfo {volume} {D80}},\ \bibinfo {pages}
  {054506} (\bibinfo {year} {2009})},\ \Eprint {http://arxiv.org/abs/0905.2160}
  {arXiv:0905.2160 [hep-lat]} \BibitemShut {NoStop}%
\bibitem [{\citenamefont {Bulava}\ \emph {et~al.}(2010)\citenamefont {Bulava}
  \emph {et~al.}}]{Bulava:2010yg}%
  \BibitemOpen
  \bibfield  {author} {\bibinfo {author} {\bibfnamefont {J.}~\bibnamefont
  {Bulava}} \emph {et~al.},\ }\href {\doibase 10.1103/PhysRevD.82.014507}
  {\bibfield  {journal} {\bibinfo  {journal} {Phys. Rev.}\ }\textbf {\bibinfo
  {volume} {D82}},\ \bibinfo {pages} {014507} (\bibinfo {year} {2010})},\
  \Eprint {http://arxiv.org/abs/1004.5072} {arXiv:1004.5072 [hep-lat]}
  \BibitemShut {NoStop}%
\bibitem [{\citenamefont {Dudek}\ \emph
  {et~al.}(2011{\natexlab{a}})\citenamefont {Dudek}, \citenamefont {Edwards},
  \citenamefont {Peardon}, \citenamefont {Richards},\ and\ \citenamefont
  {Thomas}}]{Dudek:2010ew}%
  \BibitemOpen
  \bibfield  {author} {\bibinfo {author} {\bibfnamefont {J.~J.}\ \bibnamefont
  {Dudek}}, \bibinfo {author} {\bibfnamefont {R.~G.}\ \bibnamefont {Edwards}},
  \bibinfo {author} {\bibfnamefont {M.~J.}\ \bibnamefont {Peardon}}, \bibinfo
  {author} {\bibfnamefont {D.~G.}\ \bibnamefont {Richards}}, \ and\ \bibinfo
  {author} {\bibfnamefont {C.~E.}\ \bibnamefont {Thomas}},\ }\href {\doibase
  10.1103/PhysRevD.83.071504} {\bibfield  {journal} {\bibinfo  {journal} {Phys.
  Rev.}\ }\textbf {\bibinfo {volume} {D83}},\ \bibinfo {pages} {071504}
  (\bibinfo {year} {2011}{\natexlab{a}})},\ \Eprint
  {http://arxiv.org/abs/1011.6352} {arXiv:1011.6352 [hep-ph]} \BibitemShut
  {NoStop}%
\bibitem [{\citenamefont {Dudek}\ \emph
  {et~al.}(2011{\natexlab{b}})\citenamefont {Dudek} \emph
  {et~al.}}]{Dudek:2011tt}%
  \BibitemOpen
  \bibfield  {author} {\bibinfo {author} {\bibfnamefont {J.~J.}\ \bibnamefont
  {Dudek}} \emph {et~al.},\ }\href {\doibase 10.1103/PhysRevD.83.111502}
  {\bibfield  {journal} {\bibinfo  {journal} {Phys. Rev.}\ }\textbf {\bibinfo
  {volume} {D83}},\ \bibinfo {pages} {111502} (\bibinfo {year}
  {2011}{\natexlab{b}})},\ \Eprint {http://arxiv.org/abs/1102.4299}
  {arXiv:1102.4299 [hep-lat]} \BibitemShut {NoStop}%
\bibitem [{\citenamefont {Edwards}\ \emph {et~al.}(2011)\citenamefont
  {Edwards}, \citenamefont {Dudek}, \citenamefont {Richards},\ and\
  \citenamefont {Wallace}}]{Edwards:2011jj}%
  \BibitemOpen
  \bibfield  {author} {\bibinfo {author} {\bibfnamefont {R.~G.}\ \bibnamefont
  {Edwards}}, \bibinfo {author} {\bibfnamefont {J.~J.}\ \bibnamefont {Dudek}},
  \bibinfo {author} {\bibfnamefont {D.~G.}\ \bibnamefont {Richards}}, \ and\
  \bibinfo {author} {\bibfnamefont {S.~J.}\ \bibnamefont {Wallace}},\ }\href
  {\doibase 10.1103/PhysRevD.84.074508} {\bibfield  {journal} {\bibinfo
  {journal} {Phys. Rev.}\ }\textbf {\bibinfo {volume} {D84}},\ \bibinfo {pages}
  {074508} (\bibinfo {year} {2011})},\ \Eprint {http://arxiv.org/abs/1104.5152}
  {arXiv:1104.5152 [hep-ph]} \BibitemShut {NoStop}%
\bibitem [{\citenamefont {Michael}(1985)}]{Michael:1985ne}%
  \BibitemOpen
  \bibfield  {author} {\bibinfo {author} {\bibfnamefont {C.}~\bibnamefont
  {Michael}},\ }\href {\doibase 10.1016/0550-3213(85)90297-4} {\bibfield
  {journal} {\bibinfo  {journal} {Nucl. Phys.}\ }\textbf {\bibinfo {volume}
  {B259}},\ \bibinfo {pages} {58} (\bibinfo {year} {1985})}\BibitemShut
  {NoStop}%
\bibitem [{\citenamefont {Luscher}\ and\ \citenamefont
  {Wolff}(1990)}]{Luscher:1990ck}%
  \BibitemOpen
  \bibfield  {author} {\bibinfo {author} {\bibfnamefont {M.}~\bibnamefont
  {Luscher}}\ and\ \bibinfo {author} {\bibfnamefont {U.}~\bibnamefont
  {Wolff}},\ }\href {\doibase 10.1016/0550-3213(90)90540-T} {\bibfield
  {journal} {\bibinfo  {journal} {Nucl. Phys.}\ }\textbf {\bibinfo {volume}
  {B339}},\ \bibinfo {pages} {222} (\bibinfo {year} {1990})}\BibitemShut
  {NoStop}%
\bibitem [{\citenamefont {Dudek}(2011)}]{Dudek:2011bn}%
  \BibitemOpen
  \bibfield  {author} {\bibinfo {author} {\bibfnamefont {J.~J.}\ \bibnamefont
  {Dudek}},\ }\href {\doibase 10.1103/PhysRevD.84.074023} {\bibfield  {journal}
  {\bibinfo  {journal} {Phys. Rev.}\ }\textbf {\bibinfo {volume} {D84}},\
  \bibinfo {pages} {074023} (\bibinfo {year} {2011})},\ \Eprint
  {http://arxiv.org/abs/1106.5515} {arXiv:1106.5515 [hep-ph]} \BibitemShut
  {NoStop}%
\bibitem [{\citenamefont {Edwards}\ and\ \citenamefont
  {Joo}(2005)}]{Edwards:2004sx}%
  \BibitemOpen
  \bibfield  {author} {\bibinfo {author} {\bibfnamefont {R.~G.}\ \bibnamefont
  {Edwards}}\ and\ \bibinfo {author} {\bibfnamefont {B.}~\bibnamefont {Joo}}
  (\bibinfo {collaboration} {SciDAC Collaboration}),\ }\href@noop {} {\bibfield
   {journal} {\bibinfo  {journal} {Nucl. Phys. B. Proc. Suppl.}\ }\textbf
  {\bibinfo {volume} {140}},\ \bibinfo {pages} {832} (\bibinfo {year}
  {2005})},\ \Eprint {http://arxiv.org/abs/hep-lat/0409003} {hep-lat/0409003}
  \BibitemShut {NoStop}%
\bibitem [{\citenamefont {Clark}\ \emph {et~al.}(2010)\citenamefont {Clark},
  \citenamefont {Babich}, \citenamefont {Barros}, \citenamefont {Brower},\ and\
  \citenamefont {Rebbi}}]{Clark:2009wm}%
  \BibitemOpen
  \bibfield  {author} {\bibinfo {author} {\bibfnamefont {M.~A.}\ \bibnamefont
  {Clark}}, \bibinfo {author} {\bibfnamefont {R.}~\bibnamefont {Babich}},
  \bibinfo {author} {\bibfnamefont {K.}~\bibnamefont {Barros}}, \bibinfo
  {author} {\bibfnamefont {R.~C.}\ \bibnamefont {Brower}}, \ and\ \bibinfo
  {author} {\bibfnamefont {C.}~\bibnamefont {Rebbi}},\ }\href {\doibase
  10.1016/j.cpc.2010.05.002} {\bibfield  {journal} {\bibinfo  {journal}
  {Comput. Phys. Commun.}\ }\textbf {\bibinfo {volume} {181}},\ \bibinfo
  {pages} {1517} (\bibinfo {year} {2010})},\ \Eprint
  {http://arxiv.org/abs/0911.3191} {arXiv:0911.3191 [hep-lat]} \BibitemShut
  {NoStop}%
\bibitem [{\citenamefont {Babich}\ \emph {et~al.}(2010)\citenamefont {Babich},
  \citenamefont {Clark},\ and\ \citenamefont {Joo}}]{Babich:2010mu}%
  \BibitemOpen
  \bibfield  {author} {\bibinfo {author} {\bibfnamefont {R.}~\bibnamefont
  {Babich}}, \bibinfo {author} {\bibfnamefont {M.~A.}\ \bibnamefont {Clark}}, \
  and\ \bibinfo {author} {\bibfnamefont {B.}~\bibnamefont {Joo}},\ }in\ \href
  {\doibase 10.1109/SC.2010.40} {\emph {\bibinfo {booktitle} {International
  Conference for High Performance Computing, Networking, Storage and Analysis
  (SC)}}}\ (\bibinfo {year} {2010})\ pp.\ \bibinfo {pages} {1--11},\ \Eprint
  {http://arxiv.org/abs/1011.0024} {arXiv:1011.0024 [hep-lat]} \BibitemShut
  {NoStop}%
\bibitem [{\citenamefont {Martin}\ and\ \citenamefont
  {Spearman}(1970)}]{MartinSpearman}%
  \BibitemOpen
  \bibfield  {author} {\bibinfo {author} {\bibfnamefont {A.}~\bibnamefont
  {Martin}}\ and\ \bibinfo {author} {\bibfnamefont {T.}~\bibnamefont
  {Spearman}},\ }\href@noop {} {\emph {\bibinfo {title} {Elementary Particle
  Theory}}}\ (\bibinfo  {publisher} {North-Holland Publishing Company},\
  \bibinfo {year} {1970})\BibitemShut {NoStop}%
\end{thebibliography}%

\appendix

\section{Helicity states and helicity operators}
\label{app:helicity}

Here we describe the construction of \emph{helicity operators}, using helicity states as an analogy and following the helicity state constructions of Jacob and Wick~\cite{Jacob:1959at}.  Parts of Chung~\cite{Chung:spin} and Chapter 3 of Martin and Spearman~\cite{MartinSpearman} are also relevant.  

The operator $\hat{R}_{\phi,\theta,\psi} = e^{-i \phi \hat{J}_z} e^{-i \theta \hat{J}_y} e^{-i \psi \hat{J}_z}$ rotates around the $z$-axis by $\psi$, then around the $y$-axis by $\theta$ and finally around the $z$-axis by $\phi$ (with a fixed coordinate system).  As a shorthand we will denote the (active) rotation which rotates from the $z$-axis to the direction of $\vec{p}$ [which is $(p,\theta,\phi)$ in spherical coordinates] by $\hat{R}_0$.  There are many possible choices for $\hat{R}_0$ and these lead to different phases in the definition of helicity states.  Jacob and Wick choose $\hat{R}_{\phi,\theta,\phi}$ and Chung chooses $\hat{R}_{\phi,\theta,0}$, but the choice is not important for the present discussion as long as one convention is used consistently.

Denoting a boost along the $z$-axis with magnitude $p$ by $\hat{L}_z(p)$, a \emph{helicity state} is defined by
\begin{equation}
\left|\vec{p}; J, \lambda \right> \equiv \hat{R}_0 \hat{L}_z(p) \left| J, \lambda \right> ~,
\label{equ:helstate}
\end{equation}
where $\left| J, \lambda \right >$ is a state at rest with spin $J$ and $J_z$-component $\lambda$.  From the definition, it is simple to show that under some arbitrary rotation, $\hat{R}$, these helicity states transform as\footnote{Note that there may be some phase introduced here if the rotation has some component around the $\vec{p}$ direction.}
\begin{equation}
\hat{R} \bigl|\vec{p}; J, \lambda \bigr> = \bigl| \hat{R}\vec{p}; J, \lambda \bigr> ~, \nonumber
\end{equation}
and the helicity is invariant.

In contrast, a \emph{canonical state} (using the terminology of Chung), where the spin component is measured along the $z$-axis, is defined by
\begin{eqnarray}
\left|\vec{p}; J, m \right> &\equiv& \hat{R}_0 \hat{L}_z(p) \hat{R}_0^{-1} \left| J, m \right> \nonumber \\
 &=& \hat{L}(\vec{p}) \left| J, m \right> ~,
\label{equ:canonstate}
\end{eqnarray}
where $\hat{L}(\vec{p}) = \hat{R}_0 \hat{L}_z(p) \hat{R}_0^{-1}$ is a Lorentz boost along direction $\hat{p}$ with magnitude $p$.  These states transform under rotations as
\begin{equation}
\hat{R} \bigl|\vec{p}; J, m \bigr> = \sum_{m'} \mathcal{D}^{(J)}_{m' m}(R) \bigl|\hat{R}\vec{p}; J, m' \bigr> ~, \nonumber
\end{equation}
where $\mathcal{D}^{(J)}_{m' m}(R)$ is a Wigner-$\mathcal{D}$ matrix, i.e. these states transform under rotations in the same way as $\left| J, m \right>$.

Using the definitions in Eqs.~\ref{equ:helstate} and \ref{equ:canonstate}, it is straightforward to show that helicity and canonical states are related by
\begin{equation}
\left|\vec{p}; J, \lambda \right> = \sum_{m} \mathcal{D}^{(J)}_{m \lambda}(\hat{R}_0) \left|\vec{p}; J, m \right> ~.
\end{equation}

We have shown previously\cite{Dudek:2010wm} that, in the continuum, when $\vec{p}=0$ the $\mathcal{O}^{J,M}$ constructed in Section \ref{sec:helops} overlap with only one spin:
\begin{equation}
\bigl<0 \bigl| \mathcal{O}^{J,m}(\vec{p}=\vec{0}) \bigr| \vec{p}=\vec{0}; J',m' \bigr> = Z^{[J]} \delta_{J,J'} \delta_{m,m'} ~,
\end{equation}
and so we can write
\begin{equation}
\bigl|\vec{0}; J,m \bigr> = [\mathcal{O}^{J,m}(\vec{0})]^{\dagger} \bigl| 0 \bigr>  ~.
\end{equation}
In particular, the operator, $[\mathcal{O}^{J,m}(\vec{0})]^{\dagger}$, transforms in a similar way to the state.  We can therefore carry over the discussion of canonical and helicity states and define a boosted \emph{canonical operator}, 
\begin{equation}
[\mathcal{O}^{J,m}(\vec{p})]^{\dagger} = \hat{L}(\vec{p}) [\mathcal{O}^{J,m}(\vec{0})]^{\dagger} \hat{L}^{-1}(\vec{p})  ~.
\end{equation}
A \emph{helicity operator} is then given by
\begin{equation}
[\mathbb{O}^{J,\lambda}(\vec{p})]^{\dagger} = \sum_{m} \mathcal{D}^{(J)}_{m \lambda}(R) ~ [\mathcal{O}^{J,m}(\vec{p})]^{\dagger} ~,
\end{equation}
and transforms in the same way as a helicity state.  Taking the complex conjugate we obtain
\begin{equation}
\mathbb{O}^{J,\lambda}(\vec{p}) = \sum_{m} \mathcal{D}^{(J)*}_{m \lambda}(R) ~ \mathcal{O}^{J,m}(\vec{p}) ~,
\end{equation}
which is Eq.~\ref{equ:helicityops} in Section \ref{sec:helops}.

\subsection{Parity and reflections}
\label{app:parity}

Helicity states at non-zero momentum are \emph{not} eigenstates of parity because a parity transformation, $\hat{P}$, reverses the direction of the momentum, $\vec{p} \rightarrow -\vec{p}$.  A reflection in a plane containing the momentum direction, $\hat{\Pi}$ (a parity transformation followed by a rotation to bring the momentum direction back to the original direction), preserves $\vec{p}$.  However, helicity states are also in general \emph{not} eigenstates of $\hat{\Pi}$ because under such a transformation $\lambda \rightarrow -\lambda$.  

As an example, consider a helicity state with spin $J$, rest parity $P$ and momentum, $\vec{p_z}$, directed along the $z$-axis, $\left|\vec{p_z}; J^P, \lambda \right>$; the state at rest is denoted by $\left|J^P, \lambda \right>$.  If $\hat{\Pi}_{yz}$ is a reflection in the $yz$ plane ($x \rightarrow -x$) and $\hat{P}$ is a parity transformation ($x,y,z \rightarrow -x,-y,-z$) then 
$$\hat{\Pi}_{yz} = e^{-i \pi \hat{J}_x} \hat{P} = e^{-i \pi \hat{J}_y} e^{-i \pi \hat{J}_z} \hat{P} ~.$$
In analogy to the discussion in Ref.~\cite{Jacob:1959at},
\begin{eqnarray*}
e^{-i \pi \hat{J}_y} e^{-i \pi \hat{J}_z} \left|J^P, \lambda \right> &=& \sum_{\lambda'} \mathcal{D}^{(J)}_{\lambda' \lambda}(\pi,\pi,0) \left|J^P, \lambda' \right> ~, \\
&=& (-1)^J \left|J^P, -\lambda \right> ~.
\end{eqnarray*}
Since $\hat{P}$ commutes with $\hat{R}$ and $\hat{L}_z(p)$ commutes with $\hat{\Pi}_{yz}$, we have
\begin{equation}
\hat{\Pi}_{yz} \left|\vec{p_z}; J^P, \lambda \right> = P(-1)^J \left|\vec{p_z}; J^P, -\lambda \right> ~.
\label{equ:reflect_yz}
\end{equation}

If instead we consider a reflection in the $xz$ plane ($y \rightarrow -y$), $\hat{\Pi}_{xz}$, then the discussion proceeds in a similar way~\cite{Jacob:1959at},
$$\hat{\Pi}_{xz} = e^{-i \pi \hat{J}_y} \hat{P} ~.$$
We have 
\begin{eqnarray*}
e^{-i \pi \hat{J}_y} \left|J^P, \lambda \right> &=& \sum_{\lambda'} \mathcal{D}^{(J)}_{\lambda' \lambda}(0,\pi,0) \left|J^P, \lambda' \right> ~, \\
&=& (-1)^{J-\lambda} \left|J^P, -\lambda \right> ~,
\end{eqnarray*}
and this leads to
\begin{equation}
\hat{\Pi}_{xz} \left|\vec{p_z}; J^P, \lambda \right> = P(-1)^{J-\lambda} \left|\vec{p_z}; J^P, -\lambda \right> ~.
\label{equ:reflect_xz}
\end{equation}
This has the same form as Eq.~\ref{equ:reflect_yz} but with a different phase.  If we consider an arbitrary $\vec{p}$ and an arbitrary reflection in a plane containing $\vec{p}$, the result will be of the same form as Eqs.~\ref{equ:reflect_yz} and \ref{equ:reflect_xz} but in general there will be some $\lambda$-dependent phase.

It can be seen that for $\lambda = 0$ the helicity states are eigenstates of $\hat{\Pi}$ with eigenvalue $\tilde{\eta} \equiv P(-1)^J$.  Furthermore, because $\lambda=0$ there is no additional phase depending on the choice of reflection plane.  The $\tilde{\eta}$ `parity' is a good quantum number of the zero-helicity components at finite momentum and determines which little group irrep $\lambda=0$ subduces into (Table \ref{table:subductions}).  In addition, because $\tilde{\eta}$ is related to the transformation of helicity states with $\lambda > 0$ under reflections, $\tilde{\eta}$ also appears in the subduction coefficients for $\lambda > 0$.

\section{Helicity polarisation vectors and tensors}
\label{app:polvectors}

Here we construct a representation for the polarisation vectors of a spin-one helicity state and describe the generalisation to polarisation tensors.

Taking the defining equation for $\pol^{i}(\vec{p},\lambda)$ to be
\begin{equation}
\bigl< 0 \bigl| \mathcal{O}^{i} \bigr| \vec{p}; J=1, \lambda \bigr> = Z \pol^{i}(\vec{p},\lambda) ~,
\end{equation}
if follows that
\begin{equation}
\bigl< 0 \bigl| \hat{R}^{-1}  \hat{R} \mathcal{O}^{i}  \hat{R}^{-1}  \hat{R} \bigr| p_z; J=1, \lambda \bigr> = Z \pol^{i}(p_z,\lambda) ~,
\end{equation}
and so, from the transformation properties of vectors and helicity states under rotations,
\begin{equation}
(R^{-1})^{ij} \bigl< 0 \bigl| \mathcal{O}^{j} \bigr| \vec{p}; J=1, \lambda \bigr> = Z \pol^{i}(p_z,\lambda) ~,
\end{equation}
and therefore
\begin{equation}
\bigl< 0 \bigl| \mathcal{O}^{i} \bigr| \vec{p}; J=1, \lambda \bigr> = Z R^{ij} \pol^{j}(p_z,\lambda) ~.
\end{equation}
This implies that
\begin{equation}
\label{equ:pol1}
\pol^{i}(\vec{p},\lambda) = R^{ij} \pol^{j}(p_z,\lambda) ~,
\end{equation}
and equivalently, since $R^{ij}$ is real,
\begin{equation}
\label{equ:pol2}
\pol^{i*}(\vec{p},\lambda) = R^{ij} \pol^{j*}(p_z,\lambda) ~.
\end{equation}

An explicit representation for the polarisation vectors of a spin-one helicity state with an arbitrary momentum can be constructed using Eqs.~\ref{equ:pol1} and \ref{equ:pol2} and the standard representation for $\vec{p}=\vec{p_z}$,
\begin{eqnarray}
\pol_i(\vec{p_z},m=0)    &=& \left( 0, 0 , E/M \right)  ~, \nonumber \\ 
\pol_i(\vec{p_z},m=\pm1) &=& \mp \frac{1}{\sqrt{2}} \left( 1, \pm i , 0 \right) ~, 
\end{eqnarray}
where $M$ and $E$ are, respectively, the mass and energy of the state. 

Polarisation tensors for states of higher spin can be constructed in the usual way from direct products of the spin-one polarisation vectors with the helicities coupled together using standard $SU(2)$ Clebsch-Gordan coefficients.

\section{Lattice rotation and little group conventions}
\label{app:latticerotations}

In Table \ref{table:rotations} we give the specific rotations that we use for $R_{\text{ref}}$ (Section \ref{sec:finitevol}) which rotate from $(0,0,|\vec{p}|)$ to $\vec{p}_{\text{ref}}$.  We use the same convention as in Appendix \ref{app:helicity}, namely that a rotation $R_{\phi,\theta,\psi} = e^{-i \phi \hat{J}_z} e^{-i \theta \hat{J}_y} e^{-i \psi \hat{J}_z}$ rotates around the $z$-axis by $\psi$, then around the $y$-axis by $\theta$ and finally around the $z$-axis by $\phi$ (with a fixed coordinate system).  

\begin{table*}[b]
\begin{tabular}{|c|c|c|c|c|}
\textbf{Little Group} & \textbf{$\vec{p}_{\text{ref}}$} & \textbf{$\phi$} & \textbf{$\theta$} & \textbf{$\psi$}  \\
\hline
$\Dic_4$ & $(0,0,n)$ & $0$ & $0$ & $0$ \\
$\Dic_2$ & $(0,n,n)$ & $\pi/2$ & $\pi/4$ & $-\pi/2$ \\
$\Dic_3$ & $(n,n,n)$ & $\pi/4$ & $\cos^{-1}(1/\sqrt{3})$ & $0$ \\
\hline
\end{tabular}
\caption{Rotations, $R_{\text{ref}}$, used, as described in the text.}
\label{table:rotations}
\end{table*}

In Tables \ref{table:rep_Dic4}, \ref{table:rep_Dic2} and \ref{table:rep_Dic3} we give a choice of representation matrices for, respectively, the little groups $\Dic_4$, $\Dic_2$ and $\Dic_3$.  For $\Dic_2$ and $\Dic_3$, the rotations and reflections refer to a coordinate system which has been transformed using $R_{\text{ref}}$, so that $\vec{p}$ defines the new $z$-axis.  These representations are used with the group theoretic projection formula (for example, see Appendix A of Ref.~\cite{Dudek:2010wm}) to calculate the subduction coefficients given in Table \ref{table:subductions}.

\begin{table*}[b]
\begin{tabular}{|c|c|c|c|c|c|c|c|c|}
\textbf{Irrep} & \textbf{$I$} & \textbf{$R(\pi)$} & \textbf{$R(3\pi/2)$} & \textbf{$R(\pi/2)$}
& \textbf{$\Pi$} & \textbf{$R(\pi)\Pi$} & \textbf{$R(\pi/2)\Pi$} & \textbf{$R(3\pi/2)\Pi$} \\ 
\hline
$A_1$ & 1 & 1 & 1 & 1 & 1 & 1 & 1 & 1 \\
$A_2$ & 1 & 1 & 1 & 1 & -1 & -1 & -1 & -1 \\
$E_2$ &
$\left(
\begin{smallmatrix}
 1 & 0 \\
 0 & 1
\end{smallmatrix}
\right)$ & 

$\left(
\begin{smallmatrix}
 -1 & 0 \\
 0 & -1
\end{smallmatrix}
\right)$ & 

$\left(
\begin{smallmatrix}
 0 & -i \\
 -i & 0
\end{smallmatrix}
\right)$ & 

$\left(
\begin{smallmatrix}
 0 & i \\
 i & 0
\end{smallmatrix}
\right)$ & 

$\left(
\begin{smallmatrix}
 1 & 0 \\
 0 & -1
\end{smallmatrix}
\right)$ & 

$\left(
\begin{smallmatrix}
 -1 & 0 \\
 0 & 1
\end{smallmatrix}
\right)$ & 

$\left(
\begin{smallmatrix}
 0 & -i \\
 i & 0
\end{smallmatrix}
\right)$ & 

$\left(
\begin{smallmatrix}
 0 & i \\
 -i & 0
\end{smallmatrix}
\right)$ \\
$B_1$ & 1 & 1 & -1 & -1 & 1 & 1 & -1 & -1 \\
$B_2$ & 1 & 1 & -1 & -1 & -1 & -1 & 1 & 1 \\
\hline
\end{tabular}
\caption{Choice of representation matrices for the $\Dic_4$ little group. $I$ denotes the identify transformation, $R(\phi)$ denotes a rotation around the $z$-axis by $\phi$ and $\Pi$ denotes a reflection in the $yz$ plane ($x \rightarrow -x$).}
\label{table:rep_Dic4}
\end{table*}

\begin{table*}[b]
\begin{tabular}{|c|c|c|c|c|}
\textbf{Irrep} & \textbf{$I$} & \textbf{$R(\pi)$} & \textbf{$\Pi$} & \textbf{$R(\pi)\Pi$} \\
\hline
$A_1$ & 1 & 1 & 1 & 1 \\
$A_2$ & 1 & 1 & -1 & -1 \\
$B_1$ & 1 & -1 & 1 & -1 \\
$B_2$ & 1 & -1 & -1 & 1 \\
\hline
\end{tabular}
\caption{As Table \ref{table:rep_Dic4} but for the $\Dic_2$ little group.  With these representations for $B_1$ and $B_2$ we are able to reproduce the multiplication table given in Ref.~\cite{Moore:2006ng}, but we note that there is the possibility of another convention where $B_1$ and $B_2$ are interchanged.}
\label{table:rep_Dic2}
\end{table*}

\begin{table*}[b]
\begin{tabular}{|c|c|c|c|c|c|c|}
\textbf{Irrep} & \textbf{$I$} & \textbf{$R(2\pi/3)$} & \textbf{$R(-2\pi/3)$} 
& \textbf{$R(\pi)\Pi$} & \textbf{$R(\pi/3)\Pi$} & \textbf{$R(5\pi/3)\Pi$} \\ 
\hline
$A_1$ & 1 & 1 & 1 & 1 & 1 & 1 \\
$A_2$ & 1 & 1 & 1 & -1 & -1 & -1 \\
$E_2$ &
$\left(
\begin{smallmatrix}
 1 & 0 \\
 0 & 1
\end{smallmatrix}
\right)$ & 

$\left(
\begin{smallmatrix}
 -\frac{1}{2} & \frac{i \sqrt{3}}{2} \\
 \frac{i \sqrt{3}}{2} & -\frac{1}{2}
\end{smallmatrix}
\right)$ & 

$\left(
\begin{smallmatrix}
 -\frac{1}{2} & -\frac{i \sqrt{3}}{2} \\
 -\frac{i \sqrt{3}}{2} & -\frac{1}{2}
\end{smallmatrix}
\right)$ & 

$\left(
\begin{smallmatrix}
 -1 & 0 \\
 0 & 1
\end{smallmatrix}
\right)$ & 

$\left(
\begin{smallmatrix}
 \frac{1}{2} & -\frac{i \sqrt{3}}{2} \\
 \frac{i \sqrt{3}}{2} & -\frac{1}{2}
\end{smallmatrix}
\right)$ & 

$\left(
\begin{smallmatrix}
 \frac{1}{2} & \frac{i \sqrt{3}}{2} \\
 -\frac{i \sqrt{3}}{2} & -\frac{1}{2}
\end{smallmatrix}
\right)$ \\
\hline
\end{tabular}
\caption{As Table \ref{table:rep_Dic4} but for the $\Dic_3$ little group.}
\label{table:rep_Dic3}
\end{table*}

\section{Tables of general parameterisations}
\label{app:decompositions}

Here we give general parameterisations for the infinite volume continuum overlaps of states with fermion bilinear operators using constraints of Lorentz or 3-rotation symmetry.  Tables \ref{table:paramoverlaps:0}, \ref{table:paramoverlaps:1} and \ref{table:paramoverlaps:2} give parameterisations for operators with zero, one and two vector indices respectively.  The columns labelled ``Lorentz'' give Lorentz covariant parameterisations for the overlaps, i.e. using Lorentz covariance constraints applied to our operators built from 3-vectors.  These are derived from the Lorentz covariant parameterisations given in Appendix A of Ref.~\cite{Dudek:2007wv} by setting the Lorentz indices to specific values.  The columns labelled ``3-Rotation'' give the general 3-rotation covariant parameterisations.

In Tables \ref{table:overlaps:0}-\ref{table:overlaps:2b} we give the overlap of states onto the operators $\mathbb{O}^{J,\lambda}$ defined in Eq.~\ref{equ:helicityops}.  These overlaps follow from Tables \ref{table:paramoverlaps:0}, \ref{table:paramoverlaps:1} and \ref{table:paramoverlaps:2} once the vector indices are transformed to a circular basis and then coupled together to give a definite spin as described in Section \ref{sec:helops}.  We only show overlaps where the helicity of the operator is equal to that of the state; the other overlaps are zero.

The tables give overlaps for an operator of one parity, determined from the parity of the state with which it overlaps at zero momentum.  To use the tables for operators of the opposite parity, all the state parities must be flipped; we give examples of this below.  The inclusion of a $\gamma_5$ matrix in the operator does not change the allowed overlaps, except that the parity of the states may flip as discussed below.

Note that the parameterisations are written in terms of helicity polarisation tensors, discussed in Appendix \ref{app:polvectors}, \emph{not} polarisation tensors defined in terms of the $z$-component of spin.  Repeated indices are summed over.  Although these are all fermion bilinear operators, for brevity we omit the $\bar{\psi}$ and $\psi$ in the tables.  

As an example, consider the operator $\bar{\psi} \gamma^0 \psi$ which at zero momentum overlaps only onto the $J^{PC}=0^{+-}$ state.  The Lorentz covariant parameterisations are \cite{Dudek:2007wv}
$$\bigl< 0 \bigl| \bar{\psi} \gamma^{\mu} \psi \bigr| 0^{+-}(\vec{p}) \bigr> = Z p^{\mu} ~, $$
and
$$\bigl< 0 \bigl| \bar{\psi} \gamma^{\mu} \psi \bigr| 1^{--}(\vec{p},\lambda) \bigr> = Z \pol^{\mu}(\vec{p},\lambda) ~. $$
Setting $\mu = 0$ these give
$$\bigl< 0 \bigl| \bar{\psi} \gamma^{\mu=0} \psi \bigr| 0^{+-}(\vec{p}) \bigr> = Z p^0 , $$
and
$$\bigl< 0 \bigl| \bar{\psi} \gamma^{\mu=0} \psi \bigr| 1^{--}(\vec{p},\lambda) \bigr> = Z \pol^0(\vec{p},\lambda) = Z \pol^i(\vec{p},\lambda) p^i/E , $$
where $E=p^0$ is the energy of the state and in the last line we have used $p^{\mu} \pol_{\mu}(\vec{p},\lambda) = 0$.  These two parameterisations are given in the ``L2'' column of Table \ref{table:paramoverlaps:0} where we have absorbed factors of $E$ into the arbitrary constants.

As another example, consider the operator $\bar{\psi} \gamma^5 \gamma^0 \psi$.  This is the same as the operator considered previously apart from the additional $\gamma^5$, and at zero momentum it overlaps only onto the $J^{PC}=0^{-+}$ state.  In order to use Table \ref{table:paramoverlaps:0} we must flip the parities of all the states and so the parameterisations become
$$\bigl< 0 \bigl| \bar{\psi} \gamma^5 \gamma^{\mu} \psi \bigr| 0^{-+}(\vec{p}) \bigr> = Z p^0 , $$
and
$$\bigl< 0 \bigl| \bar{\psi} \gamma^5 \gamma^{\mu} \psi \bigr| 1^{++}(\vec{p},\lambda) \bigr> = Z \pol^0(\vec{p},\lambda) = Z \pol^i(\vec{p},\lambda) p^i/E . $$

\begin{table*}[p]
\begin{tabular}{|c|c|c|c|}
\textbf{State} & \multicolumn{2}{|c|}{\textbf{Lorentz}} & \textbf{3-Rotation}  \\
$J^P$ & \textbf{$I$ (L1)} & \textbf{$\gamma^0$ (L2)} & \textbf{$I$ or $\gamma^0$} \\
\hline
$0^+$ & $Z_1$ & $Z_1$ & $Z_1$ \\
$1^-$ & -- & $Z_1 \pol_i p_i$ & $Z_1 \pol_i p_i$ \\
$2^+$ & -- & -- & $Z_1 \pol_{ii} + Z_2 \pol_{ij} p_i p_j$ \\
$3^-$ & -- & -- & $Z_1 \pol_{iij}p_j + Z_2 \pol_{ijk} p_i p_j p_k$ \\
$4^+$ & -- & -- & $Z_1 \pol_{iijj} + Z_2 \pol_{iijk} p_j p_k + Z_3 \pol_{ijkl} p_i p_j p_k p_l$ \\
\hline
\end{tabular}
\caption{General Lorentz covariant and 3-rotation covariant parameterisations for the infinite volume continuum overlaps of operators with \emph{zero vector indices} onto states up to spin-four.  $Z_i$ are arbitrary constants, $p_i$ is the 3-momentum, for brevity we omit the arguments of the polarisation tensors $\pol^{(J)}(\vec{p},\lambda)$, and repeated indices are summed over.  The parities given in the first column may all have an overall sign flip depending on which operator is considered: the parity of the state at rest overlapping with the operator must be determined and then the relative parities of the other states can be read off the table.  More explanation is given in the text.}
\label{table:paramoverlaps:0}
\end{table*}

\begin{table*}[p]
\begin{tabular}{|c|c|c|c|c|}
\textbf{State} & \multicolumn{3}{|c|}{\textbf{Lorentz}} & \textbf{3-Rotation}  \\
$J^P$ & \textbf{$O^i$ (L1)} & \textbf{$\gamma^0 D^i$ (L2)} & \textbf{$\gamma^0 \gamma^i$ (L3)} & \textbf{$O^i$, $\gamma^0 D^i$ or $\gamma^0 \gamma^i$} \\
\hline
$0^+$ & $Z_1 p^i$ & $Z_1 p^i$ & -- & $Z_1 p^i$ \\
$0^-$ & -- & -- & -- & -- \\
\hline
$1^+$ & -- & $Z_1 \epsilon^{ijk} p_j \pol_k$ & $Z_1 \epsilon^{ijk} p_j \pol_k$ & $Z_1 \epsilon^{ijk}p_j\pol_k$ \\
$1^-$ & $Z_1 \pol^i$ & $Z_{\pm} (\pol^0 p^i \pm \pol^i p^0 )$ & $Z_- ( \pol^0 p^i -  \pol^i p^0 )$ & $Z_1 \pol^i + Z_2 (\pol_j p_j) p^i$ \\
\hline
$2^+$ & -- & $Z_1 \pol^{ij}p_j$ & -- & $Z_1 \pol^{ij}p_j + Z_2 p^i \pol^{jj} + Z_3 p^i \pol^{jk} p_j p_k$ \\
$2^-$ & -- & -- & -- & $Z_1 \epsilon^{ijk}p_j \pol_{kl} p_l$ \\
\hline
$3^+$ & -- & -- & -- & $Z_1 \epsilon^{ijk}p_j \pol_{kll} + Z_2 \epsilon^{ijk}p_j \pol_{klm} p_l p_m $ \\
$3^-$ & -- & -- & -- & $Z_1 p^i \pol^{jjk}p_k + Z_2 p^i \pol^{jkl} p_j p_k p_l + Z_3 \pol^{ikl} p_k p_l + Z_4 \pol^{ijj}$ \\
\hline
$4^+$ & -- & -- & -- & $Z_1 p^i \pol^{jjkk} + Z_2 p^i \pol^{jjkl} p_k p_l + Z_3 p^i \pol^{jklm} p_j p_k p_l p_m + Z_4 \pol^{ijjk} p_k + Z_5 \pol^{ijkl} p_j p_k p_l$ \\
$4^-$ & -- & -- & -- & $Z_1 \epsilon^{ijk}p_j \pol_{kllm} p_m + Z_2 \epsilon^{ijk}p_j \pol_{klmn} p_l p_m p_n $ \\
\hline
\end{tabular}
\caption{As Table \ref{table:paramoverlaps:0} but for operators with \emph{one vector index}.  $O^i$ stands for $\gamma^i$ or $D^i$.  $\epsilon^{ijk}$ is the antisymmetric tensor.  Where $Z_{i\pm}$ appears in the decomposition it means that both $+$ and $-$ terms are allowed.}
\label{table:paramoverlaps:1}
\end{table*}

\begin{table*}[p]
\begin{tabular}{|c|c|c|c|c|}
\textbf{State} & \multicolumn{3}{|c|}{\textbf{Lorentz}} & \textbf{3-Rotation}  \\
$J^P$ & \textbf{$O^{ij}$ (L1)} & \textbf{$\gamma^0 D^i D^j$ (L2)} & \textbf{$\gamma^0 \gamma^i D^j$ (L3)} & \textbf{$O^{ij}$, $\gamma^0 D^i D^j$ or $\gamma^0 \gamma^i D^j$}  \\
\hline
$0^+$ & $Z_1 \delta^{ij} + Z_2 p^i p^j$ & $Z_1 \delta^{ij} + Z_2 p^i p^j$ & $Z_1 \delta^{ij}$ &
  $Z_1 \delta^{ij} + Z_2 p^i p^j$ \\
\hline
$0^-$ & -- & $Z_1 \epsilon^{ijk} p_k$ & $Z_1 \epsilon^{ijk} p_k$ &
  $Z_1 \epsilon^{ijk}p_k$ \\
\hline
$1^+$ & $Z_- \epsilon^{ijk} ( p_k \pol_0 -  p_0 \pol_k) $ & \begin{tabular}{c}$Z_1 \epsilon^{ikl}p^j p_k \pol_l + Z_2 \epsilon^{ijk}\pol_k$ \\ $+ Z_{\pm}(\epsilon^{jkl}p^ip_k\pol_l \pm \epsilon^{ijk}p_k \pol^l p^l$ \\ $\mp \epsilon^{ijk}\pol_k p^0 p^0)$\end{tabular} & 
  \begin{tabular}{c}$Z_1, Z_2, Z_-$ \\ terms \\ from ``L2''\end{tabular} &
  $Z_{1\pm} (\epsilon^{ikl} p_k \pol_l p^j \pm \epsilon^{jkl} p_k \pol_l p^i) + Z_2 \epsilon^{ijk} \pol_k + Z_3 \epsilon^{ijk} p_k \pol_l p_l$ \\
\hline
$1^-$ & $Z_{1\pm} (\pol^i p^j \pm \pol^j p^i)$ & \begin{tabular}{c}$Z_3 \delta^{ij} \pol^k p_k + Z_4 p^i \pol^j$ \\ $+Z_{\pm}(p^0p^0 p^j \pol^i \pm p^i p^j \pol^kp_k)$\end{tabular} & 
  \begin{tabular}{c}$Z_3,Z_-$ \\ terms \\ from ``L2''\end{tabular} &
  $Z_{1\pm} (\pol^i p^j \pm \pol^j p^i) + Z_2 p^i p^j \pol_k p_k + Z_3 \delta^{ij} \pol_k p_k$ \\
\hline
$2^+$ & $Z_1 \pol^{ij}$ & \begin{tabular}{c}$Z_8 \pol^{ik}p^jp_k$ \\ $+ Z_{\pm}(\pol^{jk} p^i p_k \pm \pol^{ij}p^0p^0)$\end{tabular} & 
  \begin{tabular}{c} $Z_-$ \\ term \\ from ``L2''\end{tabular} & 
  \begin{tabular}{c}$Z_1 \pol^{ij} + Z_2 p^i p^j \pol^{kk} + Z_{3\pm}(\pol^{ik}p^j \pm \pol^{jk}p^i)p_k + Z_4 \delta^{ij} \pol^{kk}$ \\ $+ Z_5 \delta^{ij} \pol^{kl} p_k p_l +Z_6 p^i p^j \pol^{kl} p_k p_l + Z_7 \epsilon^{ikl} \epsilon^{jmn} p_k p_m \pol_{ln}$\end{tabular} \\
\hline
$2^-$ & -- & \begin{tabular}{c}$Z_6 \epsilon^{ikl}p_k\pol^{jl}$ \\ $+ Z_{\pm}(-\epsilon^{jkl}p_k\pol^{il}p^0p^0 $ \\ $ \pm \epsilon^{ijk}\pol^{kl}p_lp^0p^0$ \\ $\mp \epsilon^{ijk}p_k \pol^{lm}p_lp_m)$\end{tabular} & 
  \begin{tabular}{c}$Z_6,Z_-$ \\ terms \\ from ``L2''\end{tabular} &
  \begin{tabular}{c}$Z_{1\pm}(\epsilon^{ikl}p_k \pol^{jl} \pm \epsilon^{jkl}p_k \pol^{il})$ \\ $ + Z_{2\pm}(\epsilon^{ikl}p_k p^j \pol^{lm}p_m \pm \epsilon^{jkl}p_k p^i \pol^{lm}p_m)$ \\ $ + Z_3 \epsilon^{ijk} \pol^{kl} p_l + Z_4 \epsilon^{ijk}p_k \pol^{ll} + Z_5 \epsilon^{ijk}p_k \pol^{lm} p_l p_m$ \end{tabular}  \\
\hline
$3^+$ & -- & -- & -- & 
  \begin{tabular}{c}$Z_{1\pm}(\epsilon^{ikl} \pol^{jlm} \pm \epsilon^{jkl} \pol^{ilm})p_k p_m + Z_{2\pm}(\epsilon^{ikl} p^j \pm \epsilon^{jkl} p^i) p_k \pol^{lmm}$ \\ $ + Z_{3\pm}(\epsilon^{ikl} p^j \pm \epsilon^{jkl} p^i) \pol^{lmn} p_k p_m p_n + Z_4 \epsilon^{ijk} \pol^{kll}$ \\ $ + Z_5 \epsilon^{ijk} \pol^{klm} p_l p_m + Z_6 \epsilon^{ijk}p_k \pol^{lmn} p_l p_m p_n + Z_7 \epsilon^{ijk}p_k \pol^{lmm} p_l$ \end{tabular} \\
\hline
$3^-$ & -- & $Z_1 \pol^{ijk}p_k$ & -- & 
  \begin{tabular}{c}$Z_1 \pol^{ijk}p_k + Z_{2\pm}(\pol^{ikk}p^j \pm \pol^{jkk}p^i) + Z_{3\pm}(\pol^{ikl}p^j \pm \pol^{jkl}p^i)p_kp_l$ \\ $ + Z_4 p^i p^j \pol^{klm}p_kp_lp_m + Z_5 p^i p^j \pol^{kll} p_k + Z_6 \delta^{ij} \pol^{klm}p_kp_lp_m$ \\ $ + Z_7 \delta^{ij} \pol^{kll}p_k + Z_8 \epsilon^{ikl} \epsilon^{jmn} \pol^{lnr} p_k p_m p_r$\end{tabular}  \\
\hline
\end{tabular}
\caption{As Table \ref{table:paramoverlaps:0} but for operators with \emph{two vector indices} and states up to spin-three.  $O^{ij}$ stands for $\gamma^i D^j$ or $D^i D^j$.}
\label{table:paramoverlaps:2}
\end{table*}

\begin{table*}[p]
\renewcommand{\arraystretch}{2}
\begin{tabular}{|c|c|c|c|c|}
\textbf{Operator} & \textbf{State} $J^P(\lambda=0)$ & \textbf{3-Rotation} & $I$ (\textbf{L1}) & $\gamma^0$ (\textbf{L2}) \\
\hline
$\mathbb{O}^{0,0}$ & $0^+$ & $Z_1$ & $Z_1$ & $Z_1$ \\
\hline
$\mathbb{O}^{0,0}$ & $1^-$ & $Z_1 \frac{pE}{m}$ & -- & $Z_1$ \\
\hline
$\mathbb{O}^{0,0}$ & $2^+$ & $\sqrt{\frac{2}{3}}\frac{p^2}{M^2} \left[ Z_1 + E^2 Z_2\right]$ & -- & -- \\ 
\hline
$\mathbb{O}^{0,0}$ & $3^-$ & $\sqrt{\frac{2}{5}}\frac{p^3E}{M^3} \left[ Z_1 + E^2 Z_2\right]$ & -- & -- \\
\hline
$\mathbb{O}^{0,0}$ & $4^+$ & $2\sqrt{\frac{2}{35}}\frac{p^4}{M^4} \left[ Z_1 + E^2 Z_2 + E^4 Z_3 \right]$ & -- & -- \\ 
\hline
\end{tabular}
\caption{Overlaps of states (up to spin-four) with the operators $\mathbb{O}^{J_{\text{op}},\lambda}$ defined in Eq.~\ref{equ:helicityops}, where these operators have been constructed from $I$ or $\gamma^0$ and using the parameterisations in Table \ref{table:paramoverlaps:0}.  As in Table \ref{table:paramoverlaps:0}, ``L1'' and ``L2'' correspond to the Lorentz covariance constraints for operators constructed from, respectively, $I$ and $\gamma^0$.  The $Z_i$ in the ``L1'' and ``L2'' columns give the terms from the ``3-Rotation'' column which are allowed under these Lorentz constraints.  Only the overlaps onto the zero helicity component of the states are shown, the overlaps with non-zero helicity are zero.  $p^2 \equiv |\vec{p}|^2$ is the square of the 3-momentum, and $M$ and $E=p^0$ are, respectively, the mass and energy of the state.}
\label{table:overlaps:0}
\end{table*}

\begin{table*}[p]
\renewcommand{\arraystretch}{2}
\begin{tabular}{|c|c|c|c|c|c|}
\textbf{Operator} & \textbf{State} $J^P$ & \textbf{3-Rotation} &  \textbf{$O^i$ (L1)} & \textbf{$\gamma^0 D^i$ (L2)} & \textbf{$\gamma^0 \gamma^i$ (L3)} \\
\hline
$\mathbb{O}^{1,\lambda}$ & $0^+$ & $\Bigl[0 ,~ ip Z_1 ,~ 0\Bigr]$ & $Z_1$ & $Z_1$ & -- \\ 
\hline
$\mathbb{O}^{1,\lambda}$ & $0^-$ & -- & -- & -- & -- \\
\hline
$\mathbb{O}^{1,\lambda}$ & $1^+$ & $\Bigl[p Z_1 ,~ 0 ,~ -p Z_1\Bigr]$ & -- & $Z_1$ & $Z_1$ \\
\hline
$\mathbb{O}^{1,\lambda}$ & $1^-$ & $\Bigl[iZ_1 ,~ i\frac{E}{M}Z_1 + i\frac{p^2E}{M} Z_2 ,~ iZ_1\Bigr]$ & $Z_1$ & $Z_1, Z_2$ & $Z_1 = -E^2 Z_2$ \\
\hline
$\mathbb{O}^{1,\lambda}$ & $2^+$ & $\Bigl[\frac{ipE}{\sqrt{2}M} Z_1,~ \sqrt{\frac{2}{3}}\frac{ip}{M^2}\left(E^2 Z_1 + p^2Z_2 + p^2E^2 Z_3 \right) ,~ \frac{ipE}{\sqrt{2}M} Z_1\Bigr]$ & -- & $Z_1$ & -- \\
\hline
$\mathbb{O}^{1,\lambda}$ & $2^-$ & $\bigl[1,0,-1\bigr] \left( \frac{p^2E}{\sqrt{2}M} \right) Z_1$ & -- & -- & -- \\ 
\hline
$\mathbb{O}^{1,\lambda}$ & $3^+$ & $\bigl[1,0,-1\bigr] \left( \frac{2p^3}{\sqrt{15}M^2} \right) \left( Z_1 + E^2Z_2 \right)$ & -- & -- & -- \\ 
\hline
$\mathbb{O}^{1,\lambda}$ & $3^-$ & \begin{tabular}{c}$(\frac{i p^2}{\sqrt{5}M^2})\Bigl[ \frac{2}{\sqrt{3}}\left(E^2 Z_3 + Z_4\right), $ \\ [-1.5ex] $\frac{\sqrt{2}E}{M} \left( p^2Z_1 + E^2 p^2 Z_2 + E^2 Z_3 + Z_4 \right), $ \\ [-1.5ex] $ \frac{2}{\sqrt{3}}\left(E^2 Z_3 + Z_4\right) \Bigr]$ \end{tabular} & -- & -- & -- \\
\hline
\end{tabular}
\caption{Overlaps of states (up to spin-three) with the operators $\mathbb{O}^{J_{\text{op}},\lambda}$ defined in Eq.~\ref{equ:helicityops}, where these operators have been \emph{constructed from operators with one vector index} and using the parameterisations in Table \ref{table:paramoverlaps:1}.  ``L1'', ``L2'' and ``L3'' correspond to the Lorentz covariance constraints for the operators in Table \ref{table:paramoverlaps:1}; the $Z_i$ in these columns refer to the terms in the ``3-Rotation'' column which are allowed under these Lorentz constraints.  Only the overlaps where the helicity of the state is the same as that of the operator are shown; the others are zero.  The basis is $\lambda = -1, 0, +1$.}
\label{table:overlaps:1}
\end{table*}

\begin{table*}[p]
\renewcommand{\arraystretch}{2}
\begin{tabular}{|c|c|c|c|c|c|}
\textbf{Operator} & \textbf{State} $J^P$ & \textbf{3-Rotation} & \textbf{$O^{ij}$ (L1)} & \textbf{$\gamma^0 D^i D^j$ (L2)} & \textbf{$\gamma^0 \gamma^i D^j$ (L3)} \\
\hline
$\mathbb{O}^{0,0}$ & $0^+$ & $\frac{1}{\sqrt{3}} (3 Z_1 + p^2 Z_2)$ &
  $Z_1, Z_2$ & $Z_1, Z_2$ & $Z_1$ \\
\hline
$\mathbb{O}^{0,0}$ & $1^-$ & $\frac{E p}{M \sqrt{3}} (2 Z_{1+} + p^2 Z_2 + 3 Z_3)$ &
  $Z_{1+}$ & $Z_{1+}, Z_2, Z_3$ & \begin{tabular}{c}$Z_3$, \\ [-1.5ex] $2 Z_{1+} = -E^2 Z_2$\end{tabular} \\
\hline
$\mathbb{O}^{0,0}$ & $2^+$ & \begin{tabular}{c}$\frac{p^2\sqrt{2}}{3M^2} \Bigl[Z_1 + p^2Z_2 + 2E^2Z_{3+}$ \\ [-1.5ex] $ + 3Z_4 + 3E^2Z_5 + E^2p^2Z_6 - M^2Z_7 \Bigr]$\end{tabular} &
  $Z_1$ & $Z_1, Z_{3+}$ & $ 2 Z_{3+} E^2 = - Z_1$ \\
\hline
$\mathbb{O}^{0,0}$ & $3^-$ & \begin{tabular}{c}$\sqrt{\frac{2}{15}}\frac{Ep^3}{M^3} \Bigl[Z_1 + 2Z_{2+} + 2E^2Z_{3+} + E^2p^2Z_4$ \\ [-1.5ex] $ + p^2Z_5 + 3E^2Z_6 + 3Z_7 - M^2 Z_8 \Bigr]$\end{tabular} & 
  -- & $Z_1$ & -- \\
\hline
\hline
$\mathbb{O}^{1,\lambda}$ & $0^+$ & -- &
  -- & -- & -- \\
\hline
$\mathbb{O}^{1,\lambda}$ & $0^-$ & $\bigl[0, i\sqrt{2}p Z_1, 0\bigr]$ &
  -- & $Z_1$ & $Z_1$ \\
\hline
$\mathbb{O}^{1,\lambda}$ & $1^+$ & $i\sqrt{2} \Bigl[-(p^2Z_{1-} - Z_2), \frac{E}{M}(Z_2 + p^2Z_3),-(p^2Z_{1-} - Z_2) \Bigr]$ &
  $Z_2 = -E^2 Z_3$ & $Z_{1-}, Z_2, Z_3$ & $Z_{1-}, Z_2, Z_3$ \\
\hline
$\mathbb{O}^{1,\lambda}$ & $1^-$ & $\sqrt{2} p Z_{1-} \bigl[1, 0, -1 \bigr]$ &
  $Z_{1-}$ & $Z_{1-}$ & $Z_{1-}$ \\
\hline
$\mathbb{O}^{1,\lambda}$ & $2^+$ & $\frac{E p^2}{M} Z_{3-} \bigl[1, 0, -1 \bigr]$ &
  -- & $Z_{3-}$ & $Z_{3-}$ \\
\hline
$\mathbb{O}^{1,\lambda}$ & $2^-$ & \begin{tabular}{c}$\frac{ip}{M} \Bigl[ -E\left(Z_{1-} + p^2Z_{2-} - Z_3 \right), $ \\ [-1.5ex] $\frac{2}{\sqrt{3}M} \bigl( -M^2Z_{1-} + p^2Z_4 + E^2Z_3+E^2p^2Z_5 \bigr), $ \\ [-1.5ex] $ -E\left(Z_{1-} + p^2Z_{2-} - Z_3 \right) \Bigr]$ \end{tabular} &
  -- & \begin{tabular}{c}$Z_{1-}$, \\ [-1.5ex] $Z_3 = -E^2 Z_5$\end{tabular} & \begin{tabular}{c} $Z_{1-}$, \\ [-1.5ex] $Z_3 = -E^2 Z_5$\end{tabular} \\
\hline
$\mathbb{O}^{1,\lambda}$ & $3^+$ & \begin{tabular}{c}$\frac{2i p^2}{\sqrt{5}M^2} \Bigl[ -\sqrt{\frac{2}{3}}\bigl( p^2 Z_{2-} - Z_4 + E^2p^2Z_{3-} + E^2 Z_{1-} - E^2 Z_5\bigr), $ \\ [-1.5ex] $\frac{E}{M} \bigl(-M^2 Z_{1-} + Z_4 + p^2 Z_7 + E^2 Z_5 + E^2 p^2 Z_6 \bigr), $ \\ [-1.5ex] $ -\sqrt{\frac{2}{3}}\bigl( p^2 Z_{2-} - Z_4 + E^2p^2Z_{3-} + E^2 Z_{1-} - E^2 Z_5 \bigr) \Bigr] $\end{tabular}  &
  -- & -- & -- \\
\hline
$\mathbb{O}^{1,\lambda}$ & $3^-$ & \begin{tabular}{c}$2\sqrt{\frac{2}{15}}\frac{p^3}{M^2} \bigl( Z_{2-} + E^2Z_{3-} \bigr) \bigl[ 1,0,-1 \bigr] $\end{tabular} &
  -- & -- & -- \\
\hline
\end{tabular}
\caption{Overlaps of states (up to spin-three) with the operators $\mathbb{O}^{J_{\text{op}},\lambda}$ defined in Eq.~\ref{equ:helicityops}, where these operators have been \emph{constructed from operators with two vector indices} and using the parameterisations in Table \ref{table:paramoverlaps:2}.   ``L1'', ``L2'' and ``L3'' correspond to the Lorentz covariance constraints for the operators in Table \ref{table:paramoverlaps:2}; the $Z_i$ in these columns refer to the terms in the ``3-Rotation'' column which are allowed under these Lorentz constraints.  Only the overlaps where the helicity of the state is the same as that of the operator are shown; the others are zero.  The basis is $\lambda = -J_{\text{op}}, \dots, +J_{\text{op}}$.  Continued in Table \ref{table:overlaps:2b}.}
\label{table:overlaps:2}
\end{table*}

\begin{table*}[p]
\renewcommand{\arraystretch}{2}
\begin{tabular}{|c|c|c|c|c|c|}
\textbf{Operator} & \textbf{State} $J^P$ & \textbf{3-Rotation} & \textbf{$O^{ij}$ (L1)} & \textbf{$\gamma^0 D^i D^j$ (L2)} & \textbf{$\gamma^0 \gamma^i D^j$ (L3)} \\
\hline
$\mathbb{O}^{2,\lambda}$ & $0^+$ & $\bigl[0,0, -\sqrt{\frac{2}{3}}p^2Z_2, 0,0 \bigr]$ &
  $Z_2$ & $Z_2$ & -- \\
\hline
$\mathbb{O}^{2,\lambda}$ & $0^-$ & -- &
  -- & -- & -- \\
\hline
$\mathbb{O}^{2,\lambda}$ & $1^+$ & $i\sqrt{2}p^2 Z_{1+} \bigl[0,1,0,-1,0\bigr]$ &
  -- & $Z_{1+}$ & $Z_{1+}$ \\
\hline
$\mathbb{O}^{2,\lambda}$ & $1^-$ & $ -\sqrt{2}p \Bigl[ 0, Z_{1+},\frac{E}{M\sqrt{3}}\left(2Z_{1+} + p^2 Z_2\right),Z_{1+},0 \Bigr]$ &
  $Z_{1+}$ & $Z_{1+}, Z_2$ & $2Z_{1+} = -E^2 Z_2$ \\
\hline
$\mathbb{O}^{2,\lambda}$ & $2^+$ & \begin{tabular}{c}$\Bigl[ -Z_1+p^2Z_7, -\frac{E}{M}\left(Z_1+p^2Z_{3+}\right), $ \\ [-1.5ex] $-\frac{1}{3M^2}\Bigl\{(M^2+2E^2)Z_1 + 2p^4Z_2 + \Bigr.$ \\ [-1.5ex] $\Bigl. 4p^2E^2Z_{3+} + 2E^2p^4Z_6 + M^2p^2Z_7 \Bigr\}, $ \\ [-1.5ex] $-\frac{E}{M}\left(Z_1+p^2Z_{3+}\right), -Z_1+p^2Z_7 \Bigl]$\end{tabular} &
  $Z_1$ & $Z_1, Z_{3+}$ & $2Z_{3+} E^2 = -Z_1$ \\
\hline
$\mathbb{O}^{2,\lambda}$ & $2^-$ & \begin{tabular}{c}$ ip \Bigl[ 2Z_{1+}, \frac{E}{M}\left(Z_{1+}+p^2Z_{2+}\right), $ \\ [-1.5ex] $0, -\frac{E}{M}\left(Z_{1+}+p^2Z_{2+}\right), -2Z_{1+} \Bigr] $\end{tabular} &
  -- & $Z_{1+}$ & $Z_{1+}$ \\
\hline
$\mathbb{O}^{2,\lambda}$ & $3^+$ & \begin{tabular}{c}$\frac{2ip^2}{\sqrt{3}M}\Bigl[ E Z_{1+}, \sqrt{\frac{2}{5}}\frac{1}{M}\bigl\{E^2 Z_{1+} + p^2Z_{2+} + E^2p^2Z_{3+} \bigr\} , $ \\ [-1.0ex] $0, -\sqrt{\frac{2}{5}}\frac{1}{M}\bigl\{E^2 Z_{1+} + p^2Z_{2+} + E^2p^2Z_{3+} \bigr\}, -E Z_{1+}\Bigr]$\end{tabular} &
  -- & -- & -- \\
\hline
$\mathbb{O}^{2,\lambda}$ & $3^-$ & \begin{tabular}{c}$-\frac{p}{\sqrt{3}M}\Bigl[ E(Z_1-p^2Z_8), \frac{2}{M}\sqrt{\frac{2}{5}}\left\{E^2 Z_1 + p^2Z_{2+} + E^2p^2 Z_{3+}\right\}, $ \\ [-1.5ex] $\frac{E}{M^2}\sqrt{\frac{1}{5}}\bigl\{(2E^2+M^2)Z_1 + 4p^2Z_{2+} + 4E^2p^2 Z_{3+} $ \\ [-1.5ex] $ + 2E^2p^4 Z_4 + 2p^4Z_5 + M^2p^2Z_8\bigr\} , $ \\ [-1.5ex] $ \frac{2}{M}\sqrt{\frac{2}{5}}\left\{E^2 Z_1 + p^2Z_{2+} + E^2p^2 Z_{3+}\right\}, E(Z_1-p^2Z_8) \Bigl] $\end{tabular} &
  -- & $Z_1$ & -- \\
\hline
\end{tabular}
\caption{Continuation of Table \ref{table:overlaps:2}.}
\label{table:overlaps:2b}
\end{table*}


\section{Robustness of the extracted spectrum}
\label{app:tests}

\begin{figure}[tb]
\includegraphics[width=0.49\textwidth]{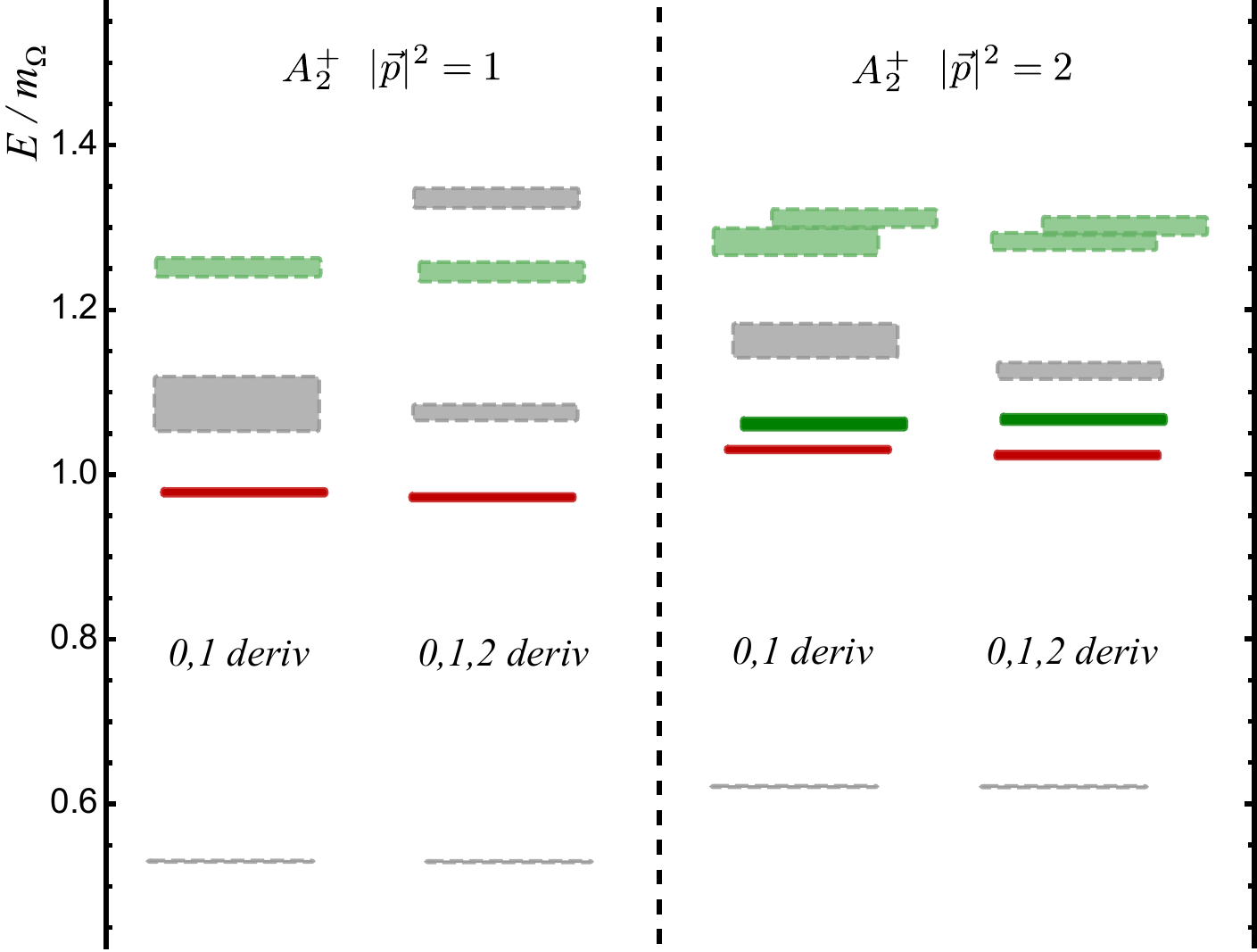}
\caption{Comparison of the lower-lying states extracted using zero and one-derivative operators, and with that basis supplemented with two-derivative operators, for the $A_2^+$ irrep with $|\vec{p}|^2=1$ ($\Dic_4$) (left panel) and $|\vec{p}|^2=2$ ($\Dic_2$) (right panel).  The box height shows the one sigma statistical uncertainty above and below the central value; the colour coding, indicating the $J^P$, is described in the text.}
\label{fig:2derivcomp}
\end{figure}

To check the robustness of our results to changing the operator basis, as well as using the full bases containing up to two-derivative operators, we also performed some analyses using only zero and one-derivative operators.  Fig.~\ref{fig:2derivcomp} shows the extracted energy levels in the $A_2^+$ irrep for $|\vec{p}|^2=1$ and $|\vec{p}|^2=2$.  All use 200 configurations, 3 time-sources and the correlators are averaged over 6 momentum directions, except for the one-derivative basis with $|\vec{p}|^2 = 2$ which is averaged over 12 momentum directions.  The one-derivative basis for this irrep contains 7 (10) operators with $|\vec{p}|^2=1$ ($|\vec{p}|^2=2$), whereas the two-derivative basis contains 20 (31) operators.  The colour coding indicates the identified $J^P$: black/grey for $J=0$, red for $J=1$ and green for $J=2$; darker shades with a solid outline correspond to positive parity and lighter shades with a dashed outline correspond to negative parity.  Orange boxes with a dotted outline indicate that the state's $J^P$ could not be unambiguously identified.

Including two-derivative operators significantly increases the size of the operator basis, reduces the statistical uncertainty on the higher extracted levels, and increases the number of states and the highest spin which can be reliably extracted (the latter point is not visible in these figures).  However, for the lower states there is no statistically significant change in the extracted energies and no significant increase in precision of the energies extracted, verifying the robustness of the energies we extract under a reduction in size of the operator basis.

\end{document}